\documentclass[fleqn,final]{unmpandathesis}
\pdfoutput=1
\usepackage[bookmarks = true, pdfpagemode = None, pdfstartview = FitH, colorlinks = true, urlcolor = blue,hyperfootnotes=false]{hyperref}
\usepackage{hypcap}
\usepackage{mathrsfs}
\usepackage{amsmath,bm}
\usepackage{amssymb}
\usepackage{amsthm}
\usepackage{graphicx}
\usepackage{pgf,pgfnodes,pgfarrows}
\usepackage{rotating}

\newcommand{\be}{\begin{equation}}
\newcommand{\ee}{\end{equation}}
\newcommand{\ben}{\begin{eqnarray}}
\newcommand{\een}{\end{eqnarray}}
\newcommand{\bF}{\begin{figure}}
\newcommand{\eF}{\end{figure}}
\newcommand{\dg}{\dagger}
\newcommand{\bes}{\begin{subequations}}
\newcommand{\ees}{\end{subequations}}
\newtheorem{theorem}{Theorem}
\newtheorem{lemma}{Lemma}

\def\spec{{\rm{Spec}}}
\newcommand{\proj}[1]{\mbox{$|#1\rangle \!\langle #1 |$}}
\newcommand{\avg}[1]{\langle #1 \rangle}

\def\ba{\mathbf{a}}
\def\bb{\mathbf{b}}
\def\bc{\mathbf{c}}
\def\bx{\mathbf{x}}
\def\mz{\mathbb{Z}}

\def\sN{\mathcal{N}}
\def\sM{\mathcal{M}}

\newcommand{\braket}[2]{\left\langle\left.#1\vphantom{#2}\right|#2\right\rangle}

\newcommand{\abs}[1]{\left\vert#1\right\vert}
\newcommand{\tr}{\text{tr}}

\definecolor{uglyBlue}{rgb}{.172,.188,.547}
\definecolor{uglyGreen}{rgb}{.07,.64,.3}

\input{Qcircuit}
\xyoption{curve}

\begin{document}

\frontmatter

\title{Studies on the Role of Entanglement in Mixed-state Quantum Computation}

\author{Animesh Datta}

\degreesubject{Ph.D., Physics}

\degree{Doctor of Philosophy \\ Physics}

\documenttype{Dissertation}

\previousdegrees{B.Tech., Electrical Engineering (Minor in
Physics),\\ Indian Institute of Technology, Kanpur 2003
}

\date{August, 2008}

\maketitle

\begin{dedication}
   To \emph{Maa}, for everything.
\end{dedication}

\begin{acknowledgments}

\renewcommand{\baselinestretch}{1.45}\selectfont

        The most significant character in any doctoral career is
that of the advisor. Carl Caves, my advisor, has been more.
Working with and under him has been a brilliant and fascinating
experience. His strategy of offering ample freedom in choosing
problems and the pace of solving them, and collaborating on them
has been his single greatest contribution to my research.
Encouraging one-liners like `The truth shall set you free' when
I've gotten negative results and confidence boosters like `A lot
of people say a lot of things' have been priceless; so has been
his dictum of writing papers with meticulous precision. His
ability to guide and advise has been phenomenal and a source of
constant inspiration.

        The contribution of the other members of the Information
Physics group have been no less. Both Ivan Deutsch and Andrew
Landahl have always been enthusiastic and encouraging of my
research, asking penetrating question in group meeting
presentations and discussing such topics on the outside. I also
thank them for being on my dissertation committee. Ivan is also to
be thanked for teaching fabulous courses on quantum optics and
atomic physics. I'm infinitely thankful to Anil Shaji for being my
surrogate advisor, with whom I have discussed almost all my
projects, and he's always been there to put me right and prevented
me from going awry so often. Other members of the group with whom
I've collaborated have taught me a lot as well. Steve Flammia,
with his random math problems and refusal ever to be frightened by
formidable mathematics has been a motivating experience.
Discussions with Sergio Boixo have also been fruitful.

        Other student members of the group have been no less
important. Seth has always been there to chat about a nagging
physics, math or programming problem. I also thank Matt, Collin,
Aaron, Pat, Kiran, Drew, Rene, Sohini, Brian, Carlos, Heather,
Brigette and Alex Tacla. Bryan Eastin and Iris have been crucial
for the random chats, on physics and often, otherwise.

        My research has benefited a lot from external collaborations and
academic visits. For these, I must thank Howard Barnum, who made
it possible for me to visit Los Alamos National Laboratory for a
summer and inspired me to think that no result was useless in
research and Michael Nielsen who sponsored a visit to the
University of Queensland. Once there, discussions with Guifre
Vidal, Michael Nielsen, Andrew Doherty, Mark DeBurgh, Nick
Menicucci, Sukhwinder Singh and others germinated ideas that have
shaped my research. Chats with W. H. Zurek, Lorenza Viola, Rolando
Somma and Leonid Gurvits from LANL, Kavan Modi and Caesar
Rodriguez from UT, Austin and Alex Monras and Emilio Bagan from
Barcelona, Robert Raussendorf, Sev, Caterina and Tzu at Waterloo
have also been enlightening.

        I am most indebted to Prof.~K.~R.~Parthasarathy with whom
I first discussed quantum computation and Prof.~P.~Ghose with whom
I did my first research in quantum mechanics. Both are to be
thanked for the immense patience they showed towards me.

        At UNM, wonderful courses by Sudhakar Prasad on E\&M, Colston
Chandler on scattering theory, Paul Parris on non-equilibrium
statistical mechanics and Cris Moore on quantum algorithms have
been very valuable. Discussions on the nature and fate of life
with Denis Seletskiy, Mark Mero, Dave McInnis, Doug Bradshaw and
Birk have kept me going, as has the assistance of Mary DeWitt,
Roxanne Littlefield, Jennie Peer, Alisa Gibson, Russ and last but
not the least, the Mountain Dew vending machine.

        Beyond that, infinite thanks are due to Mukesh for always cooking
dinner, and Navin for eating it, and Mohit, Arnab, Shailesh,
Brittany, Hari and Naresh bhai, Tom, Lisa, Riti, Deepti and Divya
for wasting time with me.

        From still farther beyond, from India, I'm indebted to my mother for
her prolonged faith and support and Bubun for her love. Lastly,
thanks beyond words for the love, support and patience of my
Priceless, beloved Baby, without whom a lot of this would have
lost its meaning.

\end{acknowledgments}

\maketitleabstract

\begin{abstract}
            In this thesis, I look at the role of quantum
entanglement in mixed-state quantum computation. The model we
consider is the DQC1 or `power of one qubit' model. I show that
there is minimal bipartite entanglement in a typical instance of
the DQC1 circuit and even put an upper bound on the possible
amount of entanglement. The upper bound does not scale with the
size of the system, making it hard to attribute the exponential
speedup in the DQC1 model to quantum entanglement. On the other
hand, this limited amount of entanglement does not imply that the
system is classically simulatable. This goes against the dictum
that quantum evolutions involving states with limited amounts of
entanglement can be simulated efficiently classically. A matrix
product state (MPS) algorithm for a typical instance of the DQC1
system requires exponential classical resources. This exposes a
gap between the amount of entanglement and the amount of purely
nonclassical correlations in a quantum system.

            This gap, I suggest, can be filled by quantum discord.
An entropic measure of quantumness that tries to capture the
quantum notion of a system being disturbed by a measurement, it is
a true measure of purely nonclassical correlations. I calculate it
in a typical instance of the DQC1 circuit and find that the amount
of discord is a constant fraction of the maximum possible discord
for a system of that size. This allows an interpretation of
quantum discord as the resource that drives mixed-state quantum
computation. I also study quantum discord as a quantity of
independent interest. Its role in the phenomenon of entanglement
distribution is studied through an easily comprehensible example.

            This thesis also contains discussions on the relation
between the complexity classes P, BQP and DQC1. Additional
material is presented on the connections between the DQC1 model,
Jones polynomials and statistical mechanics. The thesis concludes
with a discussion of a few open problems related to the DQC1
model, the quantum discord and their scope in quantum information
science.

 \clearpage 
\end{abstract}

\tableofcontents \listoffigures

\mainmatter

\chapter{Introduction}

\hfill \textsl{All great deeds, and all great thoughts have a
ridiculous beginning. Great works are often born on a street
corner or in a restaurant's revolving door.}

\hfill  - Albert Camus

\vskip1.0cm

        The field of quantum information science is largely propelled by
the discovery of quantum algorithms that promise unprecedented
advantages over the best known classical algorithms. The
algorithms of Shor, Grover and others are of more than just
academic interest. They possess the potential of affecting our way
of life in the information age. It is thus of economic,
industrial, in addition to scientific interest to comprehend the
physical principles these algorithms operate upon. These novel
algorithms harness the intrinsic structure of quantum mechanics,
believed to be the fundamental law of nature. Yet, by itself, this
is not enough to completely unravel the mysteries of quantum
computation. Part of this stems from the intrinsic mystery of
quantum mechanics as a physical theory and its counterintuitive
predictions when compared to daily life classical mechanics. For
the remainder, there are at present certain gaps in our
understanding of the working of quantum algorithms. It is some of
these gaps that we look at in this dissertation and attempt to
plug.

        The ultimate aim of quantum information science is to
build a functioning quantum computer. There have been spectacular
advances in that direction. Multiple successful implementations of
pure state algorithms in laboratories across the world have
occurred. However, most of these schemes are limited by their lack
of scalability. This is not entirely due to a lack of experimental
capabilities. Quantum systems are intrinsically hard to isolate
from their environment and manipulate. As the system gets bigger,
the challenges mount. The higher the number of degrees of freedom,
the higher the number of ways it can couple to the environment.
Thus, if we are to build a functioning quantum computer, we must
learn to operate with mixed states, the natural outcome of
environmental decoherence. This has generically been tackled by
quantum error correction, which aims at keeping quantum states as
pure as possible through their quantum evolution. It was the
notion of quantum error correction that first made the dream of a
quantum computer realistic. Excellent techniques have since been
devised to tackle an array of errors that might possibly plague a
quantum computation, and this continues to be an active area of
contemporary research. One needs to ensure that the error
correcting part of the system is itself error-free, and most
techniques of effective quantum error correction eventually rely
on concatenation of several layers of error correction. The
success of this scheme usually requisites the need to reduce
errors below a threshold by successive layers of error correction,
that is, fault tolerance for all operations. Though not entirely
sisyphean\footnote{Sisyphus, of Greek mythology, was punished by
being cursed to roll a large boulder up a hill, only to watch it
roll down again, and to repeat this throughout eternity.  The same
character lends his name to the Sisyphus method of laser cooling
of atoms.}, it is undeniable that the exercise of quantum
computation aggregated with error correction makes the whole
endeavor more challenging.

    A different approach would envisage a quantum
computer using mixed states themselves. Their robustness against
environmental decoherence and the reduction of some of the
complications of quantum error correction is a worthy enough
motivation. Additionally, one might be interested in the
computational potential of mixed quantum states by themselves.
This is where the gaps in our theoretical understanding about
quantum algorithms arrests our progress. Though some attempts have
been made, most of which rely on distilling pure states out of
them, e.g., magic state distillation~\cite{bk05}, mixed-state
quantum computation by itself is particularly poorly understood.
In this thesis, our endeavor will be to address this situation.

        In 1998, E. Knill and R. Laflamme presented the first
intrinsically mixed-state scheme of quantum
computation~\cite{kl98}. It was dubbed the `power of one qubit' or
DQC1 (deterministic quantum computation with one pure qubit)
model, as it required only one pure qubit for its operation. The
setup for this scheme is as follows:
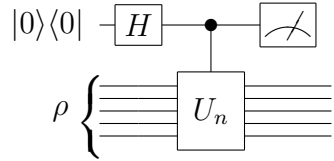
\begin{figure}[h]
\centerline{ \Qcircuit @C=.5em @R=-.5em {
    & \lstick{\ket{0}\!\bra{0}} & \gate{H} & \ctrl{1} & \meter & \push{\rule{0em}{4em}} \\
    & & \qw & \multigate{4}{U_n} & \qw & \qw \\
    & & \qw & \ghost{U_n} & \qw & \qw \\
    \lstick{\mbox{$\rho$}} & & \qw & \ghost{U_n} & \qw & \qw \\
    & & \qw & \ghost{U_n} & \qw & \qw \\
    & & \qw & \ghost{U_n} & \qw & \qw \gategroup{2}{2}{6}{2}{.6em}{\{}
}} \caption[The `power of one qubit' model]{The DQC1 circuit}
\end{figure}
The top qubit is first acted upon by a Hadamard gate, given by
$$
H = \frac{1}{\sqrt 2}\left(%
\begin{array}{cc}
  1 & 1 \\
  1 & -1 \\
\end{array}%
\right),
$$
which effects the transformations
$$
\ket{0}\xrightarrow{\;\;\;\;H\;\;\;\;}\frac{\ket{0}+\ket{1}}{\sqrt
2}\;\;\;\;\;\mbox{and}\;\;\;\;\;
\ket{1}\xrightarrow{\;\;\;\;H\;\;\;\;}\frac{\ket{0}-\ket{1}}{\sqrt
2}.
$$
The state of the system before the controlled unitary is
 $$
\frac{1}{2}\left(\proj{0} + \ket{0}\bra{1}+
\ket{1}\bra{0}+\proj{1}\right) \otimes \rho.
 $$
Next the controlled gate means that the unitary $U_n$ acts on the
lower set of $n$ qubits in the state $\rho$ if and only if the top
(control) qubit is in the state $\ket{1}$. Otherwise it does not
and $I_n$ is applied. After this gate, the state of the system is
$$
\Xi=\frac{1}{2}\left(\proj{0}\otimes \rho + \ket{0}\bra{1}\otimes
\rho U_n^{\dag}+ \ket{1}\bra{0}\otimes U_n\rho+\proj{1}\otimes
U_n\rho U_n^{\dag}\right).
 $$
This is the final state of the system and can compactly be written
as
$$
\Xi = \frac{1}{2}\left(%
\begin{array}{cc}
  \rho &  \rho U_n^{\dag} \\
  U_n\rho &  U_n\rho U_n^{\dag} \\
\end{array}%
\right).
$$
Suppose now that we make a measurement of the top qubit in the $X$
basis, $X$ being the Pauli operator
 $$
 X = \left(%
\begin{array}{cc}
  0 & 1 \\
  1 & 0 \\
\end{array}%
\right).
 $$ We leave the $n$ lower qubits untouched. The expectation of this
measurement is given, quite simply, by
$$
\avg{X}_{\Xi} =\tr[(X\otimes I_n) \Xi] =
\tr\left[\rho\left(\frac{U_n + U_n^{\dag}}{2}\right)\right] =
\mathrm{Re}[\tr(\rho U_n)].
$$
A measurement in the $Y$ basis similarly gives an expectation
value of $\mathrm{Im}[\tr(\rho U_n)]$. This is, in effect, the
DQC1 model of quantum computation.

    Suppose now that we prepared $\rho$ in a pure state $\rho
=\proj{\Phi}$ that was an eigenstate of $U_n$ such that
$U_n\ket{\Phi} = e^{i\phi}\ket{\Phi}$. Then
$$
\avg{X}_{\Xi} = \cos\phi.
$$
Then $L$ measurements of the top qubit in the $X$ basis would
provide us with an estimate of $\phi$, via $\cos\phi$ to within an
accuracy of $1/\sqrt{L}$. This is, in effect, the phase estimation
algorithm so commonly used in quantum algorithms. If, on the other
hand, $\rho = I_n/2^n$, the completely mixed state, then $L$
measurements will give us with an accuracy of $1/\sqrt{L}$, the
quantity
 $$
\avg{X}_{\Xi} = \mathrm{Re}[\tr(U_n)]/2^n.
$$
This is the real part of a quantity called the normalized trace of
the unitary matrix $U_n$, and the scheme gives us an estimate for
it with a fixed accuracy in a number of trials that does not scale
with the size of the unitary matrix. Calculation of fidelity decay
in quantum chaos, the evaluation of expectation values in
condensed matter physics and as shown very recently, even problems
in quantum metrology, quadratically signed weight enumerators and
estimation of Jones polynomials from knot theory and discrete
quantum field theories can be reduced to the evaluation of
normalized traces of particular unitary matrices.

          As unitaries conserve the purity of a state,
$\tr(\Xi^2)=\tr(\rho^2).$ For $\rho = I_n/2^n$, the states
involved in the DQC1 model are thus very closed to being maximally
mixed. Notwithstanding, this algorithm provides an exponential
speedup over the best known classical algorithm. This shows that
to harvest the benefits of exponential speedups in quantum
computation, one does not necessarily need to work with pure
states. To be able to harness this quantum advantage optimally in
the face of a decoherent environment, one needs to understand the
very simple DQC1 model better. Thus, the simple question I address
in this thesis is the following- \emph{Why does the DQC1 model
work?} This, I will attempt to answer in the subsequent chapters
and as we will see, the answer is not simple.

        Quantum entanglement is the currency within the world of
quantum information science. Over the last decade, it has come to
be believed that entanglement is the resource that facilitates
quantum advantages and speedups. This is based on evidence and
knowledge that has been accrued since the early days of quantum
information in the mid 1990's~\cite{hhhh07}. For instance,
simulations have been done that study the rise and fall in the
amount of entanglement in quantum evolutions that implement Shor's
algorithm. The step having the maximum amount of entanglement is
the so called `quantum Fourier transform', which has come be
believed as the quantum soul of the whole algorithm. Quantum
Fourier transform is a period finding technique, just like the
conventional Fourier transform. The novelty, due to Shor, was to
apply it over the field of integers, to find the period of the
modular exponentiation function, to which integer factorization
and discrete logarithm evaluation can be reduced~\cite{ekert96a}.
Grover's search algorithm has also been studied from an
entanglement perspective. It has been found that the quadratic
speedup in the algorithm is directly related to the flow of
entanglement in the algorithm.

        Preceding this line of work on quantum algorithms is the
concept of quantum teleportation~\cite{nielsen00a}. Quantum
teleportation is the ability to transfer the quantum state of a
system from one point in space to another by just classical
communication and a quantum resource. In its simplest incarnation,
a state of the form $\ket{\Psi}=\alpha\ket{0}+\beta\ket{1}$ can be
teleported from Alice to Bob by sending just two bits of classical
information \emph{if} they already share a maximally entangled
Bell state, say, $\frac{1}{\sqrt 2}(\ket{00}+\ket{11})$. Note that
$\ket{\Psi}$ is completely denoted by three real parameters. This
saving in the communication cost stems from the inherent power of
quantum mechanics. It is this power that the field of quantum
information science strives to harness for problems more realistic
than teleportation. Motivated by this, measures of entanglement
are often defined in terms of the number of Bell states that are
needed to produce or can be extracted from a certain state by
local operations and classical communications. In 2005, it was
shown by Masanes that all entangled states (including bound
entangled) can enhance the teleporting power of some quantum
state~\cite{masanes06a}.

        Ever since the work on quantum teleportation, which is a genuinely
quantum phenomenon, the research in quantum information science
has been largely dominated by quantum entanglement. Evidence for
this fact is the profuseness of papers beginning with the sentence
-- ``Entanglement is a useful resource for quantum information and
computation.'' Tragically, this realization does not advance much
the cause of quantum information science. An instance inspiring
such an opinion is the existence of quantum systems in highly
entangled states that are computationally no more powerful that
classical ones. Such a class of states are the so-called
stabilizer states. Though highly entangled, their dynamics can be
efficiently simulated classically. The twist in the story, is
however, a result by Jozsa and Linden~\cite{jozsa03a} which shows
that for any pure-state quantum computation to be exponentially
faster than its classical counterpart, the extent of entanglement
in the system must not be bounded by the system size.

            It is in the light of these circumstances that I analyze the DQC1
model in Chapter \ref{chap:dqc1}. The findings are intriguing. I
found the first example of a unitary matrix for which the DQC1
circuit showed a non-zero amount of entanglement. I then
generalized this to create a family of unitaries which had the
same behavior in terms of the entanglement content of the DQC1
state. In collaboration, I then looked for the presence of
bipartite entanglement in typical instances of the DQC1 circuit.
However, the amount of entanglement was small, in that the ratio
of entanglement present to that maximally possible goes to zero in
the limit of larger and larger systems. Since exponential speedups
are determined asymptotically in the limit of large problem size,
these results suggest that bipartite entanglement cannot be the
reason behind the success of DQC1. I was also able to provide
analytical upper bounds on the amount of entanglement (as measured
by the negativity). This bound does not scale with size of the
problem, and is the most crucial detriment in attributing the
exponential speedup in the DQC1 model to entanglement. In the
course of my studies into the entanglement of the DQC1 model, I
was never able to find a unitary matrix which showed non-trivial
negativity for $\alpha$ (polarization of the top qubit) less than
0.5. This presents a problem of the most singular nature. For
$\alpha \leq 0.5$ the DQC1 state is at best bound entangled. The
computational efficiency of the model, however, has no
discontinuity at this value. If entanglement were behind the
exponential speedup in the DQC1 model, it would imply that this
particular entanglement cannot be exploited for any other task (as
it is bound and therefore cannot be distilled to form Bell
states). This certainly would be the most peculiar state of
affairs !!

        Another facet of the result by Jozsa and Linden is that quantum
systems with small amounts of entanglement can be simulated
classically. This is one of the reasons for believing entanglement
to be a resource in the first place. This provided me with the
impetus to study the classical simulatability of the DQC1 model.
As presented in Chapter \ref{chap:mps}, I found that Matrix
Product State (MPS) algorithms are incapable of simulating typical
instances of the DQC1 system efficiently in spite of its having a
limited amount of entanglement. The degree of correlations, as
measured by the rank of the operator Schmidt decomposition, scaled
exponentially with the size of the system. This I first showed by
numerical simulation. Then, in collaboration with Guifre Vidal, I
was able to provide an analytical proof that the operator Schmidt
rank did indeed scale exponentially with the problem size. MPS
techniques are the most commonly used for simulating quantum
systems with limited amounts of entanglement. My results,
therefore, exclude a large family of algorithms from simulating
the DQC1 model. Simultaneously, it shows the claim made at the
beginning of this paragraph to be invalid in the case of mixed
quantum systems.

        The results of Chapters \ref{chap:dqc1} and \ref{chap:mps}
leave us with a very uneasy realization. A typical instance of the
DQC1 model has very little entanglement, and yet it cannot be
simulated classically. The natural conclusion is that there must
be more to quantum computation that just entanglement. This is the
scenario I investigate in Chapter \ref{chap:discord}. I suggest
that quantum discord\footnote{This
quantity~\cite{ollivier01a,henderson01a} was brought to my
attention by Anil Shaji.}, a notion of quantum correlations more
general than entanglement, is possibly that missing element which
captures the power of a quantum computation completely. In
particular, I show that across the most important split in the
DQC1 circuit, which is shown to be separable in Chapter
\ref{chap:dqc1}, there is a non-zero amount of discord. The
analytic calculations done by me were for the set of random
Haar-distributed unitaries. This set is, however, not efficiently
implementable in terms of any gate set universal for quantum
computation, but sets of pseudo-random unitaries exist which are
efficiently implementable and have eigenvalue statistics that
mimic that of the Haar-distributed unitaries. I used these for
numerical confirmation of my analytic results. Finally, the ratio
of this discord to the maximum possible does not vanish in the
limit of large systems. Thus I present quantum discord as the
potential resource behind quantum advantages in computation.

        Quantum discord turns out to be an interesting and useful quantity
in its own right, with applications wider than just to mixed state
quantum computation. I have preliminary evidence that it plays a
role in quantum communication, too, at least in the phenomenon of
distributing entanglement. The details of this inference and the
role of discord in entanglement distribution is presented in
Section~\ref{S:cirac}.

        In addition to studying the underpinnings of the DQC1
model, I have studied an interesting connection between the DQC1
model and knot theory. The DQC1 model is equivalent to
approximately evaluating the Jones polynomial of the trace closure
of a braid. Rather, this problem is actually complete for the
complexity class DQC1. The problem of approximately evaluating the
Jones polynomial of the \emph{plat} closure of a braid is
BQP-complete. It is believed that the class DQC1 actually lies in
territory betwixt P and BQP, as shown in
Fig~\ref{complexityclass}. Since evaluation of the Jones
polynomial is equivalent to evaluating the partition function of
certain spin models like the Ising model, and different kind of
braid closures correspond to different boundary conditions, this
problem shows us a path for understanding the relations between
complexity classes through physical models and theories like
statistical mechanics. Thus, studying the DQC1 model and its
correspondence to Jones polynomials promises to be a very fruitful
one, and I discuss this in a little more detail in
Appendix~\ref{S:jones}.

\begin{figure}
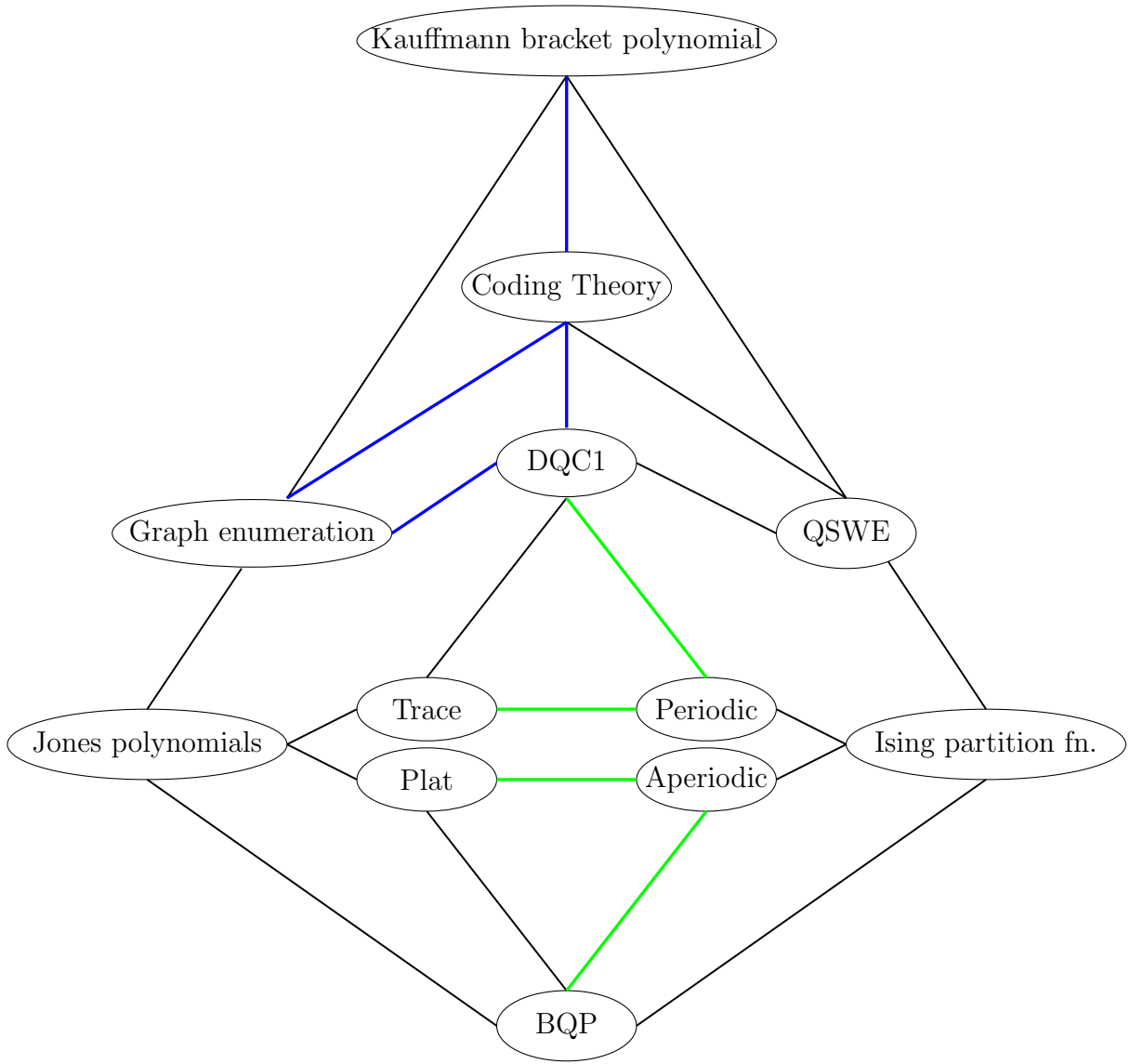

\centerline{
 \begin{pgfpicture}
        \pgfellipse[stroke]{\pgfpoint{0cm}{4cm}}{\pgfxy(3.0,0)}{\pgfxy(0,0.5)}
        \pgfellipse[stroke]{\pgfpoint{-6cm}{-6cm}}{\pgfxy(2,0)}{\pgfxy(0,0.5)}
        \pgfellipse[stroke]{\pgfpoint{0cm}{-2cm}}{\pgfxy(1,0)}{\pgfxy(0,0.48)}
        \pgfellipse[stroke]{\pgfpoint{4cm}{-3cm}}{\pgfxy(1,0)}{\pgfxy(0,0.50)}
        \pgfellipse[stroke]{\pgfpoint{6cm}{-6cm}}{\pgfxy(2,0)}{\pgfxy(0,0.5)}
        \pgfellipse[stroke]{\pgfpoint{-2cm}{-6.5cm}}{\pgfxy(1,0)}{\pgfxy(0,0.45)}
        \pgfellipse[stroke]{\pgfpoint{-2cm}{-5.5cm}}{\pgfxy(1,0)}{\pgfxy(0,0.45)}
        \pgfellipse[stroke]{\pgfpoint{2cm}{-6.5cm}}{\pgfxy(1,0)}{\pgfxy(0,0.45)}
        \pgfellipse[stroke]{\pgfpoint{2cm}{-5.5cm}}{\pgfxy(1,0)}{\pgfxy(0,0.45)}
        \pgfellipse[stroke]{\pgfpoint{0cm}{-10cm}}{\pgfxy(1,0)}{\pgfxy(0,0.5)}
        \pgfellipse[stroke]{\pgfpoint{0cm}{0.5cm}}{\pgfxy(1.5,0)}{\pgfxy(0,0.5)}
        \pgfellipse[stroke]{\pgfpoint{-4.5cm}{-3cm}}{\pgfxy(2,0)}{\pgfxy(0,0.48)}
    \pgfsetlinewidth{0.75pt}
        \pgfxyline(-4,-6)(-3,-6.5)          
        \pgfxyline(-4,-6)(-3,-5.5)          
        \pgfxyline(4,-6)(3,-6.5)            
        \pgfxyline(4,-6)(3,-5.5)            
        \pgfxyline(0,-9.5)(-2,-6.95)        
        \pgfxyline(0,-2.5)(-2,-5.05)        
        \pgfxyline(1,-2)(3,-3)              
        \pgfxyline(-1,-10)(-6,-6.5)         
        \pgfxyline(1,-10)(6,-6.5)           
        \pgfxyline(-4.65,-3.5)(-6,-5.5)      
        \pgfxyline(0,3.5)(-4,-2.5)          
        \pgfxyline(0,3.5)(4,-2.5)           
        \pgfxyline(4.6,-3.4)(6,-5.5)           
        \pgfxyline(0,0)(4,-2.5)             
        \pgfputat{\pgfxy(0,-2)}{\pgfbox[center,center]{DQC1}}
        \pgfputat{\pgfxy(4,-3)}{\pgfbox[center,center]{QSWE}}
        \pgfputat{\pgfxy(0,0.5)}{\pgfbox[center,center]{Coding Theory}}
        \pgfputat{\pgfxy(0,4)}{\pgfbox[center,center]{Kauffmann bracket polynomial}}
        \pgfputat{\pgfxy(-6,-6)}{\pgfbox[center,center]{Jones polynomials}}
        \pgfputat{\pgfxy(0,-10)}{\pgfbox[center,center]{BQP}}
        \pgfputat{\pgfxy(-4.5,-3)}{\pgfbox[center,center]{Graph enumeration}}
        \pgfputat{\pgfxy(-2,-5.5)}{\pgfbox[center,center]{Trace}}
        \pgfputat{\pgfxy(-2,-6.5)}{\pgfbox[center,center]{Plat}}
        \pgfputat{\pgfxy(2,-5.5)}{\pgfbox[center,center]{Periodic}}
        \pgfputat{\pgfxy(2,-6.5)}{\pgfbox[center,center]{Aperiodic}}
        \pgfputat{\pgfxy(6,-6)}{\pgfbox[center,center]{Ising partition fn.}}
\pgfsetlinewidth{1.25pt}
    \color{green}
        \pgfxyline(0,-9.5)(2,-6.95)          
        \pgfxyline(0,-2.5)(2,-5.05)          
        \pgfxyline(-1,-6.5)(1,-6.5)          
        \pgfxyline(-1,-5.5)(1,-5.5)          
\pgfsetlinewidth{1.25pt}
    \color{blue}
        \pgfxyline(0,3.5)(0,1)
        \pgfxyline(0,0)(0,-1.5)
        \pgfxyline(-2.5,-3)(-1,-2)          
        \pgfxyline(0,0)(-4,-2.5)            
      \end{pgfpicture}}
\caption[DQC1 amongst physics, mathematics and computer science]{A
schematic that displays the rich interdisciplinary nature of
quantum computation in general, and DQC1 in particular. Black
lines show the connections already known in the literature. Green
lines are the connections first made in this thesis. The blue line
show some of the possible connections that can be made and the
potential for exploring the DQC1 model from several independent
points of view. `Trace' and `Plat' refer to closures of braids
whose Jones polynomials are of interest, while `Periodic' and
`Aperiodic' refer to boundary conditions for Ising models. `Graph
enumeration' refers to the class of graph theoretic problems like
the evaluation of the Tutte, the dichromatic, or the chromatic
polynomial. The Kauffman bracket polynomial is a knot invariant
which is a generalization of the Jones polynomial, and is related
to partition functions of more esoteric statistical mechanical
models like the vertex model and the IRF (interactions round a
face) model. For further details, see Appendix~\ref{A:knot}.}
\label{relations}
\end{figure}

        The above explorations by no means exhaust the class of
problems a mixed-state quantum computer or the DQC1 model can
solve (See Fig~\ref{relations} for some additional possibilities).
One of the more exotic applications is through the quadratically
signed weight enumerators (QSWEs). We will mention this as well in
brief in Appendix~\ref{S:qswe} and not go into any details of it
in this thesis.

\newpage

\section{List of Publications}

It is unlikely that any physics PhD thesis in this age of instant
electronic collaboration  is a collation of entirely original
results. So it is with this dissertation. In addition to the
background material taken from the literature, several new results
presented in this dissertation were arrived at in collaboration
with other researchers. Several of these have been published in
refereed journals. This dissertation contains only my research on
mixed-state quantum computation and the DQC1 model. In addition to
this research, I have worked on two other topics during my
graduate career at UNM, \textit{viz}, quantum metrology and the
abstract theory of entanglement.  I provide below the list of my
publications based on work done as a PhD student, classified
according to which of the three topics they pertain to.  For the
papers on mixed-state quantum computation, I give which chapter
each corresponds to in this dissertation.

\textbf{Mixed-State Quantum Computation }
\begin{enumerate}

\item Quantum discord and the power of one qubit,
\textcolor{blue}{Animesh Datta}, Anil Shaji, Carlton M. Caves,
\emph{Phys. Rev. Lett}, \textbf{100}, 050502 (2008). (Chapter
\ref{chap:discord})

\item Entanglement and the Power of One Qubit,
\textcolor{blue}{Animesh Datta}, Steven T. Flammia, Carlton M.
Caves, \emph{Phys. Rev. A}, \textbf{72}, 042316 (2005). (Chapter
\ref{chap:dqc1})

\item Role of entanglement and correlations in mixed-state quantum
computation, \textcolor{blue}{Animesh Datta}, Guifre Vidal,
\emph{Phys. Rev. A}, \textbf{75}, 042310 (2007). (Chapter
\ref{chap:mps})
\\
\\
\textbf{Quantum Metrology}

\item Quantum Metrology: Dynamics vs Entanglement, Sergio Boixo,
\textcolor{blue}{Animesh Datta}, Steven T. Flammia, Matthew J.
Davis, Anil Shaji Carlton M. Caves, Submitted to \emph{Physical
Review Letters}.

\item Quantum-limited metrology with product states, Sergio Boixo,
\textcolor{blue}{Animesh Datta}, Steven T. Flammia, Anil Shaji,
Emilio Bagan, Carlton M. Caves, \emph{Phys. Rev. A}, \textbf{77},
012317 (2008).

\item On Decoherence in Quantum Clock Synchronization, Sergio
Boixo, Carlton M. Caves, \textcolor{blue}{Animesh Datta}, Anil
Shaji, \emph{Laser Physics: Special Issue on Quantum Information
and Quantum Computation}, \textbf{16}, 1525 (2006).
\\
\\
\textbf{Quantum Entanglement}

\item Doubly constrained bounds on the entanglement of formation,
\textcolor{blue}{Animesh Datta}, Steven T. Flammia, Anil Shaji,
Carlton M. Caves, arXiv:quant-ph/0608086.

\item Constrained bounds on measures of entanglement,
\textcolor{blue}{Animesh Datta}, Steven T. Flammia, Anil Shaji,
Carlton M. Caves, \emph{Phys. Rev. A}, \textbf{75}, 062117 (2007)

\end{enumerate}

The remainder of the Introduction explains in brief the results
obtained in work not included in this dissertation.

        The study of entanglement in a mixed quantum state led to some
work on the general theory of quantum entanglement, which though
not a part of this dissertation is listed below. This project
considered entanglement measures constructed from two positive,
but not completely positive, maps on density operators that were
used as constraints in placing bounds on the entanglement of
formation, the tangle, and the concurrence of $4\times N$ mixed
states. The norm-based entanglement measures constructed from
these two maps, called negativity and phi-negativity,
respectively, lead to two sets of bounds on the entanglement of
formation, the tangle, and the concurrence. These bounds were
compared, and we identified the sets of $4\times N$ density
operators for which the bounds from one constraint are better than
the bounds from the other.  Extensions of our results to bipartite
states of higher dimensions and with more than two constraints
were also discussed.

        Another area of my research has been quantum metrology. The lure
of beating the shot-noise limit without using entanglement led my
collaborators and myself to study various quantum clock
synchronization protocols. We showed that the proposed schemes for
beating the standard quantum limit without entanglement require as
much resources as those using entanglement. We later turned to
surpassing even the Heisenberg limit and proposed a scheme that
involved product input states and measurements. Our results also
provided evidence that the entanglement generated in the process
was not responsible for the enhancement in precision. The next
step involved proposing a demonstration of such a protocol using
two-component Bose-Einstein condensates in $^{87}\mathrm{Rb}$.
This is the first realistic laboratory proposal of beating the
$1/n$ limit in metrology, certainly one using BECs

\chapter{Background\label{chap:background}}

\hfill \textsl{Take nothing on its looks; take everything on
evidence. There's no better rule.}

\hfill  -Charles Dickens in \textsl{Great Expectations}

\vskip1.0cm

        The intent of this chapter is to present research that is
not entirely original, but nonetheless vital for the appreciation
of the results appearing in the following chapters. Almost all of
it can be found in the literature, and this will be pointed out at
appropriate spots. This chapter will set the context and relevance
of the DQC1 model. Subsequently we will outline the basics of
entanglement theory and its relevance in quantum computation.
Later we will present arguments and motivations for considering a
quantity such as quantum discord and \emph{its} relevance to
quantum information science. There is some original research in
this chapter, in the later parts of Sec~\ref{introdiscord}. All of
Sec~\ref{S:cirac} is original, dealing with quantum discord in
quantum communication. Though not a central theme of this thesis,
this role is vital to our understanding of quantum discord.

    We begin this chapter with the requirements for a scalable quantum
computation. We put the DQC1 model in perspective with the rest of
quantum computation. We discuss the relevance of the DQC1 model in
several problems of physical importance and its place in
computational complexity theory. We will briefly discuss the role
of entanglement in all this and realize its shortcomings. Then we
will move on to study a quantity that we think might be able to
explain phenomena that entanglement fails at. This quantity,
quantum discord, will be discussed in detail. We will explain how
it is evaluated, and work out an example. Then we will study a
phenomenon, called entanglement distribution, through a simple
example and show, for the first time, the role of discord in its
success.

        Without further ado, let us begin by what are postulated
to be the requirements for building a fully controllable and
scalable quantum computer. Known as DiVincenzo's criteria
\cite{divincenzo00}, they are as follows:
 \begin{enumerate}
\item Be scalable system in terms of well defined qubits (2-level
quantum systems)

\item Be initializable into a simple quantum state such as
$\ket{00\cdots 000}$

\item Have decoherence times long compared to gate operation times

\item Have a universal set of quantum gates

\item Permit high efficiency, single qubit measurements
 \end{enumerate}
    However, fully controllable and scalable quantum computers are likely many
years from realization.  This motivates study and development of
somewhat less ambitious quantum information processors, defined as
devices that fail to satisfy one or more of DiVincenzo's five
criteria for a quantum computer~\cite{divincenzo00} listed above.
An example of such a quantum information processor is a
mixed-state quantum system, which fails to pass DiVincenzo's
second requirement, that the system be prepared in an initial pure
state.

The prime example of a mixed-state quantum information processor
is provided by liquid-state NMR experiments in quantum information
processing~\cite{jones01}.  Current NMR experiments, which operate
with initial states that are highly mixed thermal states, use a
technique called pseudo-pure-state synthesis to process the
initial thermal state and thereby to simulate pure-state quantum
information processing. This technique suffers from an exponential
loss of signal strength as the number of qubits per molecule
increases and thus is not scalable. There is a different technique
for processing the initial thermal state, called algorithmic
cooling~\cite{sv99}, which pumps entropy from a subset of qubits
into the remaining qubits, leaving the special subset in a pure
state and the remaining qubits maximally mixed. Algorithmic
cooling provides an in-principle method for making liquid-state
NMR---or any qubit system that begins in a thermal
state---scalable, in essence by providing an efficient algorithmic
method for cooling a subset of the initially thermal qubits to a
pure state, thereby satisfying DiVincenzo's second criterion.

Knill and Laflamme~\cite{kl98} proposed a related mixed-state
computational model, which they called DQC1, in which there is
just one initial pure qubit, along with $n$ qubits in the
maximally mixed state. Although provably less powerful than a
pure-state quantum computer~\cite{asv00}, DQC1 can perform some
computational tasks efficiently for which there are no known
polynomial time classical algorithms. Knill and Laflamme referred
to the power of this mixed-state computational model as the
``power of one qubit.'' In particular, a DQC1 quantum circuit can
be used to evaluate, with fixed accuracy independent of $n$, the
normalized trace, $\tr(U_n)/2^n$, of any $n$-qubit unitary
operator $U_n$ that can be implemented efficiently in terms of
quantum gates~\cite{kl98,lcnv02}.  In Sec.~\ref{S:classical} we
consider whether there might be efficient classical algorithms for
estimating the normalized trace, and we conclude that this is
unlikely.

        The efficient quantum algorithm for estimating the normalized
trace provides an exponential speedup over the best known
classical algorithm for simulations of some quantum processes. It
is known how to reduce the problem of measuring the average
fidelity decay of a quantum map under small perturbations to that
of evaluating the normalized trace of a unitary matrix
\cite{pklo04}. The DQC1 algorithm for this instance provides an
exponential speedup. However, in testing the integrability of the
classical limit of a quantum system, the same basic algorithm
provides a square-root speedup \cite{plmp03}. Evaluating the local
density of states, however, can be executed with an exponential
advantage using only one pure qubit \cite{elpc04}. There has also
been an attempt to exploit the DQC1 system in quantum metrology
\cite{bs07}.

\begin{figure}[!h]
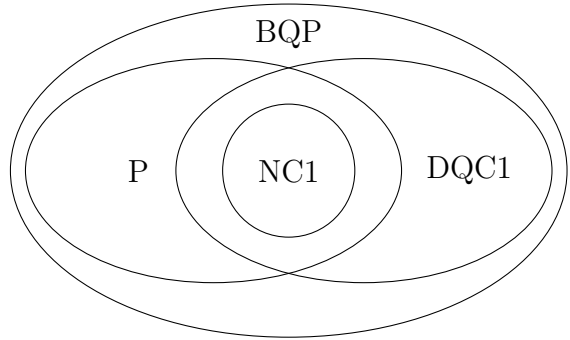

\centerline{
 \begin{pgfpicture}
        \pgfellipse[stroke]{\pgforigin}{\pgfxy(3.7,0)}{\pgfxy(0,2.2)}
        \pgfellipse[stroke]{\pgfpoint{1cm}{0cm}}{\pgfxy(2.5,0)}{\pgfxy(0,1.48)}
        \pgfellipse[stroke]{\pgfpoint{-1cm}{0cm}}{\pgfxy(2.5,0)}{\pgfxy(0,1.48)}
        \pgfcircle[stroke]{\pgfxy(0,0)}{25pt}
        \pgfputat{\pgfxy(0,0)}{\pgfbox[center,center]{NC1}}
        \pgfputat{\pgfxy(2.4,0)}{\pgfbox[center,center]{DQC1}}
        \pgfputat{\pgfxy(-2,0)}{\pgfbox[center,center]{P}}
        \pgfputat{\pgfxy(0,1.8)}{\pgfbox[center,center]{BQP}}
      \end{pgfpicture}}
\caption[Polynomial hierarchy with P, BQP and DQC1]{The relative
hierarchy of the complexity classes P, BQP and DQC1.}
\label{complexityclass}
\end{figure}

        A very interesting application of the DQC1 model has been to
topological quantum computation. It has been shown that the
evaluation of the plat closure of the Jones polynomial of the same
braid is BQP-complete \cite{fklw03,fkw02,flw02}. An explicit
quantum algorithm for this task was presented in 2006 \cite{ajl06}
in the language of quantum circuits. This technique has now been
extended to more general topological invariants like the HOMFLYPT
polynomial and the Kauffman polynomial \cite{wy06}. Recently, it
was shown that the DQC1 model can be used to evaluate the Jones
polynomial of the trace closure of a braid efficiently
\cite{sj07}. Indeed the problem of evaluating the Jones polynomial
for the trace closure of a braid is complete for the complexity
class DQC1. The notion of DQC1 as a quantum complexity class was
developed in \cite{s06}. Since then, one of the important
questions has been the relation of this class to other well known
classes like P and BQP. As has already been shown, DQC1 is less
powerful than BQP, a full scale pure state quantum computer
\cite{asv00}. Still, DQC1 can efficiently simulate classical
circuits of logarithmic depth \cite{asv00}, a class known as NC1,
which is contained in P, the class of problems that can be solved
on a classical polynomial in polynomial time. On the other hand,
as we show in Section~\ref{S:classical}, it is unlikely that DQC1
is contained in P. This leads to a very intriguing picture of the
relative powers of these complexity classes in the polynomial
hierarchy as shown in Fig~\ref{complexityclass}.

        This figure has a number of intriguing question lurking
underneath. The least of these is a proof of the location of DQC1
in this figure. Also, if this figure is correct, as we believe at
present, then DQC1 is really a class that lies between quantum and
classical computation. One would also like to understand the
physical principle that governs this relative positioning. Such a
sort of understanding will tell us of the particular aspects of
nature that help us beat classical computers using quantum
mechanics. We have been pursuing this program with some success.
We describe this is Appendix~\ref{A:knot}.

    Study of the power of one qubit is motivated partly by NMR
experiments, but our primary motivation in this thesis is to
investigate the role of entanglement in quantum computation, using
DQC1 as a theoretical test bed for the investigation.  For
pure-state quantum computers, Jozsa and Linden \cite{jozsa03a}
have shown that exponential speedup over a classical computer
requires that entanglement not be restricted to blocks of qubits
of fixed size as problem size increases. Entanglement whose extent
increases with problem size is thus a necessary prerequisite for
the exponential speedup achieved by a pure-state quantum computer.
On the other hand, the Gottesman-Knill theorem \cite{nielsen00a}
demonstrates that global entanglement is far from sufficient for
exponential speedup.  While this means that the role of
entanglement is not entirely understood for pure-state quantum
computers, far less is known about the role of entanglement in
mixed-state quantum computers. When applied to mixed-state
computation, the Jozsa-Linden proof does not show that
entanglement is a requirement for exponential speedup. Indeed,
prior to the work reported in this dissertation, it has not
previously been shown that there is any entanglement in the DQC1
circuits that provide an exponential speedup over classical
algorithms.

            Evidently, to invoke the above results, one must first
calculate the entanglement in a quantum computation. This exercise
of calculating the entanglement in a general quantum state is
highly non-trivial. Indeed, even the simpler task of proving the
separability of a general quantum state is NP-Hard \cite{g03}. For
a general mixed quantum state, the task is further hardened by the
lack of a unique measure of entanglement. This is partly due to
the fact that entanglement, instead of being a physical notion, is
a mathematical construct. Though it has been identified with
physical tasks that can be thought of as operational
interpretations of entanglement, this has not simplified the task
of actually calculating the numerical value of entanglement. The
computational task is not eased by the definition of mixed-state
measures through convex-roof extensions of pure state measures,
most commonly the von-Neumann entropy of the reduced density
matrix. One exception is the negativity, which is formally easy to
define and computationally feasible to calculate, a rare feature
for an entanglement measure. It is this measure that we use in
this thesis, after elaborating upon it in the next section. For an
extensive review on the theory of entanglement, the reader is
referred to the recent article by the Horodecki family
\cite{hhhh07}.

\section{Entanglement: Properties of negativity}
\label{S:negativity}

In this section we briefly review properties of negativity as an
entanglement measure, focusing on those properties that we need in
the subsequent analysis (for a thorough discussion of negativity,
see Ref.~\cite{vw02}).

Let $A$ be an operator in the joint Hilbert space of two systems,
system~1 of dimension $d_1$ and system~2 of dimension $d_2$.  The
partial transpose of $A$ with respect to an orthonormal basis of
system~2 is defined by taking the transpose of the matrix elements
of $A$ with respect to the system-2 indices.  A partial transpose
can also be defined with respect to any basis of system~1. Partial
transposition preserves the trace, and it commutes with taking the
adjoint.

The operator that results from partial transposition depends on
which basis is used to define the transpose, but these different
partial transposes are related by unitary transformations on the
transposed system and thus have the same eigenvalues and singular
values. Moreover, partial transposition on one of the systems is
related to partial transposition on the other by an overall
transposition, which also preserves eigenvalues and singular
values.  Despite the nonuniqueness of the partial transpose, we
can talk meaningfully about its invariant properties, such as its
eigenvalues and singular values. Similar considerations show that
the eigenvalues and singular values are invariant under local
unitary transformations.

The singular values of an operator $O$ are the eigenvalues of
$\sqrt{O^\dag O}\equiv|O|$ (or, equivalently, of $\sqrt{O
O^\dag}$).  Any operator has a polar decomposition $O=T|O|$, where
$T$ is a unitary operator.  Writing $|O|=W^\dag SW$, where $W$ is
the unitary that diagonalizes $|O|$ and $S$ is the diagonal matrix
of singular values, we see that any operator can be written as
$O=VSW$, where $V=TW^\dag$ and $W$ are unitary operators.

We denote a partial transpose of $A$ generically by $\breve A$. We
write the eigenvalues of $\breve A$ as $\lambda_j(\breve A)$ and
denote the singular values by $s_j(\breve A)$.  If $A$ is
Hermitian, so is $\breve A$, and the singular values of $\breve
A$, i.e., the eigenvalues of $|\breve A|$, are the magnitudes of
the eigenvalues, i.e, $s_j(\breve A)=|\lambda_j(\breve A)|$.

If a joint density operator~$\rho$ of systems 1 and 2 is
separable, its partial transpose $\breve\rho$ is a positive
operator.  This gives the Peres-Horodecki entanglement
criterion~\cite{p96,hhh96}: if $\breve\rho$ has a negative
eigenvalue, then $\rho$ is entangled (the converse is not
generally true).  The magnitude of the sum of the negative
eigenvalues of the partial transpose, denoted by
\begin{equation}
\sN(\rho)\equiv-\sum_{\lambda_j(\breve\rho)<0}\lambda_j(\breve\rho)\;,
\end{equation}
is a measure of the amount of entanglement.  Partial transposition
preserves the trace, so $\tr(\breve\rho)=1$, from which we get
\begin{equation}
1+2\sN(\rho)=\sum_j|\lambda_j(\breve\rho)|=
\tr|\breve\rho|\equiv\sM(\rho)\;,
\end{equation}
where $\sM(\rho)$ is a closely related entanglement measure.  The
quantity $\sN(\rho)$ was originally called the
\emph{negativity\/}~\cite{vw02}; we can distinguish the two
measures by referring to $\sM(\rho)$ as the \emph{multiplicative
negativity}, a name that emphasizes one of its key properties and
advantages over $\sN(\rho)$.  In this thesis, however, we use the
multiplicative negativity exclusively and so refer to it simply as
``the negativity''.

The negativity $\sM(\rho)$ equals one for separable states, and it
is an entanglement monotone~\cite{vw02}, meaning that (i)~it is a
convex function of density operators and (ii)~it does not increase
under local operations and classical communication.  The
negativity has the property of being multiplicative in the sense
that the $\sM$ value for a state that is a product of states for
many pairs of systems is the product of the $\sM$ values for each
of the pairs.  By the same token, $\log\sM(\rho)$, called the
\emph{log-negativity\/}, is additive, but the logarithm destroys
convexity so the log-negativity is not an entanglement
monotone~\cite{vw02}.  For another point of view on the
monotonicity of the log-negativity, see Ref.~\cite{Plenio05}.

The minimum value of the negativity is one, but we need to know
the maximum value to calibrate our results.  Convexity guarantees
that the maximum value is attained on pure states.  We can find
the maximum~\cite{SLee03} by considering the Schmidt decomposition
of a joint pure state of systems~1 and 2,
\begin{equation}
|\psi\rangle=\sum_{j=1}^d\sqrt{\mu_j}|j,j\rangle\;,
\end{equation}
where $d=\min(d_1,d_2)$.  Taking the partial transpose of $\rho$
relative to the Schmidt basis of system~2 gives
\begin{equation}
\breve\rho =\sum_{j,k=1}^d\sqrt{\mu_j\mu_k}|j,k\rangle\langle
k,j|\;,
\end{equation}
with eigenvectors and eigenvalues
\begin{eqnarray}
|j,j\rangle\;,&&\mbox{eigenvalue $\mu_j$,}\nonumber\\
{1\over\sqrt2}(|j,k\rangle\pm|k,j\rangle)\;, &&\mbox{eigenvalue
$\pm\sqrt{\mu_j\mu_k}$,}\quad j<k.
\end{eqnarray}
This gives a negativity
\begin{equation}
\sM(\psi)= 1+2\sum_{j<k}\sqrt{\mu_j\mu_k}
=\sum_{j,k=1}^d\sqrt{\mu_j\mu_k}
=\biggl(\sum_{j=1}^d\sqrt{\mu_j}\biggr)^2\;.
\end{equation}
The concavity of the square root implies
$\sum_j\sqrt{\mu_j}\le\sqrt d$, with equality if and only if
$\mu_j=1/d$ for all $j$, i.e., $|\psi\rangle$ is maximally
entangled. We end up with
\begin{equation}
1\le\sM(\rho)\le d\;.
\end{equation}

\section{Beyond Entanglement: Quantum Discord}\label{introdiscord}

    Characterizing and quantifying the information-processing
capabilities offered by quantum phenomena like entanglement,
superposition, and interference is one of the primary objectives
of quantum information theory. Amongst these, entanglement has
defined much of the research in quantum information science for
almost a decade now. In spite of some progress
\cite{ekert98a,jozsa03a,kendon06a}, the precise role of
entanglement in quantum information processing remains an open
question
\cite{bennett99c,braunstein99a,vidal03b,kenigsberg06a,datta07a}.
It is quite well established that entanglement is essential for
certain kinds of quantum-information tasks like teleportation  and
super-dense coding. In these cases, it is also known that the
quantum enhancement must come from (pre-shared) entanglement
between parts of the system.  It is not known, however, if all
information-processing tasks that can be done more efficiently
with a quantum system than with a comparable classical system
require entanglement as a resource. Indeed, there are several
instances where we see a quantum advantage in the absence or near
absence of entanglement. One is quantum cryptography \cite{bb84},
where many protocols involve quantum states that are not
entangled. These quantum cryptography protocols are provably more
secure and thus better than the best known classical cryptographic
techniques. This certainly cannot be attributed to entanglement.
The second example we present is that of the DQC1 model, the theme
of this dissertation. As this thesis shows, in Chapter
\ref{chap:dqc1}, the system has very little entanglement - and
even that little vanishes asymptotically - yet it provides an
exponential speedup. These examples drive home one point:
\emph{all of quantum information science cannot be reduced to a
study of quantum entanglement.}

        This however leaves us in a lurch. What then explains
all the quantum advantages~? Certainly, it has something to do
with the structure and inherent non-locality of quantum mechanics.
Interestingly, it is known that non-locality and entanglement are
not equivalent features \cite{bgs05}. Entanglement is the feature
that having complete information about parts of a quantum system
does not imply complete information of the whole system. This
characterization goes all the way back to Schr\"{o}dinger
\cite{s35}. Not surprisingly, this is not the only
characterization of a quantum system. There are other ones. For
instance, the collapse of one part of a subsystem after a
measurement of another is another feature unique to quantum
systems. One can conceive of defining a quantity that captures
this feature. Such a quantity is the quantum discord
\cite{ollivier01a}.

        Quantum discord has been related to the superposition
principle and the vanishing of discord shown to be a criterion for
the preferred, effectively classical states of a system, i.e., the
pointer states \cite{ollivier01a}. In addition, it has been used
in analyzing the powers of a quantum Maxwell's demon
\cite{zurek03b} and also in the study of pure quantum states as a
resource \cite{horodecki05a}.

        We start with a discussion of quantum discord, its definition and
its relevance in quantum information theory. Consider the
following two-qubit separable state
\begin{equation}
\label{eq:state1} \rho =\frac{1}{4}\Big[(\proj{+} \otimes
\proj{0}) \; + \; (\proj{-} \otimes \proj{1}) + (\proj{0}
\otimes\proj{-}) \;  + \;(\proj{1} \otimes \proj{+}) \Big],
\end{equation}
in which four nonorthogonal states of the first qubit are
correlated with four nonorthogonal states of the second qubit.
Such correlations cannot exist in any classical state of two bits.
The extra correlations the quantum state can contain compared to
an equivalent classical system with two bits could reasonably be
called quantum correlations.  Entanglement is certainly a kind of
quantum correlation, but it is not the only kind.  In other words,
separable quantum states can have correlations that cannot be
captured by a probability distribution defined over the states of
an equivalent classical system.

Quantum discord attempts to quantify all quantum correlations
including entanglement.  It must be emphasized here that the
discord supplements the measures of entanglement that can be
defined on the system of interest.  It aims to capture all the
nonclassical correlations present in a system, those that can be
identified as entanglement and then some more.

The information-theoretic measure of correlations between two
systems $A$ and $B$ is the mutual information, $\mathcal{I}(A:B) =
H(A) + H(B)- H(A,B)$. If $A$ and $B$ are classical systems whose
state is described by a probability distribution $p(A,B)$, then
$H(\cdot)$ denotes the Shannon entropy, $H({\bm p})\equiv -\sum_j
p_j \log p_j$, where $\bm p$ is a probability vector. If $A$ and
$B$ are quantum systems described by a combined density matrix
$\rho_{AB}$, then $H(\cdot)$ stands for the corresponding von
Neumann entropy, $H (\rho) \equiv -\tr (\rho \log \rho)$.

For classical probability distributions, Bayes's rule leads to an
equivalent expression for the mutual information,
$\mathcal{I}(A:B) = H(B) - H(B|A)$, where the conditional entropy
$H(B|A)$ is an average of Shannon entropies for $B$, conditioned
on the alternatives for $A$.  For quantum systems, we can regard
this form for $\mathcal{I}(A:B)$ as defining a conditional
entropy, but it is not an average of von Neumann entropies and is
not necessarily nonnegative~\cite{cerf99b}.

    Another way of generalizing the classical conditional entropy to
the quantum case is to recognize that classically $H(B|A)$
quantifies the ignorance about the system $B$ that remains if we
make measurements to determine $A$.  When $A$ is a quantum system,
the amount of information we can extract about it depends on the
choice of measurement.  If we restrict to projective measurements
(\emph{a priori}, we could consider POVMs, but we will show later
that rank one projectors suffice) described by a complete set of
orthogonal projectors, $\{\Pi_j\}$, corresponding to outcomes~$j$,
then the state of $B$ after a measurement is given by
\begin{equation}
\label{redstates}
 \rho_{B|j} =
\tr_M\bigl(\Pi_j\rho_{AB}\Pi_j\bigr)/p_j, \;\;
p_j=\tr_{A,B}\bigl(\rho_{AB}\Pi_j\bigr).
\end{equation}
A quantum analogue of the conditional entropy can then be defined
as $\tilde{H}_{\{\Pi_j\}}(B|A) \equiv \sum_j p_j
H(\rho_{B|j})\ge0$.  Since $\rho_B=\sum_jp_j\rho_{B|j}$, the
concavity of von Neumann entropy implies that $H(B)\ge\tilde
H_{\{\Pi_j\}}(B|A)$.  We can now define an alternative quantum
version of the mutual information,
\begin{equation}
 \label{quantJ}
\mathcal{J}_{\{\Pi_j\}}(A:B) \equiv H(B) - \tilde H_{\{\Pi_j\}}
(B|A)\ge 0.
\end{equation}
Performing projective measurements onto a complete set of
orthogonal states of $A$ effectively removes all nonclassical
correlations between $B$ and $A$.  In the post-measurement state,
mutually orthogonal states of $A$ are correlated with at most as
many states of $B$. It is easy to see that these sorts of
correlations can be present in an equivalent classical system. If
on the other hand, the system $B$ is bigger than $A$, the
correlations are certainly non-classical, or quantum.

    The value of $\mathcal{J}_{\{\Pi_j\}}(A:B)$ in Eq.~(\ref{quantJ})
depends on the choice of $\{\Pi_j\}$.  We want
$\mathcal{J}_{\{\Pi_j\}}(A:B)$ to quantify {\em all\/} the
classical correlations in $\rho_{AB}$, so we maximize
$\mathcal{J}_{\{\Pi_j\}}(A:B)$ over all $\{\Pi_j\}$ and define a
measurement-independent mutual information $\mathcal{J}(A:B)
\equiv H(B) - \tilde H(B|A)\ge0$, where $\tilde H(B|A)\equiv
\min_{\{\Pi_j\}}\sum_j p_j H(\rho_{B|j})$ is a
measurement-independent conditional information. Henderson and
Vedral~\cite{henderson01a} investigated how $\mathcal{J}(A:B)$
quantifies classical correlations.  The criterion they postulated
for a measure of classical correlations $\mathcal{C}$
were\cite{henderson01a}
 \begin{enumerate}
 \item $\mathcal{C}=0$ for product states
 \item $\mathcal{C}$ is invariant under local unitary
transformations. This is because any change of basis should not
affect the correlation between two subsystems.
 \item $\mathcal{C}$ is non-increasing under local operations. If the two subsystems evolve independently
then the correlation between them cannot increase.
 \item $\mathcal{C}=H(\rho_A)=H(\rho_B)$ for pure states.
 \end{enumerate}
One can then define a measure of purely quantum correlations as
the difference of the total correlations in a system and
$\mathcal{C}$. Taking the quantum mutual information to be a
measure of total correlations in a system, the purely quantum
correlations can be measured by
 \begin{equation}
\label{discorddef}
 \mathcal{D}(A,B) \equiv \mathcal{I}(A:B)-\mathcal{J}(A:B) =  \tilde H(B|A)-H(B|A).
\end{equation}
This quantity was first called quantum discord by Ollivier and
Zurek in \cite{ollivier01a}. Another operational approach to
identifying the nature of correlations is presented in
\cite{gpw05}. The discord of the state~(\ref{eq:state1}) is
$\frac{3}{4}\log\frac{4}{3}=0.311$.

        This thesis is the first to propound a role for quantum
discord in quantum computation and information. It is, therefore,
imperative that we present all the known properties of discord
here. This is the content of the next section. We will present
several properties of quantum discord, some of which are present
in prior literature. We will provide appropriate indications at
requisite spots.

\subsection{Properties of Quantum Discord}
\label{S:discordprops}

    The first property that we present is that the discord is
nonnegative and is zero for states with only classical
correlations~\cite{henderson01a,ollivier01a}. Thus a nonzero value
of $\mathcal{D}(A,B)$ indicates the presence of nonclassical
correlations~\cite{ollivier01a}.

\begin{theorem}
 \label{discordispositove} Quantum discord is always positive,
\textit{\textit{i.e.}}, $\mathcal{D}(A,B)\geq 0.$
\end{theorem}

\noindent {\it Proof: } Consider the joint state $\rho_{AB}$
subject to one dimensional orthogonal measurements
$\Pi_j=\proj{e_j}$ on $B$, extended to arbitrary (at most
$dim(B)^2$) dimensions. Then
 $$
 p_j\,\rho_{A|j} =
\tr_{B}(\rho_{AB}\Pi_j)=\bra{e_j}\rho_{AB}\ket{e_j},\;\;\;
 p_j = \tr_{B}(\rho_{B}\Pi_j)=\bra{e_j}\rho_{AB}\ket{e_j}.
 $$
Note that the measurement is made on the system $B$, while in the
definition of discord, it was on $A$. Discord is not symmetric
under the exchange of the subsystems, but this is not a concern as
we just as well have proved the result for $\mathcal{D}(B,A)$.

Suppose now that a system $C$ interacts with $B$ so as to make the
desired measurement
($U\ket{e_j}\otimes\ket{0}=\ket{e_j}\otimes\ket{f_j}$), leaving
the state
 \be
 \label{extrho}
 \rho'_{ABC}=\sum_{j,k}\bra{e_j}\rho_{AB}\ket{e_k}\otimes\ket{e_j}\bra{e_k}\otimes\ket{f_j}\bra{f_k}.
 \ee
If the eigendecomposition of $ \rho_{AB} =\sum_l
\lambda_l\proj{r_l} $, then
 \ben
 \rho'_{ABC}&=&\sum_{j,k,l}\lambda_l\langle \mathbb{I}_A,e_j\proj{r_l}\mathbb{I}_A,e_k\rangle \otimes\ket{e_j}\bra{e_k}\otimes\ket{f_j}\bra{f_k}\nonumber\\
            &=& \sum_{l}\lambda_l\proj{e_l,r_l,f_l}\nonumber
\een whereby
 $$
 H(\rho'_{ABC})=H(\rho_{AB}).
 $$
Also, from Eq. \ref{extrho},
 \bes
\label{partials}
 \ben
 \rho'_{AB}&=&\sum_{j}p_j\rho_{A|j}\otimes\proj{e_j},\;\;\;\mbox{so}\;\;\;H(\rho'_{AB})=H({\bm p})+\sum_j p_jH(A|j),\\
 \rho'_{BC}&=&\sum_{j,k}\proj{e_j}\rho_B\proj{e_k}\otimes\ket{f_j}\bra{f_k},\;\;\;\mbox{so}\;\;\;H(\rho'_{BC})=H(\rho_B),\\
  \rho'_{B}&=&\sum_{j}p_j\proj{e_j},\;\;\;\mbox{so}\;\;\;H(\rho'_{B})=H({\bm p}).
 \een
 \ees
Now use the strong subadditivity of the von-Neumann entropy which
is
 \be
 \label{ssa}
 H(\rho'_{ABC}) + H(\rho'_{B}) \leq H(\rho'_{AB}) + H(\rho'_{BC}).
 \ee
Eqs. \ref{partials} reduces this to
 \be
 H(\rho_{AB}) + H({\bm p}) \leq H({\bm p})+\sum_j p_jH(A|j) + H(\rho_B),
 \ee
whereby
 \be
 \tilde H_{\{\Pi_j\}}(A|B) \equiv \sum_j p_jH(A|j)\geq H(\rho_{AB})- H(\rho_B)
\equiv H(A|B).
 \ee
This, being true for all measurements, also holds for the minimum.
So
 $$
\mathcal{D}(A,B) = \min_{\{\Pi_j\}}\tilde H_{\{\Pi_j\}}(A|B) -
H(A|B) \geq 0.
 $$ \qed

Having proved that the quantum discord is always nonnegative, it
is worthwhile to seek the condition for it being zero. The reason
is that the set of states with zero discord is exactly those which
have no nonclassical correlations in it. This is the aim of the
next theorem. It is evident that the condition for zero discord
can be reduced to that of the equality in strong subadditivity in
Eq \ref{ssa}. To that end, we will employ a result of Hayden
\textit{et al}.~\cite{hjpw04} which we present below for
completeness.

 \begin{lemma}[\cite{hjpw04}]
 \label{ssaeq}
A state $\rho'_{ABC}$ on $\mathcal{H}_A\otimes
\mathcal{H}_B\otimes \mathcal{H}_C$ satisfies strong subadditivity
(Eq. \ref{ssa}) with equality if and only if there is a
decomposition of system $B$ as
 $$
 \mathcal{H}_B = \bigoplus_j \mathcal{H}_{B^L_j}\otimes \mathcal{H}_{B^R_j}
 $$
into a direct sum of tensor products such that
 $$
\rho'_{ABC} = \bigoplus_j q_j \rho_{AB^L_j}\otimes \rho_{AB^R_j}
  $$
with states $\rho_{AB^L_j}$ on $\mathcal{H}_A\otimes
\mathcal{H}_{B^L_j}$ and $\rho_{AB^R_j}$ on
$\mathcal{H}_{B^R_j}\otimes\mathcal{H}_C$, and a probability
distribution $\{q_j\}.$
\end{lemma}

\begin{theorem}
 \label{zerodiscord}  $\mathcal{D}(A,B)= 0$ if and only if
the state $\rho_{AB}$ is block diagonal in its own eigenbasis,
that is
 $$
 \rho_{AB} = \sum_{j}P_j\rho_{AB}P_j
 $$
where $\rho_{AB}=\sum_j \tau_j P_j$, with $\{\tau\}$ a probability
distribution.
\end{theorem}

\noindent{\it Proof:} The decomposition of the Hilbert space of
$B$ can be written as
 $$
 \mathbb{I}_B = \sum_{\alpha}\Pi_{\alpha} =\sum_{\alpha}\Pi_{\alpha
L}\otimes \Pi_{\alpha R},
 $$
and $\Pi_{\alpha}\Pi_{\beta}=\delta_{\alpha\beta}\Pi_{\alpha}.$ In
our case, the state $\rho'_{ABC}$ is invariant under the exchange
of $B$ and $C$ relative to the measurement basis, here denoted by
$\ket{E_{\alpha j}}$ and $\ket{F_{\alpha j}}$. Thus using
Lemma~\ref{ssaeq}, we conclude that it must have the form
 $$
\rho'_{ABC} = \sum_{\alpha}q_{\alpha} \rho_{A|\alpha}\otimes
\rho^{\alpha}_{BC},
 $$
where $\rho^{\alpha}_{BC} =
\Pi_{\alpha}\rho^{\alpha}_{BC}\Pi_{\alpha},$ with $\Pi_{\alpha}$
being projectors of the form
 $$
 \Pi_{\alpha} = \sum_{j}\proj{E_{\alpha j}}\otimes \proj{F_{\alpha
j}}.
 $$
Thus
 $$
 \rho^{\alpha}_{BC} = \sum_{j,k}\rho^{\alpha}_{jk}\ket{E_{\alpha j}}\bra{E_{\alpha
k}}\otimes \ket{F_{\alpha j}}\bra{F_{\alpha k}}
 $$
and
 $$
 \rho'_{AB} = \sum_{\alpha}q_{\alpha} \rho_{A|\alpha}\otimes
\rho^{\alpha}_{B} = \sum_{\alpha,j}
\rho^{\alpha}_{jj}q_{\alpha}\rho_{A|\alpha}\otimes \ket{E_{\alpha
j}}\bra{E_{\alpha j}}.
 $$
Undoing the measurement,
$U\ket{e_j}\otimes\ket{0}=\ket{e_j}\otimes\ket{f_j}$, gives
 $$
\rho_{AB}=\bra{0_C}U^{\dag}\rho'_{ABC}U\ket{0_C}=\sum_{\alpha}q_{\alpha}
\rho_{A|\alpha}\otimes \rho^{\alpha}_{B}.
 $$
Diagonalizing $\rho^{\alpha}_B=\sum_j \lambda^{\alpha}_j
\proj{\lambda^{\alpha}_j}$, we get
 $$
 \rho_{AB}=\sum_{\alpha,j}\lambda^{\alpha}_j
q_{\alpha}\rho_{A|\alpha}\proj{\lambda^{\alpha}_j}.
 $$
Relabelling, we have that the discord is zero if and only if
 \be
\rho_{AB}=\sum_{j} p_j \rho_{A|j}\otimes \proj{\lambda_j}
 \ee
in the basis that diagonalizes $\rho_B.$ The $\alpha$ subspaces
take into account that if the states $\rho_{A|j}$ are the same for
different $j$, then we can attain zero discord by using any
measurement in the subspace spanned by those values of $j$.

  Diagonalising $\rho_{A|j} = \sum_k \mu_{jk}\proj{\mu_{jk}}$, we
get that a state has zero discord if and only if
 \be
 \label{treebasis}
\rho_{AB} = \sum_{jk} p_j\mu_{jk}\proj{\mu_{jk},\lambda_j}.
 \ee
Thus the eigenbasis of $\rho_{AB}$ has a tree product structure
$\ket{\mu_{jk}}\otimes\ket{\lambda_j}$. From Eq. \ref{treebasis},
it is also evident that a state has zero discord if and only if it
is block diagonal in its eigenbasis, that is,
 \be
\mathcal{D}(A,B) =0 \;\;\;\mbox{iff} \;\;\;\rho_{AB} =
\sum_jP_j\rho_{AB}P_j
 \ee
with
 $$
 P_j = \sum_k \proj{\mu_{jk},\lambda_j}.
 $$
which is the statement of the theorem. \qed

It is now a valid question to ask for the maximum possible value
of discord. The next theorem addresses just this point.

\begin{theorem}
 \label{maxdiscord}
The value of quantum discord is upper bounded by the von-Neumann
entropy of the measured subsystem, \textit{i.e.},
$\mathcal{D}(A,B)\leq H(A)$.
\end{theorem}

\noindent{\it Proof:} Start with the eigendecomposition
 $$
\rho_{AB}=\sum_a p_a\Pi_a,
$$
which yields
 $$
p_j\rho_{B|j}= \sum_a p_a p_{j|a}\rho_{B|a,j},
$$
where
$$
\rho_{B|a,j}=\tr_A(\Pi_j\rho_B\Pi_a)/p_{j|a},
$$
is a pure state of $B$. It follows from the pure-state
decomposition
$$
\rho_{B|j}=\sum_a p_{a|j}\rho_{B|a,j},
$$
that $H(\{a\}|j)\ge H(\rho_{B|j})$.  Thus
 \ben
H(A,B)&=&H(\{a\}) \ge H(\{a\}|j) \nonumber\\
  &=&    \sum_j p_j H(\{a\}|j) \nonumber\\
 &\ge&  \tilde H_{\{\Pi_j\}}(B|A) \nonumber\\
 &\ge& \tilde H(B|A),
 \een
from which the marginal entropy $H(A)$ follows as the upper bound
on discord. \qed

    When the joint state $\rho_{AB}$ is pure, $H(A,B)$ and $\tilde
H(B|A)$ are zero, $H(B)=H(A)=-H(B|A)$, and the discord is equal to
its maximum value, $H(A)$, which is a measure of entanglement for
bipartite pure states. In other words, for pure states all
nonclassical correlations characterized by quantum discord can be
identified as entanglement as measured by the marginal entropy.

        Coming back to the question of mixed quantum states, we
find that the actual evaluation of quantum discord involves the
minimization of an entropic quantity over the set of all POVMS, as
shown in Eq~\ref{discorddef}. This minimization, though perhaps
easier than the one involved in the convex roof optimization
needed for certain measures of entanglement, is nevertheless not
trivial. Some simplification is in order as we show in the
following two theorems.
\begin{theorem}
 \label{concavediscord}
The quantity $\mathcal{D}(A,B)=H(B|A)-\tilde H_{\mathcal{E}}(B|A)$
is a concave function on the set of POVMs $\mathcal{E}$.
\end{theorem}
\noindent{\it Proof:} We start by stating that for quantum states
the von Neumann entropy is concave, that is,
 \be
  H(\lambda \rho_1 + (1-\lambda)\rho_2) \geq  \lambda H(\rho_1) + (1-\lambda)H(\rho_2)
   \ee
for all $0 \leq \lambda \leq 1$. To study the concavity of the
classical measure $\tilde H(B|A)$, let introduce Positive Operator
Valued Measurements (POVMs) given by a set of quantum operations
 \be
 \mathcal{A}_j = \sum_{\alpha}A_{j\alpha}\odot A_{j\alpha}^{\dag},
 \ee
with
  \be
  \sum_{\alpha}A_{j\alpha}^{\dag}A_{j\alpha}=E_j\;\;\;\mbox{and}\;\;\;\sum_j
E_j = \mathbb{I}. \nonumber
  \ee
Not confusing the POVM elements $A_{j\alpha}$ with the system $A$
on which they are executed, the post-measurement state of $B$ is
 \ben
 \rho_{B|j} &=& \tr_{A}(\mathcal{A}_j(\rho_{AB}))\Big/p_j = \sum_{\alpha}\tr_{A}(A_{j\alpha}\rho_{AB} A_{j\alpha}^{\dag})\Big/p_j \nonumber\\
            &=& \tr_{A}(\rho_{AB}\sum_{\alpha}A_{j\alpha}^{\dag}A_{j\alpha})\Big/p_j \nonumber\\
            &=& \tr_{A}(\rho_{AB}E_j)\Big/p_j.
 \een
The post-measurement state is thus completely determined by the
POVM. Now let us call the POVM $\mathcal{E}=\{E_j\}$ and define
the functional $F(\rho_B,\mathcal{E})$ as
 \be
 F(\rho_B,\mathcal{E}) \equiv \tilde H(B|A) = \sum_j p_{j} H(\rho_{B|j}).
 \ee
Let $\mathcal{C}$ and $\mathcal{D}$ be two POVMs and $\mathcal{E}
= \lambda\,\mathcal{C} + (1-\lambda)\, \mathcal{D}$ their
combination. $\mathcal{E}$ is certainly a POVM since the set of
POVMs is convex. This is, in effect, tantamount to
 \be
E_j = \lambda\, C_j + (1-\lambda)\,D_j \nonumber.
 \ee
Then it is simple to show that
 $$
 p_{E_j}\,\rho_{S|E_j} = \lambda\, p_{C_j}\,\rho_{B|C_j} + (1-\lambda)\,p_{D_j}\,\rho_{B|D_j},\;\;\;
 p_{E_j} = \lambda\,p_{C_j} + (1-\lambda)\,p_{D_j}.
 $$
Using these two relations and the fact that the von-Neumann
entropy is concave, it can be seen that
 \ben
 F(\rho_B,\mathcal{E}) &=& \sum_{j}p_{E_j} H(\rho_{B|E_j}) \nonumber\\
                    &\geq& \sum_j\, \lambda p_{C_j}H(\rho_{B|C_j}) + (1-\lambda)\, p_{D_j}H(\rho_{B|D_j})  \nonumber\\
                    &=& \lambda\, F(\rho_B,\mathcal{C}) +(1-\lambda)\,F(\rho_B,\mathcal{D}).
 \een
This shows that the objective function which is to be minimized to
obtain the discord is concave over the convex set of POVMs . \qed

Thus the minima that provides the value of discord will always be
attained at the extremal points of the set of POVMs. The extreme
points of the set of POVMs on a $D$ dimensional system is no more
than $D^2$ dimensional; a necessary and sufficient condition for
its extremality being that the eigenvectors of the POVM elements
be linearly independent as operators~\cite{dpp05a}. It consists of
projectors of all ranks. Fortunately, the next theorem will
provide a considerable simplification.

\begin{theorem}
 \label{rankone}
The minimum of the quantity $\mathcal{D}(A,B)=H(B|A)-\tilde
H_{\mathcal{E}}(B|A)$ is always attained for a rank-1 POVM.
\end{theorem}
\noindent{\it Proof:} We start by supposing a POVM on system $A$,
whose elements can be fine-grained as
  $$
 E_j = \sum_{k}E_{jk}.
  $$
Then
 $$
p_{jk}\rho_{B|jk} =\tr_{A}(\rho_{AB}E_{jk}),\;\;\;p_{jk} =
\tr(\rho_{AB}E_{jk}).
 $$
Evidently, $\sum_k p_{jk}=p_j$ whereby we can define $ p_{k|j} =
p_{jk}/p_j. $ Also,
 \be
\rho_{B|j}= \tr_A(\rho_{AB}E_j)/p_j = \sum_k \frac{p_{jk}}{p_j}
\tr_A(\rho_{AB}E_{jk})/p_{jk}=\sum_k p_{k|j}\rho_{B|jk}.
 \ee
Now,
 \ben
 \sum_j p_jH(\rho_{B|j})&=&\sum_j p_jH\left(\sum_k p_{k|j}\rho_{B|jk}\right)\nonumber\\
                        &\geq& \sum_{j,k} p_jp_{k|j}H(\rho_{B|jk}) \nonumber\\
                        &=& \sum_{j,k}p_{jk} H(\rho_{B|jk}).
 \een
Since any POVM element can be written in terms of its
eigendecomposition, the minimum conditional entropy, and therefore
the discord is always attained on a rank-1 POVM. \qed

A rank-1 POVM is extremal if and only if its elements are linearly
independent \cite{dpp05a}. Even after these simplifications, the
minimization over all rank one POVMs remains a hard task for
general Hilbert spaces. We are not aware of the exact
computational complexity of the problem, which might be
formidable, but one can envisage casting this as a semi-definite
program. This particular aspect is still under investigation by
us. Accordingly, this thesis deals only with instances where the
minimization is done over a two dimensional Hilbert space.

        In light of the present scenario, we conclude the
discussion on the evaluation of quantum discord by explicitly
working out an example. The state we consider is the well known $2
\times 4$ bound entangled state due to Horodecki \cite{h97},
written in the standard product basis as
 \be
 \label{E:Horodstate}
 \rho = \frac{1}{1+7p}\left(%
\begin{array}{cccccccc}
  p & 0 & 0 & 0 & 0 & p & 0 & 0 \\
  0 & p & 0 & 0 & 0 & 0 & p & 0 \\
  0 & 0 & p & 0 & 0 & 0 & 0 & p \\
  0 & 0 & 0 & p & 0 & 0 & 0 & 0 \\
  0 & 0 & 0 & 0 & \frac{1+p}{2} & 0 & 0 & \frac{\sqrt{1-p^2}}{2} \\
  p & 0 & 0 & 0 & 0 & p & 0 & 0 \\
  0 & p & 0 & 0 & 0 & 0 & p & 0 \\
  0 & 0 & p & 0 & \frac{\sqrt{1-p^2}}{2} & 0 & 0 & \frac{1+p}{2} \\
\end{array}%
\right)\;\;\;\;\;\;\mbox{for}\;\;\;\; 0 \leq p \leq 1.
 \ee
 That this state is bound entangled for all $0\leq p\leq 1$can be concluded from its
positive partial transpose, and its entanglement is hidden in a
very subtle way. The calculation of the discord begins with the
evaluation of the von-Neumann entropy of $\rho$. Its spectrum is
 \ben
 \bm{\lambda}&=&\Bigl\{0,0,0,\frac{p}{1 + 7 p}, \frac{2p}{1 + 7 p},\frac{2p}{1 + 7 p}, \nonumber\\ && \hspace{0.0cm}
 \frac{1 + 9 p + 14 p^2 - \sqrt{1 + 12 p + 23 p^2 - 70 p^3 + 98 p^4}}{2(1 + 14 p + 49 p^2)}, \nonumber\\ && \hspace{0.0cm}
 \frac{1 + 9 p + 14 p^2 + \sqrt{1 + 12 p + 23 p^2 - 70 p^3 + 98 p^4}}{2 (1 + 14 p + 49 p^2)}\,\Bigr\},
 \een
from which $ H(\rho)= H(\bm{\lambda}) =
-\tr(\bm{\lambda}\log(\bm{\lambda}))$.  The reduced density matrix
of the smaller subsystem is
 \be
\rho_M = \tr_S(\rho) = \left(%
\begin{array}{cc}
  \frac{4p}{1+7p} & 0 \\
  0 & \frac{1+3p}{1+7p} \\
\end{array}%
\right),
 \ee
whose von-Neumann entropy is trivially obtained. As is clear by
now, we have chosen to make our measurement on the 2 dimensional
subsystem. These projectors are generally defined as
 \be
 \label{proj}
 \Pi_1 = \frac{\mathbb{I}_2 + \bm{a}\cdot\bm{\sigma}}{2}\otimes \mathbb{I}_4,\;\;\;\;\;\;\;\; \Pi_2 = \frac{\mathbb{I}_2 - \bm{a}\cdot\bm{\sigma}}{2}\otimes
\mathbb{I}_4,
 \ee
with $a_1 = \sin\theta\cos\phi, a_2 = \sin\theta\sin\phi, a_3=
\cos\theta$. Using these, we calculate
 \bes
 \ben
 p_1 &=& \tr(\Pi_1\rho\Pi_1)=\frac{1 + 7 p - \cos\theta + p\cos\theta}{2+14p}, \\
 p_2 &=& \tr(\Pi_2\rho\Pi_2)=\frac{1 + 7 p + \cos\theta - p\cos\theta}{2+14p},
 \een
 \ees
and
 \ben
 \label{rhored1}
 \rho_1 &=& \frac{\tr_M(\Pi_1\rho\Pi_1)}{p_1} = \frac{1}{4(1+7p)p_1}\;\times\\
    && \!\!\!\!\! \left(%
\begin{array}{cccc}
  1+3p-\cos\theta(1-p) & 2p\sin\theta e^{i\phi} & 0 & \sqrt{1-p^2}(1-\cos\theta) \\
  2p\sin\theta e^{-i\phi} & 4p & 2p\sin\theta e^{i\phi} & 0 \\
  0 & 2p\sin\theta e^{-i\phi} & 4p  & 2p\sin\theta e^{i\phi} \\
   \sqrt{1-p^2}(1-\cos\theta) & 0 & 2p\sin\theta e^{-i\phi} & 1+3p-\cos\theta(1-p) \\
\end{array}%
\right), \nonumber
 \een
 \ben
 \label{rhored2}
\rho_2 &=& \frac{\tr_M(\Pi_2\rho\Pi_2)}{p_2}=\frac{1}{4(1+7p)p_2}\;\times  \\
    && \!\!\!\!\!\left(%
\begin{array}{cccc}
  1+3p+\cos\theta(1-p) & -2p\sin\theta e^{i\phi} & 0 & \sqrt{1-p^2}(1+\cos\theta) \\
  -2p\sin\theta e^{-i\phi} & 4p & -2p\sin\theta e^{i\phi} & 0 \\
  0 & -2p\sin\theta e^{-i\phi} & 4p  & -2p\sin\theta e^{i\phi} \\
   \sqrt{1-p^2}(1+\cos\theta) & 0 & -2p\sin\theta e^{-i\phi} & 1+3p+\cos\theta(1-p) \\
\end{array}%
\right) \nonumber.
 \een

\begin{figure}
\begin{center}
\includegraphics[scale=1.0]{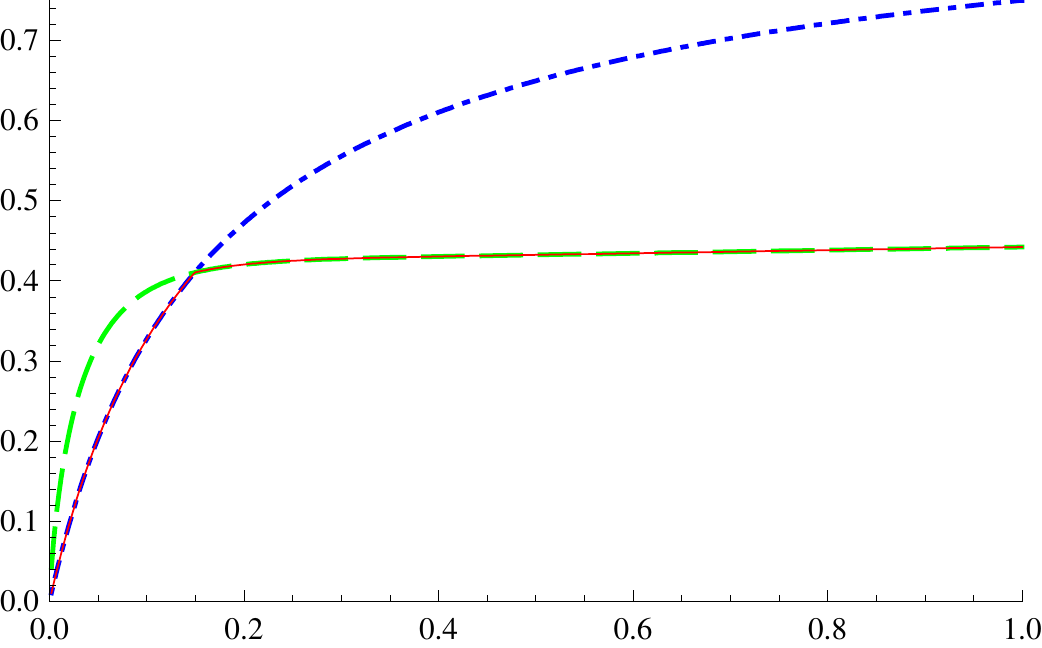}
\caption[Quantum discord in a $2\times 4$ bound entangled state]{
The green dashed line is the value of the discord when the
projector (Eq~\ref{proj}) is parameterized by $\theta =
\pi/2,\phi=0$. The blue dot-dashed line is when the projector is
given by $\theta = 0,\phi=0$. The solid red line, which is the
smaller of the two, is the quantum discord in the $2 \times 4$
bound entangled Horodecki state of Eq~\ref{E:Horodstate} obtained
by explicit numerical minimization. The expression for the discord
is given by Eq~\ref{E:discordhorod}. We see how the optimal
projector changes from the former to the latter at $p=1/7$. This
exemplifies how the optimal measurement can change even for the
same family of states and therefore is tricky to obtain
analytically.} \label{F:boundhorod}
\end{center}
\end{figure}

    This state provides a very good example when optimizing over the
POVM though seemingly tractable analytically, actually leads to
erroneous conclusions. This is due to the fact that the optimal
measurement that minimizes the conditional entropy changes with
the value of $p$. We can however do it numerically without much
ado. The value of discord in this case is
 \be
 \label{E:discordhorod}
\mathcal{D} =
H_2\left(\frac{4p}{1+7p}\right)+\tr(\bm{\lambda}\log(\bm{\lambda}))
+\min_{(\theta,\phi)}\left[p_1 H(\rho_1)+p_2 H(\rho_2)\right],
 \ee
$H_2(\cdot)$ being the binary Shannon entropy. We plot the result
of the numerical minimization as the green line in
Fig~\ref{F:boundhorod}. The plot was obtained in a couple of
minutes on a 1.83 GHz Dell XPS laptop. In addition to
demonstrating the feasibility of the minimization in small
instances, Fig~\ref{F:boundhorod} shows one very important
feature. This state is bound entangled, meaning it has
entanglement that cannot be distilled \cite{hhhh07}. Quantifying
the amount of entanglement in a quantum state is generally hard,
harder still for bound entangled states. As we have now shown,
quantum discord can be used to used to quantify non-classical
correlations in such states, which further makes discord a useful
tool in the study of quantum states. Additionally, we are going to
demonstrate an analytical technique that might be useful in the
minimization as well its caveats. Let us now proceed towards that.

       It is worthwhile to point out at the outset that the optimal measurements
that give us the actual value of discord are \emph{not} given by
the eigenvectors of the reduced density matrices. Indeed, an
independent measure of non-classicality has been proposed where
measurements are made on both the subsystems, and that too in the
eigenbasis of their respective subsystems \cite{Luo08}. In the
minimization of quantum discord, the hard part  is often using the
explicit analytical expressions of the eigenvalues of the states
in Eqs~\ref{rhored1}, \ref{rhored2}. Here is a handy trick that is
often helpful. Consider $p=1$. In that case the Eqs~\ref{rhored1},
\ref{rhored2} simplify drastically to

 \be
 \label{rho1p1}
 \rho_1 = \left(%
\begin{array}{cccc}
  \frac{1}{4} & \frac{\sin\theta e^{i\phi}}{8} & 0 & 0 \\
  \frac{\sin\theta e^{-i\phi}}{8} & \frac{1}{4} & \frac{\sin\theta e^{i\phi}}{8} & 0 \\
  0 & \frac{\sin\theta e^{-i\phi}}{8} & \frac{1}{4} & \frac{\sin\theta e^{i\phi}}{8} \\
  0 & 0 & \frac{\sin\theta e^{i\phi}}{8} & \frac{1}{4} \\
\end{array}%
\right),
 \ee
and
 \be
\label{rho2p1}
 \rho_2 = \left(%
\begin{array}{cccc}
    \frac{1}{4} & -\frac{\sin\theta e^{i\phi}}{8} & 0 & 0 \\
  -\frac{\sin\theta e^{-i\phi}}{8} & \frac{1}{4} & -\frac{\sin\theta e^{i\phi}}{8} & 0 \\
  0 & \frac{\sin\theta e^{-i\phi}}{8} & \frac{1}{4} & -\frac{\sin\theta e^{i\phi}}{8} \\
  0 & 0 & -\frac{\sin\theta e^{i\phi}}{8} & \frac{1}{4} \\
\end{array}%
\right),
 \ee
both of which have spectrum
 \ben
\label{E:eigvec}
 \bm{\lambda}(\rho_1)&=& \bm{\lambda}(\rho_1)\nonumber\\
&=&
\Bigl\{\frac{4-\sqrt{6-2\sqrt5}\sin\theta}{16},\frac{4+\sqrt{6-2\sqrt5}\sin\theta}{16}, \nonumber\\
 &&\hspace{0.0cm} \frac{4-\sqrt{6+2\sqrt5}\sin\theta}{16},\frac{4+\sqrt{6+2\sqrt5}\sin\theta}{16}\Bigl\}.\een
Simultaneously, $p_1=p_2=1/2.$ Then the conditional quantum
entropy is just $H(\bm{\lambda}(\rho_1))$, the Shannon entropy of
the vector in Eq \ref{E:eigvec}, which is minimized when the
distribution is most disparate, that is, $\theta = \pi/2.$ We note
that $\phi$, having dropped out of the problem, can be set to
zero. We now have at least one point where we have carried out the
minimization explicitly. One might now be tempted to believe that
this point gives the minimum for all $p$, and we would have
succeeded in the minimization without doing it explicitly.
Fig~\ref{F:boundhorod} shows the catch. The red line is the
expression for discord obtained when setting $\theta = \pi/2,
\phi=0$. We see that it matches with the actual minima for a range
of $p$, but not all of it. The region where they diverge is $p \in
\{0,1/7\}.$ Interestingly enough, in this region, the optimal
choice of the projector is given by $\theta=0,\phi=0$. This is
shown by the blue line, which we see lies exactly on top of the
green line obtained by explicit numerical minimization in $p \in
\{0,1/7\}$.

        To reemphasize the point made at the beginning of the last
paragraph, we present the eigenvectors of $\mathbf{E}(\rho_1)$ and
$\mathbf{E}(\rho_2)$ of $\rho_1$ and $\rho_2$ in Eqs. \ref{rho1p1}
and \ref{rho2p1} respectively. For all values of $\theta$ and
$\phi$,
 \be
\mathbf{E}(\rho_1) = \left\{\left(%
\begin{array}{c}
  e^{3i\phi} \\
  -\frac{\sqrt{5}-1}{2}e^{2i\phi} \\
  -\frac{\sqrt{5}-1}{2}e^{i\phi} \\
  1 \\
\end{array}%
\right),
\left(%
\begin{array}{c}
  -e^{3i\phi} \\
  -\frac{\sqrt{5}-1}{2}e^{2i\phi} \\
  -\frac{\sqrt{5}-1}{2}e^{i\phi} \\
  1 \\
\end{array}%
\right),
\left(%
\begin{array}{c}
  -e^{3i\phi} \\
  \frac{\sqrt{5}+1}{2}e^{2i\phi} \\
  -\frac{\sqrt{5}+1}{2}e^{i\phi} \\
  1 \\
\end{array}%
\right),
\left(%
\begin{array}{c}
  e^{3i\phi} \\
  \frac{\sqrt{5}+1}{2}e^{2i\phi} \\
  \frac{\sqrt{5}+1}{2}e^{i\phi} \\
  1 \\
\end{array}%
\right)\right\},
 \ee
and
 \be
\mathbf{E}(\rho_2) = \left\{\left(%
\begin{array}{c}
  -e^{3i\phi} \\
  -\frac{\sqrt{5}-1}{2}e^{2i\phi} \\
  \frac{\sqrt{5}-1}{2}e^{i\phi} \\
  1 \\
\end{array}%
\right),
\left(%
\begin{array}{c}
  e^{3i\phi} \\
  -\frac{\sqrt{5}-1}{2}e^{2i\phi} \\
  -\frac{\sqrt{5}-1}{2}e^{i\phi} \\
  1 \\
\end{array}%
\right),
\left(%
\begin{array}{c}
  e^{3i\phi} \\
  \frac{\sqrt{5}+1}{2}e^{2i\phi} \\
  \frac{\sqrt{5}+1}{2}e^{i\phi} \\
  1 \\
\end{array}%
\right),
\left(%
\begin{array}{c}
  -e^{3i\phi} \\
  \frac{\sqrt{5}+1}{2}e^{2i\phi} \\
  -\frac{\sqrt{5}+1}{2}e^{i\phi} \\
  1 \\
\end{array}%
\right)\right\},
 \ee
and the ordering of the eigenvectors correspond to the eigenvalues
in Eq \ref{E:eigvec}. As is evident, these is no $\theta$
dependence in these vectors. Accordingly, it is no surprise that
they cannot provide any information about the optimal measurement
that appears in the calculation of discord.

        As mentioned before, and reiterated by this example, there is a
gap between discord and entanglement. Certainly, zero discord
states are separable. The converse is, in general \emph{not} true
for mixed quantum states. In light of this, a valid question is
whether there is a bound on the value of discord for separable
states. This result can be used indirectly to detect an entangled
state. In the following section we inquire about the value of
quantum discord for separable states.

\subsection{Discord for separable states}\label{sepdiscord}

        Let us consider a separable state
 $$
 \rho_{AB}=\sum_i p_i\; \rho^A_i \otimes \rho^B_i,
 $$
whereby
 $$
 \rho_{A}=\sum_i p_i\; \rho^A_i \;\;\;\mbox{and} \;\;\;\rho_{B}=\sum_i p_i\;
 \rho^B_i.
 $$
The quantum discord of a quantum state is defined as
 $$
 \mathcal{D} = H(\rho_A) -H(\rho_{AB}) + \min_{\{\Pi_j\}}\sum_j q_j\;H(\rho_{B|j}),
 $$
where
 $$
\rho_{B|j} = \frac{\tr_A\bigl(\Pi_j\rho_{AB}\Pi_j\bigr)}{q_j},
\qquad q_j=\tr_{A,B}\bigl(\rho_{AB}\Pi_j\bigr).
 $$
It is known that for a separable state
 \be
 \label{sepent}
 H(\rho_{AB}) \geq H(\rho_{A}) \;\;\;\mbox{and}\;\;\;H(\rho_{AB}) \geq
 H(\rho_{B}),
 \ee
whereby
 \ben
 \mathcal{D} &=& H(\rho_A) -H(\rho_{AB}) + \min_{\{\Pi_j\}}\sum_j q_j\;H(\rho_{B|j}) \nonumber \\
             &\leq& \min_{\{\Pi_j\}}\sum_j q_j\;H(\rho_{B|j}) \nonumber \\
             &\leq& \min_{\{\Pi_j\}}\;H\left(\sum_j q_j \rho_{B|j}\right),
 \een
where the last inequality follows from the concavity of $H$. Now
 \ben
 \sum_j q_j \rho_{B|j} &=& \sum_j \tr_A\left(\Pi_j\rho_{AB}\Pi_j\right)
                       = \sum_{j}\tr_A\left(\Pi_j\;\rho_{AB}\right) \nonumber\\
                       &=& \tr_A\biggl(\bigl(\sum_j\Pi_j\bigr)\rho_{AB}\biggr)
                       = \rho_B.
 \een
  Therefore, we have,
   $$
  \mathcal{D} \leq H(\rho_B).
   $$
Of course, one can choose not to invoke the separability condition
first, and rather say that (since $\mathcal{J} \geq 0$)
  \be
  \mathcal{D} \leq H(\rho_A) + H(\rho_B) - H(\rho_{AB}),
  \ee
and then employ (\ref{sepent}) to conclude
  \be
  \mathcal{D} \leq \min\bigl(H(A),H(B),\mathcal{I}(A:B)\bigr),
  \ee
  which is probably a more sensible upper bound for separable
  states. This, in fact, leads to the conclusion that $\mathcal{D}=0$
  for pure separable states.

\section{Quantum discord in quantum communication}
\label{S:cirac}

        The centerpiece of this thesis is undoubtedly mixed-state quantum
information. Nonetheless, we have an instance from quantum
communication where quantum discord might have a role to play. We
will not delve into a general analysis of this; rather we will
point out the role of quantum discord through a simple example. We
also thank W. H. Zurek for pointing out the problem.

        The phenomenon we have in mind is the distribution of
entanglement. The significance of this task cannot be
overemphasized in quantum information science. To be able to
entangle two distant particles is at the heart of quantum
communication and computation. That this can be achieved via a
third mediating particle which remains separable at all times from
the other two was shown in \cite{cvdc03}. Let us, in very brief,
demonstrate this claim.

        We start with a tripartite state $\rho_{abc}$, where $a$
and $b$ denote the systems of Alice and Bob, while $c$ is the
mediator particle or channel, which starts out in Alice's
possession. In our particular example, taken from \cite{cvdc03},
 \be
\rho_{abc} =
\frac{1}{6}\sum_{k=0}^3\ket{\Psi_k,\Psi_{-k},0}\bra{\Psi_k,\Psi_{-k},0}
+ \frac{1}{6}\sum_{i=0}^1\ket{i,i,1}\bra{i,i,1},
 \ee
where $$\ket{\Psi_k}=\frac{\ket{0}+e^{ik\pi/2}\ket{1}}{\sqrt 2}.$$
This state is separable between all three parties as it is a
convex combination of product states. In the standard product
basis,
 \be
 \rho_{abc} = \frac{1}{6}\left(%
\begin{array}{cccccccc}
  1 & 0 & 0 & 0 & 0 & 0 & 1 & 0 \\
  0 & 1 & 0 & 0 & 0 & 0 & 0 & 0 \\
  0 & 0 & 1 & 0 & 0 & 0 & 0 & 0 \\
  0 & 0 & 0 & 0 & 0 & 0 & 0 & 0 \\
  0 & 0 & 0 & 0 & 1 & 0 & 0 & 0 \\
  0 & 0 & 0 & 0 & 0 & 0 & 0 & 0 \\
  1 & 0 & 0 & 0 & 0 & 0 & 1 & 0 \\
  0 & 0 & 0 & 0 & 0 & 0 & 0 & 1 \\
\end{array}%
\right).
 \ee

        Firstly, Alice performs a controlled NOT ($\mathrm{CNOT}$) gate
operation between $a$ and $c$ (with $a$ as the control), producing
the state
 \be
\sigma_{abc} =
\frac{1}{6}(\ket{000}+\ket{111})(\bra{000}+\bra{111}) +
\frac{1}{6}\sum_{i,j,k=0}^1\beta_{ijk}\ket{ijk}\bra{ijk},
 \ee
with all $\beta$'s being 0 except for
$\beta_{001}=\beta_{010}=\beta_{101}=\beta_{110}=1.$ The particle
$c$ is then sent over to Bob. Since the $\mathrm{CNOT}$ was
executed between $a$ and $c$, the state must be separable with
respect to the $b$-$ac$ split. $\sigma_{ab}$ and $\sigma_{bc}$ are
also separable. Also, the state is separable under the $c$-$ab$
split as the state is invariant under the exchange of $b$ and $c$.
Likewise, $\sigma_{ac}$ is separable. The $a$-$bc$ split is more
interesting though. For this we write in the standard product
basis,
 \be
 \sigma_{abc} = \frac{1}{6}\left(%
\begin{array}{cccccccc}
  1 & 0 & 0 & 0 & 0 & 0 & 0 & 1 \\
  0 & 1 & 0 & 0 & 0 & 0 & 0 & 0 \\
  0 & 0 & 1 & 0 & 0 & 0 & 0 & 0 \\
  0 & 0 & 0 & 0 & 0 & 0 & 0 & 0 \\
  0 & 0 & 0 & 0 & 0 & 0 & 0 & 0 \\
  0 & 0 & 0 & 0 & 0 & 1 & 0 & 0 \\
  0 & 0 & 0 & 0 & 0 & 0 & 1 & 0 \\
  1 & 0 & 0 & 0 & 0 & 0 & 0 & 1 \\
\end{array}%
\right).
 \ee
Then the spectrum of the partial transpose of $\sigma_{abc}$ with
respect to $a$ is
 $$
\left\{-\frac{1}{6},\frac{1}{6},\frac{1}{6},\frac{1}{6},\frac{1}{6},\frac{1}{6},\frac{1}{6},\frac{1}{6}\right\},
 $$
which has a negative eigenvalue. Hence, the $a$-$bc$ split
actually contains entanglement.

        Once Bob receives the particle, he does a $\mathrm{CNOT}$
between $b$ and $c$, with the former as the control. The resulting
state is
 \be
 \tau_{abc} = \frac{1}{3}\ket{\Phi^+}\bra{\Phi^+} \otimes
\ket{0}\bra{0} + \frac{2}{3} \mathbb{I}_{ab}\otimes
\ket{1}\bra{1},
 \ee
where $\ket{\Phi^+} = (\ket{00}+\ket{11})/\sqrt 2$ is the
maximally entangled state. In the standard product basis,
 \be
 \tau_{abc} = \frac{1}{6}\left(%
\begin{array}{cccccccc}
  1 & 0 & 0 & 0 & 0 & 0 & 1 & 0 \\
  0 & 1 & 0 & 0 & 0 & 0 & 0 & 0 \\
  0 & 0 & 0 & 0 & 0 & 0 & 0 & 0 \\
  0 & 0 & 0 & 1 & 0 & 0 & 0 & 0 \\
  0 & 0 & 0 & 0 & 0 & 0 & 0 & 0 \\
  0 & 0 & 0 & 0 & 0 & 1 & 0 & 0 \\
  1 & 0 & 0 & 0 & 0 & 0 & 1 & 0 \\
  0 & 0 & 0 & 0 & 0 & 0 & 0 & 1 \\
\end{array}%
\right).
 \ee
The particle $c$ is still separable from the other two particles,
which we have managed to get entangled. However, the $a$-$bc$
split and the $b$-$ac$ split have the same amount of entanglement
as in the $a$-$bc$ split of $\sigma_{abc}$.  One uses this
entanglement to get a Bell state by measuring $c$ in the standard
basis. On average, we can thus use this $1$ ebit of entanglement
$1/3$ of the time for any desired purpose.

            Having demonstrated our claim, we now invest in a
little introspection. If entanglement is to behave as a resource,
one would expect some nature of conservation law to hold. At
least, we should be able to argue that the entanglement we
generated was a result of some form of expenditure. We performed
two $\mathrm{CNOT}$ gates, which generates some entanglement in
the intermediate states. However, we are at a loss to take this
any further due to a lack of our ability quantifying entanglement,
particularly in entropy units. We will now show that quantum
discord will provide a natural accounting of the resources and a
more soothing resolution.

            Let us consider measurements on $c$, as that is the
most contentious party in our protocol. The projectors are
\be
 \Pi_1 =\mathbb{I}_{ab}\otimes\frac{\mathbb{I}_c + \bm{\alpha}\cdot\bm{\sigma}}{2},\;\;\;\;\;\;\;\;
 \Pi_2 = \mathbb{I}_{ab}\otimes\frac{\mathbb{I}_c - \bm{\alpha}\cdot\bm{\sigma}}{2},
 \ee
 with $\alpha_1 = \sin\theta\cos\phi, \alpha_2 = \sin\theta\sin\phi, \alpha_3= \cos\theta$ as
usual.

        We will start with the state $\rho_{abc}.$ A simple calculation
shows
 $$
 p_1 = \frac{3+\cos\theta}{6},\;\;\;\;\;p_2 = \frac{3-\cos\theta}{6},
 $$
and that the spectrum of the reduced operators are
 \ben
\bm{\lambda}(\rho_{ab}^1) &=&
\left\{0,0,0,0,\frac{1}{2},\frac{1+\cos\theta}{2(3+\cos\theta)},\frac{1+\cos\theta}{2(3+\cos\theta)},\frac{1-\cos\theta}{2(3+\cos\theta)}
\right\},\\
\bm{\lambda}(\rho_{ab}^2) &=&
\left\{0,0,0,0,\frac{1}{2},\frac{1+\cos\theta}{2(3-\cos\theta)},\frac{1-\cos\theta}{2(3-\cos\theta)},\frac{1-\cos\theta}{2(3-\cos\theta)}
\right\}.
 \een
We immediately see that the minimization in this case is a single
variable affair ($\phi$ having dropped out), and this quickly
leads to (for $\theta = 0$)
 \be
 \label{E:disrho}
\mathcal{D}(\rho_{abc}) =0.
 \ee
Next, let us calculate the discord for the state $\sigma_{abc}$.
We find
 $$
 p_1 = p_2= \frac{1}{2},
 $$
and that the spectrum of the reduced operators is
 \be
\bm{\lambda}(\sigma_{ab}^1) =\bm{\lambda}(\sigma_{ab}^2) =
\left\{0,0,0,0,\frac{1}{3}\cos^2\frac{\theta}{2},\frac{1}{3}\sin^2\frac{\theta}{2},\frac{2+\sin\theta}{6},\frac{2-\sin\theta}{6}\right\}.
 \ee
Again, the minimization involves involves one variable, whose
minimum is attained for $\theta =0$, as
 \be
 \label{E:dissigma}
\mathcal{D}(\sigma_{abc}) = \frac{1}{3}.
 \ee
   Finally, a similar calculation for $\tau_{abc}$ provides
 \be
 \label{E:distau}
\mathcal{D}(\tau_{abc}) =0.
 \ee
 The three results, in Eqs~\ref{E:disrho}, \ref{E:dissigma},
\ref{E:distau} lead us to very striking conclusions, \emph{viz}.,
 \begin{enumerate}
 \item The amount of discord generated is exactly equal to
the average amount of entanglement we can extract in the next
stage of the protocol.
 \item The fact there is no discord at the end of the protocol is
circumstantial evidence that the discord is somehow `converted'
into entanglement.
 \end{enumerate}
This might be construed to imply that discord is a more
fundamental resource for quantum information than entanglement. In
our view, however, further investigation is called for before such
conclusions are drawn. Indeed, a genuine measure of tripartite
entanglement may help clear up the resource nature of entanglement
in the protocol. Another line of investigation would be to
demonstrate the qualitative and/or quantitative role of discord in
the general scheme of entanglement distribution \cite{cvdc03}, of
which we presented just an example, or in other mixed-state
protocols involving entanglement. A third line of research would
be to arrive at an independent operational interpretational for
quantum discord. An interpretation of this nature will demonstrate
the power, utility and resourcefulness of quantum discord.

        With these statements, we will move on to our study of
mixed-state quantum computation in Chapters~\ref{chap:dqc1} and
\ref{chap:mps}.

\chapter{Entanglement in the DQC1 model\label{chap:dqc1}}

\hfill \textsl{Why then, can one desire too much of a good thing
?}

\hfill  -William Shakespeare in \textsl{As You Like It}

\vskip1.0cm

        The purpose of this chapter is to investigate the existence of and
amount of entanglement in the DQC1 circuit that is used to
estimate the normalized trace.  The DQC1 model consists of a
\emph{special qubit\/} (qubit~0) in the initial state
$|0\rangle\langle0|={1\over2}(I_1+Z)$, where $Z$ is a Pauli
operator, along with $n$ other qubits in the completely mixed
state, $I_n/2^n$, which we call the \emph{unpolarized qubits}. The
circuit consists of a Hadamard gate on the special qubit followed
by a controlled unitary on the remaining qubits~\cite{lcnv02}:
\vspace{-1em}
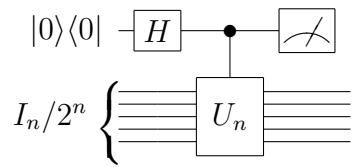
\begin{figure}[h]
\centerline{ \Qcircuit @C=.5em @R=-.5em {
    & \lstick{\ket{0}\!\bra{0}} & \gate{H} & \ctrl{1} & \meter & \push{\rule{0em}{4em}} \\
    & & \qw & \multigate{4}{U_n} & \qw & \qw \\
    & & \qw & \ghost{U_n} & \qw & \qw \\
    \lstick{\mbox{$I_n/2^n$}} & & \qw & \ghost{U_n} & \qw & \qw \\
    & & \qw & \ghost{U_n} & \qw & \qw \\
    & & \qw & \ghost{U_n} & \qw & \qw \gategroup{2}{2}{6}{2}{.6em}{\{}
}} \caption[The DQC1 circuit]{The DQC1 circuit\label{E:circuit}}
\end{figure}
After these operations, the state of the $n+1$ qubits becomes
\begin{equation}
\rho_{n+1}= {1\over2N} \Bigl(|0\rangle\langle0|\otimes
I_n+|1\rangle\langle1|\otimes I_n +|0\rangle\langle1|\otimes
U_n^\dagger +|1\rangle\langle0|\otimes U_n\Bigr) = \frac{1}{2N}
    \begin{pmatrix}
        I_n & \, U_n^\dag \\
        \, U_n & I_n
    \end{pmatrix} \;,
\label{E:rhoout}
\end{equation}
where $N=2^n$.  The information about the normalized trace of
$U_n$ is encoded in the expectation values of the Pauli operators
$X$ and $Y$ of the special qubit, i.e., $\langle X\rangle={\rm
Re}[\tr(U_n)]/2^n$ and $\langle Y\rangle=-{\rm Im}[\tr(U_n)]/2^n$.

To read out the desired information, say, about the real part of
the normalized trace, one runs the circuit repeatedly, each time
measuring $X$ on the special qubit at the output.  The measurement
results are drawn from a distribution whose mean is the real part
of the normalized trace and whose variance is bounded above by 1.
After $L$ runs, one can estimate the real part of the normalized
trace with an accuracy $\epsilon\sim1/\sqrt L$.  Thus, to achieve
accuracy $\epsilon$ requires that the circuit be run
$L\sim1/\epsilon^2$ times. More precisely, what we mean by
estimating with fixed accuracy is the following: let $P_e$ be the
probability that the estimate is farther from the true value than
$\epsilon\,$; then the required number of runs is
$L\sim\ln(1/P_e)/\epsilon^2$.  That the number of runs required to
achieve a fixed accuracy does not scale with number of qubits and
scales logarithmically with the error probability is what is meant
by saying that the DQC1 circuit provides an efficient method for
estimating the normalized trace.

Throughout much of our analysis, we use a generalization of the
DQC1 circuit, in which the initial pure state of the special qubit
is replaced by the mixed state ${1\over2}(I_1+\alpha Z)$, which
has polarization $\alpha$,
\vspace{-1em}
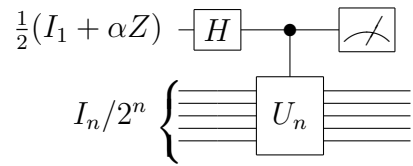
\begin{figure}[h]
\centerline{ \Qcircuit @C=.5em @R=-.5em {
    & \lstick{{1\over2}(I_1+\alpha Z)} & \gate{H} & \ctrl{1} & \meter & \push{\rule{0em}{4em}} \\
    & & \qw & \multigate{4}{U_n} & \qw & \qw \\
    & & \qw & \ghost{U_n} & \qw & \qw \\
    \lstick{\mbox{$I_n/2^n$}} & & \qw & \ghost{U_n} & \qw & \qw \\
    & & \qw & \ghost{U_n} & \qw & \qw \\
    & & \qw & \ghost{U_n} & \qw & \qw \gategroup{2}{2}{6}{2}{.6em}{\{}
}} \caption[The DQC1 circuit with arbitrary polarization]{The DQC1
circuit with arbitrary polarization
\label{E:circuitalpha}}
\end{figure}
giving an overall initial state
\begin{equation}
\rho_i={1\over2N}(I_1+\alpha Z)\otimes I_n
={1\over2N}\!\left[I_{n+1}+ \alpha
\begin{pmatrix}
I_n&0\\
0&-I_n
\end{pmatrix}
\right]\;.
\end{equation}
We generally assume that $\alpha\ge0$, except where we explicitly
note otherwise.  After the circuit is run, the system state
becomes
\begin{equation}
\rho_{n+1}(\alpha)= {1\over2N}\!\left[I_{n+1}+ \alpha
\begin{pmatrix}
0&U_n^\dag\\
U_n&0
\end{pmatrix}
\right]= \frac{1}{2N}
    \begin{pmatrix}
        I_n &  \alpha U_n^\dag \\
        \alpha U_n & I_n
    \end{pmatrix} \;.
\label{E:rhooutalpha}
\end{equation}
The effect of subunity polarization is to reduce the expectation
values of $\langle X\rangle$ and $\langle Y\rangle$ by a factor of
$\alpha$, thereby making it more difficult to estimate the
normalized trace.  Specifically, the number of runs required to
estimate the normalized trace becomes
$L\sim\ln(1/P_e)/\alpha^2\epsilon^2$.  Reduced polarization
introduces an additional overhead, but as long as the special
qubit has nonzero polarization, the model still provides an
efficient estimation of the normalized trace.  What we are dealing
with is really the ``power of even the tiniest fraction of a
qubit.''

For $n+1$ qubits, \emph{all\/} states contained in a ball of
radius $r_{n+1}$ centered at the completely mixed state are
separable~\cite{braunstein99a,gb03} (distance is measured by the
Hilbert-Schmidt norm).  Unitary evolution leaves the distance from
the completely mixed state fixed, so at all times during the
circuit in Fig.~(\ref{E:circuitalpha}), the system state is a
fixed distance $\sqrt{\tr(\rho_i-I_{n+1}/2N)^2}=\alpha
2^{-(n+1)/2}$ from the completely mixed state.  This suggests that
with $\alpha$ small enough, there might be an exponential speedup
with demonstrably separable states.  This suggestion doesn't pan
out, however, because the radius of the separable ball decreases
exponentially faster than $2^{-(n+1)/2}$.  The best known lower
bound on $r_n$ is $2\times 6^{-n/2}$~\cite{gb04}; for the system
state to be contained in a ball given by this lower bound, we need
$\alpha\le 2\times 3^{-(n+1)/2}$. The exponential decrease of
$\alpha$ means that an exponentially increasing number of runs is
required to estimate the normalized trace with fixed accuracy.
More to the point, the possibility that the actual radius of the
separable ball might decrease slowly enough to avoid an
exponential number of runs is ruled out by the existence of a
family of $n$-qubit entangled states found by D\"ur~\emph{et
al.}~\cite{dur99}, which establishes an upper bound on $r_n$ that
goes as $2\times 2^{-n}$ for large $n$, implying that $\alpha\le
2\times 2^{-(n+1)/2}$ if the system state is to be in the ball
given by the upper bound.  These considerations do not demonstrate
the impossibility of an exponential speedup using separable
states, but they do rule out the possibility of finding such a
speedup within the maximal separable ball about the completely
mixed state.

We are thus motivated to look for entanglement in states of the
form in Eq.~(\ref{E:rhooutalpha}), for at least some unitary
operators $U_n$. Initial efforts in this direction are not
encouraging.  It is clear from the start that the marginal state
of the $n$ unpolarized qubits remains completely mixed, so these
qubits are not entangled among themselves. Moreover, in the state
in Eq.~(\ref{E:rhooutalpha}), as was shown in Ref.~\cite{pklo04},
the special qubit is unentangled with the $n$ unpolarized qubits,
no matter what $U_n$ is used.  To see this, one plugs the
eigendecomposition of the unitary, $U_n=\sum_j
e^{i\phi_j}|e_j\rangle\langle e_j|$, into the expression for
$\rho_{n+1}(\alpha)$.  This gives a separable decomposition
\begin{equation}
\rho_{n+1}(\alpha)= {1\over2N}\sum_j (|a_j\rangle\langle
a_j|+|b_j\rangle\langle b_j|) \otimes|e_j\rangle\langle e_j|\;,
\end{equation}
where
$|a_j\rangle=\cos\theta|0\rangle+e^{i\phi_j}\sin\theta|1\rangle$
and
$|b_j\rangle=\sin\theta|0\rangle+e^{i\phi_j}\cos\theta|1\rangle$,
with $\sin2\theta=\alpha$.

No entanglement of the special qubit with the rest and no
entanglement among the rest---where then are we to find any
entanglement?  We look for entanglement relative to other
divisions of the qubits into two parts.  In such bipartite
divisions the special qubit is grouped with a subset of the
unpolarized qubits.  To detect entanglement between the two parts,
we use the Peres-Horodecki partial transpose
criterion~\cite{p96,hhh96}, and we quantify whatever entanglement
we find using a closely related entanglement monotone which we
call the \emph{multiplicative negativity\/}~\cite{vw02}.  The
Peres-Horodecki criterion and the multiplicative negativity do not
reveal all entanglement---they can miss what is called bound
entanglement---but we are nonetheless able to demonstrate the
existence of entanglement in states of the form in
Eq.~(\ref{E:rhoout}) and Eq.~(\ref{E:rhooutalpha}). For
convenience, we generally refer to the multiplicative negativity
simply as the negativity.  The reader should note, as we discuss
in Sec.~\ref{S:negativity}, that the term ``negativity'' was
originally applied to an entanglement measure that is closely
related to, but different from the multiplicative negativity.

The amount of entanglement depends, of course, on the unitary
operator~$U_n$ and on the bipartite division.  We present three
results in this regard.  First, in Sec.~\ref{S:examples}, we
construct a family of unitaries $U_n$ such that for $\alpha>1/2$,
$\rho_{n+1}(\alpha)$ is entangled for all bipartite divisions that
put the first and last unpolarized qubits in different parts, and
we show that for all such divisions, the negativity is
$(2\alpha+3)/4$ for $\alpha\ge1/2$ ($5/4$ for $\alpha=1$),
independent of~$n$.  Second, in Sec.~\ref{S:random}, we present
numerical evidence that the state $\rho_{n+1}$ of
Eq.~(\ref{E:rhoout}) is entangled for typical unitaries, i.e.,
those created by random quantum circuits.  For $n+1=5,\ldots,10$,
we find average negativities between 1.155 and just above 1.16 for
the splitting that puts $\left\lfloor n/2\right\rfloor$ of the
unpolarized qubits with qubit~0.  Third, in Sec.~\ref{S:bounds},
we show that for all unitaries and all bipartite divisions of the
$n+1$ qubits, the negativity of $\rho_{n+1}(\alpha)$ is bounded
above by the constant $\sqrt{1+\alpha^2}$ ($\sqrt2\simeq1.414$ for
$\alpha=1$), independent of $n$.  Thus, when $n$ is large, the
negativity achievable by the DQC1 circuit~(\ref{E:circuit})
becomes a vanishingly small fraction of the maximum negativity,
$\sim2^{n/2}$, for roughly equal bipartite divisions.

The layout of this chapter is as follows.  In
Sec.~\ref{S:classical} we examine the classical problem of
estimating the normalized trace of a unitary.  In
Sec.~\ref{S:negativity} we reviewed the pertinent properties of
the negativity. We apply these to obtain our three key results in
Secs.~\ref{S:examples}--\ref{S:bounds}.  We conclude in
Sec.~\ref{S:conclusion} and prove a brief Lemma in
Appendix~\ref{A:appendix}. Throughout we use $\breve A$ to stand
for the partial transpose of an operator $A$ relative to a
particular bipartite tensor-product structure, and we rely on
context to make clear which bipartite division we are using at any
particular point in this chapter.

\section{Classical evaluation of the trace}\label{S:classical}

Every fair judgement on the power of quantum computation to solve
a mathematical problem must be made relative to the best known
classical algorithm to solve the same. In that spirit, this
section briefly outlines a classical method for estimating the
trace of a unitary operator that can be implemented efficiently in
terms of quantum gates, and we indicate why this appears to be a
problem that is exponentially hard in the number of qubits.

The trace of a unitary matrix $U_n\equiv U$ is the sum over the
diagonal matrix elements of~$U$:
\begin{equation}
\label{E:sum1} \tr(U)=\sum_{\ba} \bra{\ba}U\ket{\ba}\;.
\end{equation}
Here $\ba$ is a bit string that specifies a computational-basis
state of the $n$ qubits.  By factoring $U$ into a product of
elementary gates from a universal set and inserting a resolution
of the identity between all the gates, we can write $\tr(U)$ as a
sum over the amplitudes of Feynman paths.  A difficulty with this
approach is that the sum must be restricted to paths that begin
and end in the same state.  We can circumvent this difficulty by
preceding and succeeding $U$ with a Hadamard gate on all the
qubits.  This does not change the trace, but does allow us to
write it as
\begin{equation}
\label{E:sum2} \tr(U)=\sum_{\ba,\bb,\bc} \bra{\ba}H^{\otimes
n}\ket{\bb}\bra{\bb}U\ket{\bc}\bra{\bc}H^{\otimes n}\ket{\ba}
={1\over2^n}\sum_{\ba,\bb,\bc}
(-1)^{\ba\cdot(\bb+\bc)}\bra{\bb}U\ket{\bc}\;.
\end{equation}
Now if we insert a resolution of the identity between the
elementary gates, we get $\tr(U)$ written as an unrestricted sum
over Feynman-path amplitudes, with an extra phase that depends on
the initial and final states.

Following Dawson~\emph{et al.}~\cite{dhhmno05}, we consider two
universal gate sets: (i)~the Hadamard gate $H$, the $\pi/4$ gate
$T$, and the controlled-NOT gate and (ii)~$H$ and the Toffoli
gate.  With either of these gate sets, most of the Feynman paths
have zero amplitude.  Dawson~\emph{et al.}~\cite{dhhmno05}
introduced a convenient method, which we describe briefly now, for
including only those paths with nonzero amplitude.  One associates
with each wire in the quantum circuit a classical bit value
corresponding to a computational basis state.  The effect of an
elementary gate is to change, deterministically or stochastically,
the bit values at its input and to introduce a multiplicative
amplitude.  The two-qubit controlled-NOT gate changes the input
control bit $x$ and target bit $y$ deterministically to output
values $x$ and $y\oplus x$, while introducing only unit
amplitudes.  Similarly, the three-qubit Toffoli gates changes the
input control bits $x$ and $y$ and target bit $z$
deterministically to $x$, $y$, and $z\oplus xy$, while introducing
only unit amplitudes.  The $T$ gate leaves the input bit value $x$
unchanged and introduces a phase $e^{ix\pi/4}$.  The Hadamard gate
changes the input bit value $x$ stochastically to an output value
$y$ and introduces an amplitude $(-1)^{xy}/\sqrt2$.

The classical bit values trace out the allowed Feynman paths, and
the product of the amplitudes introduced at the gates gives the
overall amplitude of the path.  In our application of evaluating
the trace~(\ref{E:sum2}), a path is specified by $n$ input bit
values (which are identical to the output bit values), $n$ random
bit values introduced by the initial Hadamard gates, and $h$
random bit values introduced at the $h$ Hadamard gates required
for the implementation of $U$.  This gives a total of $2n+h$ bits
to specify a path and thus $2^{2n+h}$ allowed paths.  We let $\bx$
denote collectively the $2n+h$ path bits.

If we apply the gate rules to a Hadamard-Toffoli circuit, the only
gate amplitudes we have to worry about are the $\pm1/\sqrt2$
amplitudes introduced at the Hadamard gates.  There being no
complex amplitudes, the trace cannot be complex.  Indeed, for this
reason, achieving universality with the $H$-Toffoli gate set
requires the use of a simple encoding, and we assume for the
purposes of our discussion that this encoding has already been
taken into account.  With all this in mind, we can write the
trace~(\ref{E:sum2}) as a sum over the allowed paths,
\begin{equation}
\tr(U)=\frac{1}{2^{n+h/2}}\sum_{\bx}(-1)^{\psi(\bx)}\;.
\end{equation}
Here $\psi(\bx)$ is a polynomial over $\mz_2$, specifically, the
mod-2 sum of the products of input and output bit values at each
of the Hadamard gates.  The downside is that a string of Toffoli
gates followed by a Hadamard can lead to a polynomial that is high
order in the bit values.  As pointed out by Dawson~\emph{et
al.}~\cite{dhhmno05}, we can deal with this problem partially by
putting a pair of Hadamards on the target qubit after each Toffoli
gate, thus replacing the quadratic term in the output target bit
with two new random variables and preventing the quadratic term
from iterating to higher order terms in subsequent Toffoli gates.
In doing so, we are left with a cubic term in $\psi(\bx)$ from the
amplitude of the first Hadamard.  The upshot is that we can always
make $\psi(\bx)$ a cubic polynomial.

Notice now that we can rewrite the trace as
\begin{equation}
\tr(U) =\frac{1}{2^{n+h/2}}\left[
\begin{pmatrix}
\mbox{number of $\bx$ such}\\ \mbox{that $\psi(\bx)=0$}
\end{pmatrix}-
\begin{pmatrix}
\mbox{number of $\bx$ such}\\ \mbox{that $\psi(\bx)=1$}
\end{pmatrix}\right]\;,
\end{equation}
thus reducing the problem of evaluating the trace exactly to
counting the number of zeroes of the cubic polynomial $\psi(\bx)$.
This is a standard problem from computational algebraic geometry,
and it is known that counting the number of zeroes of a general
cubic polynomial over any finite field is \#{\bf P} complete
\cite{ek90}.  It is possible that the polynomials that arise from
quantum circuits have some special structure that can be exploited
to give an efficient algorithm for counting the number of zeroes,
but in the absence of such structure, there is no efficient
classical algorithm for computing the trace exactly unless the
classical complexity hierarchy collapses and \emph{all\/} problems
in \#{\bf P} are efficiently solvable on a classical computer.

Of course, it is not our goal to compute the trace exactly, since
the quantum circuit only provides an efficient method for
estimating the normalized trace to fixed accuracy.  This suggests
that we should estimate the normalized trace by sampling the
amplitudes of the allowed Feynman paths.  The normalized trace,
\begin{equation}
{\tr(U)\over2^n}=\frac{1}{2^{2n+h}}\sum_{\bx}2^{h/2}(-1)^{\psi(\bx)}\;,
\end{equation}
which lies between $-1$ and $+1$, can be regarded as the average
of $2^{2n+h}$ quantities whose magnitude, $2^{h/2}$, is
exponentially large in the number of Hadamard gates.  To estimate
the average with fixed accuracy requires a number of samples that
goes as $2^h$, implying that this is not an efficient method for
estimating the normalized trace.  The reason the method is not
efficient is pure quantum mechanics, i.e., that the trace is a sum
of amplitudes, not probabilities.

If we apply the gate rules to a Hadamard-$T$-controlled-NOT
circuit, the bit value on each wire in the circuit is a mod-2 sum
of appropriate bit values in $\bx$, but now we have to worry about
the amplitudes introduced by the Hadamard and $T$ gates.  The
trace~(\ref{E:sum2}) can be written as
 \begin{equation}
  \tr(U)={1\over2^{n+h/2}}
  \sum_{\bx}e^{i(\pi/4)\chi(\bx)}(-1)^{\phi(\bx)}\;.
  \label{E:sum3}
 \end{equation}
Here $\phi(\bx)$ is a polynomial over $\mz_2$, obtained as the
mod-2 sum of the products of input and output bit values at each
of the Hadamard gates.  Since the output value is a fresh binary
variable and the input value is a mod-2 sum of bit values in
$\bx$, $\phi(\bx)$ is a purely quadratic polynomial over $\mz_2$.
The function $\chi(\bx)$ is a mod-8 sum of the input bit values to
all of the $T$ gates.  Since these input bit values are mod-2 sums
of bit values in $\bx$, $\chi(\bx)$ is linear in bit values, but
with an unfortunate mixture of mod-2 and mod-8 addition.  We can
get rid of this mixture by preceding each $T$ gate with a pair of
Hadamards, thus making the input to the every $T$ gate a fresh
binary variable.  With this choice, $\chi(\bx)$ becomes a mod-8
sum of appropriate bit values from~$\bx$.

We can rewrite the sum~(\ref{E:sum3}) in the following way:
\begin{equation}
 \tr(U)={1\over2^{n+h/2}}
 \sum_{j=0}^7 e^{i\frac{\pi}{4}j}\!\! \left[\!\!
\begin{pmatrix}
\mbox{number of $\bx$ such that}\\
\mbox{$\chi(\bx)=j$ and $\phi(\bx)=0$}
\end{pmatrix}\!\!
- \!\!
\begin{pmatrix}
\mbox{number of $\bx$ such that}\\
\mbox{$\chi(\bx)=j$ and $\phi(\bx)=1$}
\end{pmatrix}\!\!
\right].
\end{equation}
Thus the problem now reduces to finding simultaneous (binary)
solutions to the purely quadratic $\mz_2$ polynomial $\phi(\bx)$
and the purely linear $\mz_8$ polynomial $\chi(\bx)$.  One has to
be careful here to note that we are only interested in binary
solutions, so we are not solving $\chi(\bx)=j$ over all values in
$\mz_8$.  The number of solutions of a purely quadratic polynomial
over $\mz_2$ can be obtained trivially~\cite{ek90}, but the
constraint over $\mz_8$ means that one must count the number of
solutions over a mixture of a field and a ring.  The complexity
class for this problem is not known, but given the equivalence to
counting the number of solutions of a cubic polynomial over
$\mz_2$, it seems unlikely that there is an efficient classical
algorithm.  Moreover, an attempt to estimate the normalized trace
by sampling allowed paths obviously suffers from the problem
already identified above.

\section{Entanglement in the DQC1 circuit} \label{S:examples}

        In this section we present our main results. As outlined
earlier, we will use the negativity as measure of entanglement.
The negativity is the sum of the singular values of $\breve\rho$.
For states of the form we are interested in, given by
Eq.~(\ref{E:rhooutalpha}), the negativity is determined by the
singular values of the partial transpose of the unitary
operator~$U_n$. To see this, consider any bipartite division of
the qubits. Performing the partial transpose on the part that does
not include the special qubit, we have
\begin{equation}
\breve\rho_{n+1}(\alpha)= \frac{1}{2N}
    \begin{pmatrix}
        I_n &  \alpha\breve U_n^\dag \\
        \alpha\breve U_n & I_n
    \end{pmatrix} \;,
\label{E:breverhooutalpha}
\end{equation}
where $\breve U_n$ is the partial transpose of $U_n$ relative to
the chosen bipartite division.  Notice that if we make our
division between the special qubit and all the rest, then $\breve
U_n=U_n^T$ is a unitary operator, and $\breve\rho_{n+1}(\alpha)$
is the quantum state corresponding to using $U_n^T$ in the
circuit~(\ref{E:circuitalpha}); this shows that for this division,
the negativity is 1, consistent with our earlier conclusion that
the special qubit is not entangled with the other qubits.  For a
general division, we know there are unitaries $V$ and $W$ such
that $\breve U_n=VSW$, where $S$ is the diagonal matrix of
singular values $s_j(\breve U_n)$.  This allows us to write
\begin{equation}
\breve\rho_{n+1}(\alpha)=
    \begin{pmatrix}
        W^\dagger & 0 \\
        0 & V
    \end{pmatrix}
    \frac{1}{2N}
    \begin{pmatrix}
        I_n &  \alpha S \\
        \alpha S & I_n
    \end{pmatrix}
    \begin{pmatrix}
        W & 0 \\
        0 & V^\dagger
    \end{pmatrix}\;,
\end{equation}
showing that $\breve\rho_{n+1}(\alpha)$ is a unitary
transformation away from the matrix in the middle and thus has the
same eigenvalues. The block structure of the middle matrix makes
it easy to find these eigenvalues, which are given by $[1\pm\alpha
s_j(\breve U_n)]/2N$. This allows us to put the negativity in the
form
\begin{equation}
\sM\bigl(\rho_{n+1}(\alpha)\bigr)={1\over2N} \sum_{j=1}^N|1+\alpha
s_j(\breve U_n)|+|1-\alpha s_j(\breve U_n)| = {1 \over N}
\sum_{j=1}^N \max\bigl(\abs{\alpha} s_j(\breve U_n), 1\bigr)\;,
\label{E:Msingular}
\end{equation}
which is valid for both positive and negative values of $\alpha$.
An immediate consequence of Eq.~(\ref{E:Msingular}) is that
$\sM\bigl(\rho_{n+1}(\alpha)\bigr)=\sM\bigl(\rho_{n+1}(-\alpha)\bigr)$,
as one would expect.  Since $\rho_{n+1}(\alpha)$ is a mixture of
$\rho_{n+1}(+1)=\rho_{n+1}$ and $\rho_{n+1}(-1)$, convexity tells
us immediately that
$\sM\bigl(\rho_{n+1}(\alpha)\bigr)\le\sM\bigl(\rho_{n+1}\bigr)$,
i.e., that a mixed input for the special qubit cannot increase the
negativity over that for a pure input.  More generally, we have
that the negativity cannot decrease at any point as $\alpha$
increases from 0 to~1.

As the first result of this section, we construct a family of
unitaries $U_n$ that produce global entanglement in the DQC1
circuit~(\ref{E:circuit}). For $\alpha=1$, the negativity produced
by this family is equal to $5/4$, independent of $n$, for all
bipartite divisions that put the first and last unpolarized qubits
in different parts.  We conjecture that this is the maximum
negativity that can be achieved in a circuit of the
form~(\ref{E:circuit}).

Before the measurement, the output state of the
circuit~(\ref{E:circuitalpha}) is given by
Eq.~(\ref{E:rhooutalpha}). To construct the unitaries $U_n$, we
first introduce a two-qubit unitary matrix
\begin{equation}
\label{E:U2}
    U_{2} \equiv \begin{pmatrix}
        A_1 & C_1 \\
        D_1 & B_1 \\
    \end{pmatrix} \;,
\end{equation}
where $A_1$, $B_1$, $C_1$, and $D_1$ are single-qubit ($2\times2$)
matrices that must satisfy $A_1^\dagger A_1+D_1^\dagger D_1=
B_1^\dagger B_1+C_1^\dagger C_1=I_1$ and $A_1^\dagger
C_1+D_1^\dagger B_1=0$ to ensure that $U_2$ is unitary.  The
$n$-qubit unitary $U_n$ is then defined by
\begin{eqnarray}
    U_n &\equiv&
    \begin{pmatrix}
        I_{n-2} \otimes A_1 & X_{n-2} \otimes C_1 \\
        X_{n-2} \otimes D_1 & I_{n-2} \otimes B_1
    \end{pmatrix} \nonumber \\
    &=&|0\rangle\langle0|\otimes I_{n-2}\otimes A_1
    +|1\rangle\langle1|\otimes I_{n-2}\otimes B_1 \nonumber \\
    &&\phantom{|}
    +|0\rangle\langle1|\otimes X_{n-2}\otimes C_1
    +|1\rangle\langle0|\otimes X_{n-2}\otimes D_1
    \;.
    \label{E:Un}
\end{eqnarray}
Here we use $X_1$, $Y_1$, and $Z_1$ to denote single-qubit Pauli
operators.  A subscript $k$ on the identity operator or a Pauli
operator denotes a tensor product in which that operator acts on
each of $k$ qubits.  If we adopt the convention that $X_0 = I_0 =
1$, then $U_n$ reduces to $U_2$ when $n=2$.  It is easy to design
a quantum circuit that realizes $U_n$.  The structure of the
circuit is illustrated by the case of $U_4$:
\begin{figure}[h]
\centerline{
    \Qcircuit @C=.5em @R=0.5em {
    & \lstick{\mbox{1st qubit}} & \ctrl{2} & \ctrl{3} & \multigate{1}{U_2} & \ctrl{3} & \ctrl{2} & \qw \\
    & \lstick{\mbox{4th qubit}} & \qw      & \qw      & \ghost{U_n}        & \qw    & \qw & \qw \\
    & \lstick{\mbox{2nd qubit}} & \targ    & \qw      & \qw                & \qw      & \targ & \qw \\
    & \lstick{\mbox{3rd qubit}} & \qw      & \targ    & \qw                & \targ    & \qw & \qw                    }
} \caption[Structure of the unitary $U_4$]{Unitary $U_4$ that
generates entanglement in DQC1 circuit for all $n$}
\label{E:Ucircuit}
\end{figure}
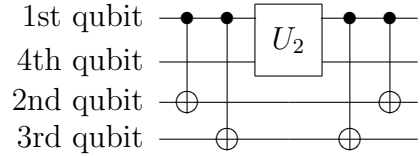
In general, the two-qubit unitary $U_2$, acting on the first and
last qubits, is bracketed by controlled-NOT gates from the first
qubit, acting as control, to each of the other qubits, except the
last, as targets.

Because $I_1$ and $X_1$ are invariant under transposition, it is
clear from the form of $U_n$ that in the state in
Eq.~(\ref{E:rhooutalpha}), all qubits, except~0, 1, and~$n$, are
invariant under transposition.  We can use this fact to find the
negativity for all bipartite divisions. First consider any
bipartite division that puts qubits 1 and $n$ in the same part.
There are two possibilities.  If the special qubit is in the same
part as qubits~1 and $n$, then partial transposition on the other
part leaves $\rho_{n+1}(\alpha)$ unchanged, so the negativity
is~1.  If the special qubit is not in the same part as~1 and $n$,
then partial transposition on the part that includes~1 and $n$ is
the same as partial transposition of all the unpolarized qubits, a
case we already know to have negativity equal to 1.  We conclude
that any bipartite division that puts~1 and $n$ in the same part
has negativity equal to 1.

Turn now to bipartite divisions that put qubits~1 and $n$ in
different parts.  There are two cases to consider: (i)~the special
qubit is in the same part as qubit~1, and (ii)~the special qubit
is in the same part as qubit~$n$.  In case~(i), partial
transposition of the part that contains qubit~$n$ gives
\begin{equation}
\breve\rho_{n+1}(\alpha)={1\over2N}
    \begin{pmatrix}
        I_n                 & \alpha \breve U_n^\dagger \\
        \alpha \breve U_n   & I_n
    \end{pmatrix}
\;\;\mbox{with}\;\; \breve U_n=
\begin{pmatrix}
        I_{n-2} \otimes A_1^T & X_{n-2} \otimes C_1^T \\
        X_{n-2} \otimes D_1^T & I_{n-2} \otimes B_1^T
    \end{pmatrix}\;.
\end{equation}
The structure of $\breve U_n$ comes about due to the transposition
over the last qubit in Eq.~(\ref{E:Un}). In case~(ii), partial
transposition of the part that contains qubit~1 gives (here the
transposition is over the first qubit in Eq.~(\ref{E:Un}))
\begin{equation}
\breve\rho_{n+1}(\alpha)={1\over2N}
    \begin{pmatrix}
        I_n                 & \alpha \breve U_n^\dagger \\
        \alpha \breve U_n   & I_n
    \end{pmatrix}
\;\;\mbox{with} \;\;\breve U_n=
\begin{pmatrix}
        I_{n-2} \otimes A_1 & X_{n-2} \otimes D_1 \\
        X_{n-2} \otimes C_1 & I_{n-2} \otimes B_1
    \end{pmatrix}\;.
\end{equation}

The basic structure of $\breve\rho_{n+1}(\alpha)$ is the same in
both cases.  Without changing the spectrum, we can reorder the
rows and columns to block diagonalize $\breve\rho_{n+1}(\alpha)$
so that there are $N/4$ blocks, each of which has the form
\begin{equation}
{1\over2N}
\begin{pmatrix}
    I_2                 & \alpha \breve U_2^\dagger \\
    \alpha \breve U_2   & I_2
\end{pmatrix}
={4\over N}\breve\rho_3(\alpha)\;, \label{E:breverho2}
\end{equation}
where $\breve\rho_3(\alpha)$ is the appropriate partial transpose
of the three-qubit output state.  Thus the spectrum of
$\breve\rho_{n+1}(\alpha)$ is the same as the spectrum of
$\breve\rho_3(\alpha)$, except that each eigenvalue is reduced by
a factor of $4/N$.  In calculating the negativity, since each
eigenvalue is ($N/4$)-fold degenerate, the reduction factor of
$4/N$ is cancelled by a degeneracy factor of $N/4$, leaving us
with the fundamental result of our construction,
\begin{equation}
\sM\bigl(\rho_{n+1}(\alpha)\bigr)=\sM\bigl(\rho_3(\alpha)\bigr)\;.
\end{equation}
This applies to both cases of bipartite splittings that we are
considering, showing that all divisions have the same negativity
as the corresponding $n=2$ construction.

We now specialize to a particular choice of $U_2$ given by
\begin{equation}
A_1=
    \begin{pmatrix}
        0 & 0 \\ 0 & 1
    \end{pmatrix}\;,
\quad B_1=
    \begin{pmatrix}
        1 & 0 \\ 0 & 0
    \end{pmatrix}\;,
\quad C_1=
    \begin{pmatrix}
        0 & 1 \\ 0 & 0
    \end{pmatrix}\;,
\quad\mbox{and}\quad D_1=
    \begin{pmatrix}
        0 & 0 \\ 1 & 0
    \end{pmatrix}\;.
\label{E:ABCD}
\end{equation}
For this choice, the two cases of bipartite division lead to the
same partial transpose.  The spectrum of
\begin{equation}
\breve\rho_3(\alpha)={1\over8}
\begin{pmatrix}
I_1 & 0 & \alpha A_1 & \alpha D_1 \\
0 & I_1 & \alpha C_1 & \alpha B_1  \\
\alpha A_1 & \alpha D_1 & I_1 & 0\\
\alpha C_1 & \alpha B_1 & 0 & I_1
\end{pmatrix}
\end{equation}
is
\begin{equation}
\spec(\breve\rho_3\bigl(\alpha)\bigr)= {1\over
8}(1+2\alpha,1,1,1,1,1,1-2\alpha)\;,
\end{equation}
giving a negativity equal to 1 for $\alpha\le1/2$ and a negativity
\begin{equation}
\sM\bigl(\rho_{n+1}(\alpha)\bigr)=\sM\bigl(\rho_3(\alpha)\bigr)=
{1\over4}(2\alpha+3) \quad\mbox{for}\quad \alpha\ge1/2\;.
\end{equation}
This result shows definitively that the circuit~ can produce
entanglement, at least for $\alpha>1/2$.  We stress that the
negativity achieved by this family of unitaries is independent of
$n\ge2$.

For $\alpha=1$, the negativity achieved by this family reduces to
$5/4$.  For large $n$, this amount of negativity is a vanishingly
small fraction of the maximum possible negativity, $\sim2^{n/2}$,
for roughly equal divisions of the qubits.  This raises the
question whether it is possible for other unitaries to achieve
larger negativities.  A first idea might be to find two-qubit
unitaries $U_2$ that yield a higher negativity
$\sM(\rho_3)=\sM(\rho_{n+1})$ when plugged into the construction
of this section, but the bounds we find in Sec.~\ref{S:bounds}
dispose of this notion, since they show that $5/4$ is the maximum
negativity that can be achieved for $n=2$.  Another approach would
be to generalize the construction of this section in a way that is
obvious from the circuit in Fig.~(\ref{E:Ucircuit}), i.e., by
starting with a $k$-qubit unitary in place of the two-qubit
unitary of Fig.~(\ref{E:Ucircuit}).  Numerical investigation of
the case $k=3$ has not turned up negativities larger than $5/4$.
We conjecture that $5/4$ is the maximum negativity that can be
achieved by states of the form~(\ref{E:rhoout}).  Though we have
not been able to prove this conjecture, we show in the next
section that typical unitaries for $n+1\le10$ achieve negativities
less than $5/4$ and in the following section that the negativity
is rigorously bounded by $\sqrt2$.

We stress that we are not suggesting that the construction of this
section, with $U_2$ given by Eq.~(\ref{E:ABCD}), achieves the
maximum negativity for all values of $\alpha$, for that would mean
that we believed that the negativity cannot exceed 1 for
$\alpha\le1/2$, which we do not.  Although we have not found
entanglement for $\alpha\le 1/2$, we suspect there are states with
negativity greater than 1 as long as $\alpha$ is large enough that
$\rho_{n+1}(\alpha)$ lies outside the separable ball around the
maximally mixed state, i.e., $\alpha\ge2^{(n+1)/2}r_{n+1}$.  The
bound of Sec.~\ref{S:bounds} only says that
$\sM\bigl(\rho_{n+1}(\alpha)\bigr)\le\sqrt{1+\alpha^2}$, thus
allowing negativities greater than 1 for all values of $\alpha$
except $\alpha=0$.  Moreover, since the negativity does not detect
bound entanglement, there could be entangled states that have a
negativity equal to~1.

\section{The average negativity of a random unitary} \label{S:random}

Having constructed a family of unitaries that yields a DQC1 state
with negativity $5/4$, a natural question to ask is, ``What is the
negativity of a typical state produced by the circuit in
Fig.~(\ref{E:circuit})?''  To address this question, we choose the
unitary operator in the circuit in Fig.~(\ref{E:circuit}) at
random and calculate the negativity.  Of course, one must first
define what it means for a unitary to be ``typical'' or ``chosen
at random''. The natural measure for defining this is the Haar
measure, which is the unique left-invariant measure for the group
${\sf U}(N)$ \cite{Conway90}.  The resulting ensemble of unitaries
is known as the Circular Unitary Ensemble, or CUE, and it is
parameterized by the Hurwitz decomposition \cite{hurwitz1897}.
Although this is an exact parameterization, implementing it
requires computational resources that grow exponentially in the
size of the unitary~\cite{emerson03a}.  To circumvent this, a
pseudo-random distribution that requires resources growing
polynomially in the size of the unitary was formulated and
investigated in Ref.~\cite{emerson03a}. This is the distribution
from which we draw our random unitaries, and we summarize the
procedure for completeness.

We first define a random ${\sf SU}(2)$ unitary as
\begin{equation}
    R(\theta,\phi,\chi) =
    \begin{pmatrix}
    e^{i \phi} \cos\theta & e^{i \chi} \sin\theta \\
    -e^{-i \chi} \sin\theta & e^{-i \phi} \cos\theta \\
    \end{pmatrix} \;,
\end{equation}
where $\theta$ is chosen uniformly between $0$ and $\pi/2$, and
$\phi$ and $\chi$ are chosen uniformly between $0$ and $2\pi$.  A
random unitary applied to each of the $n$ qubits is then
\begin{equation}
    {\sf R} = \bigotimes_{i=1}^{n} R(\theta_i, \phi_i, \chi_i) \;,
\end{equation}
where a separate random number is generated for each variable at
each value of $i$.  Now define a mixing operator ${\sf M}$ in
terms of nearest-neighbor $Z\otimes Z$ couplings as
\begin{equation}
    {\sf M} =
    \exp\left(i \frac{\pi}{4} \sum_{j=1}^{n-1} Z^{(j)} \otimes Z^{(j+1)} \right) \;.
\end{equation}
The pseudo-random unitary is then given by
\begin{equation}
    {\sf R}_j{\sf M}{\sf R}_{j-1}\cdots{\sf M}{\sf R}_2{\sf M}{\sf R}_1 \;,
\end{equation}
where $j$ is a positive integer that depends on $n$, and each
${\sf R}_k$ is chosen randomly as described above.  For a given
$n$, the larger $j$ is, the more accurately the pseudo-random
unitary distribution resembles the actual CUE.  From the results
in Ref.~\cite{emerson03a}, $j=40$ gives excellent agreement with
the CUE for unitary operators on at least up to 10 qubits, so this
is what we use in our calculations.

\begin{figure}
\includegraphics[scale=.75]{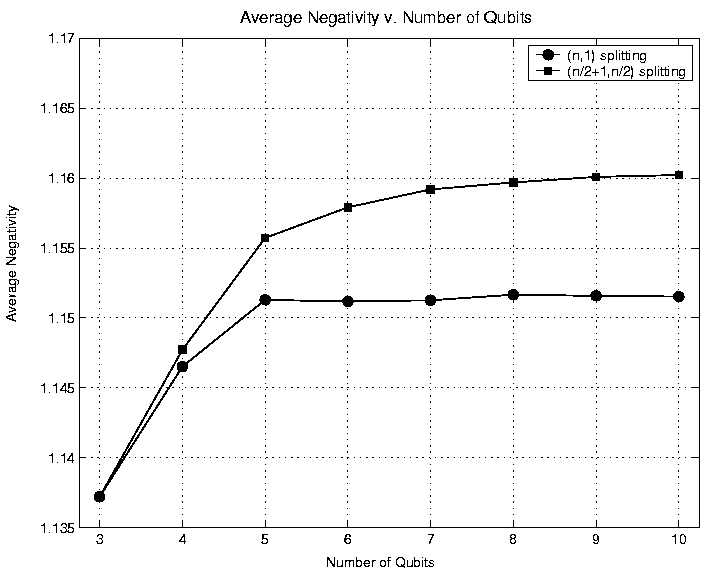}
\caption[Mean of the negativity in a DQC1 circuit]{Average
negativity of the state $\rho_{n+1}$ of
Eq.~(\protect\ref{E:rhoout}) ($\alpha=1$) for a randomly chosen
unitary $U_n$ for two different bipartite splittings, $(n,1)$ and
$\left(\left\lfloor n/2 \right\rfloor +1, \left\lceil n/2
\right\rceil\right)$.  The $(n,1)$ splitting appears to reach an
upper bound quickly, whereas the other splitting is still rising
slowly at 10 qubits.} \label{F:randomm}
\end{figure}
\begin{figure}
\includegraphics[scale=.75]{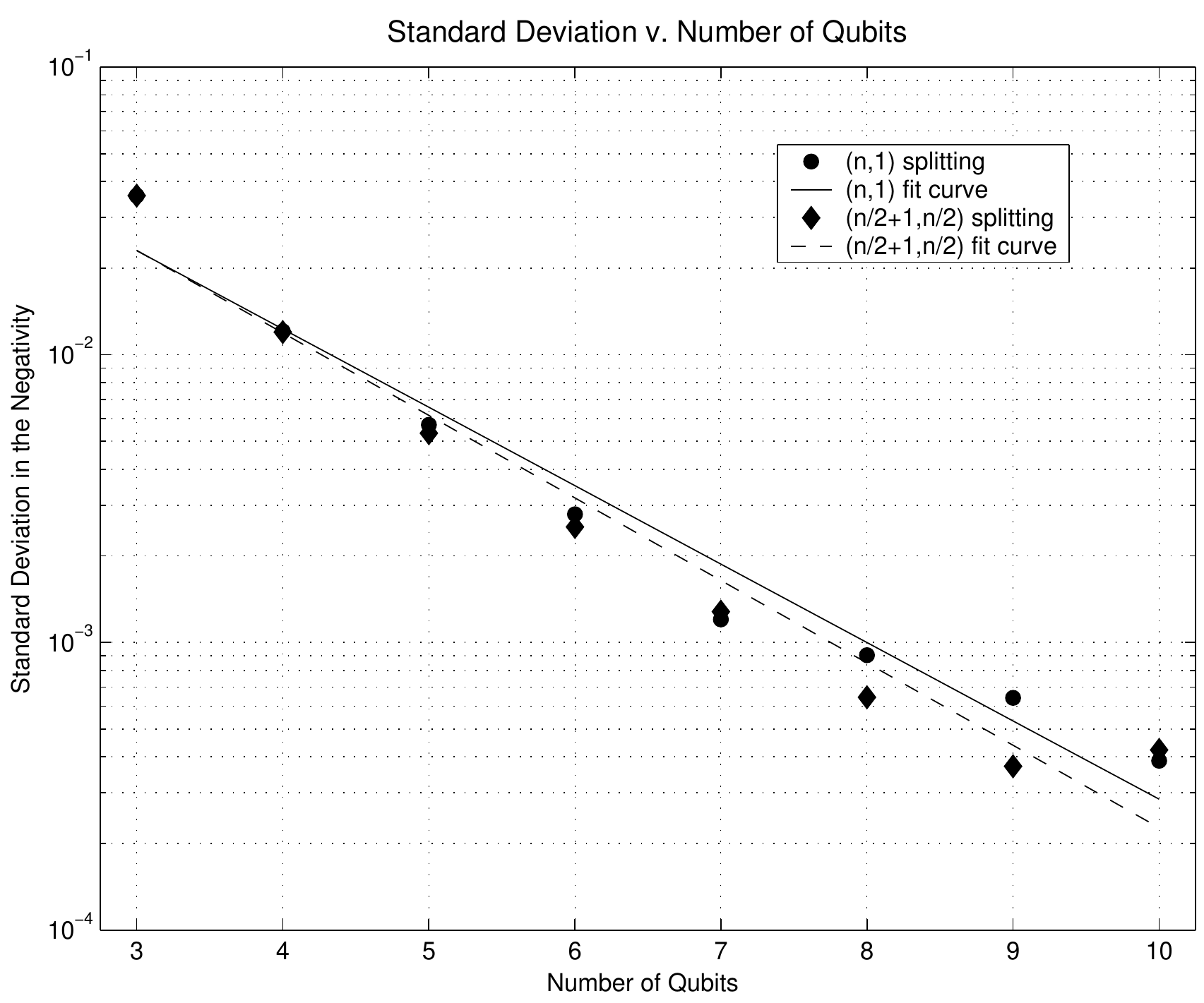}
\caption[Variance in the negativity in a DQC1 circuit]{Semi-log
plot of the standard deviation in the negativity of the randomly
chosen state $\rho_{n+1}$. The fit curves show that the standard
deviation is decaying exponentially, so that for large numbers of
qubits, almost all unitaries give the same negativity.}
\label{F:randomv}
\end{figure}
Due to boundary effects, not all bipartite splittings that put $k$
unpolarized qubits in one part are equivalent. Nevertheless, we
consider only bipartite divisions that split the qubits along
horizontal lines placed at various points in the circuit of
Eq.~(\ref{E:circuit}).  We refer to the division that groups the
last $k$ qubits together as the $(n+1-k,k)$ splitting.  For
$\alpha=1$, we calculate the average negativity and standard
deviation of a pseudo-random state $\rho_{n+1}$ for two different
bipartite splittings, $(n,1)$ and $(\left\lfloor n/2 \right\rfloor
+1, \left\lceil n/2 \right\rceil )$. These results are plotted in
Figs.~\ref{F:randomm} and ~\ref{F:randomv}.

\begin{figure}
\includegraphics[scale=0.75]{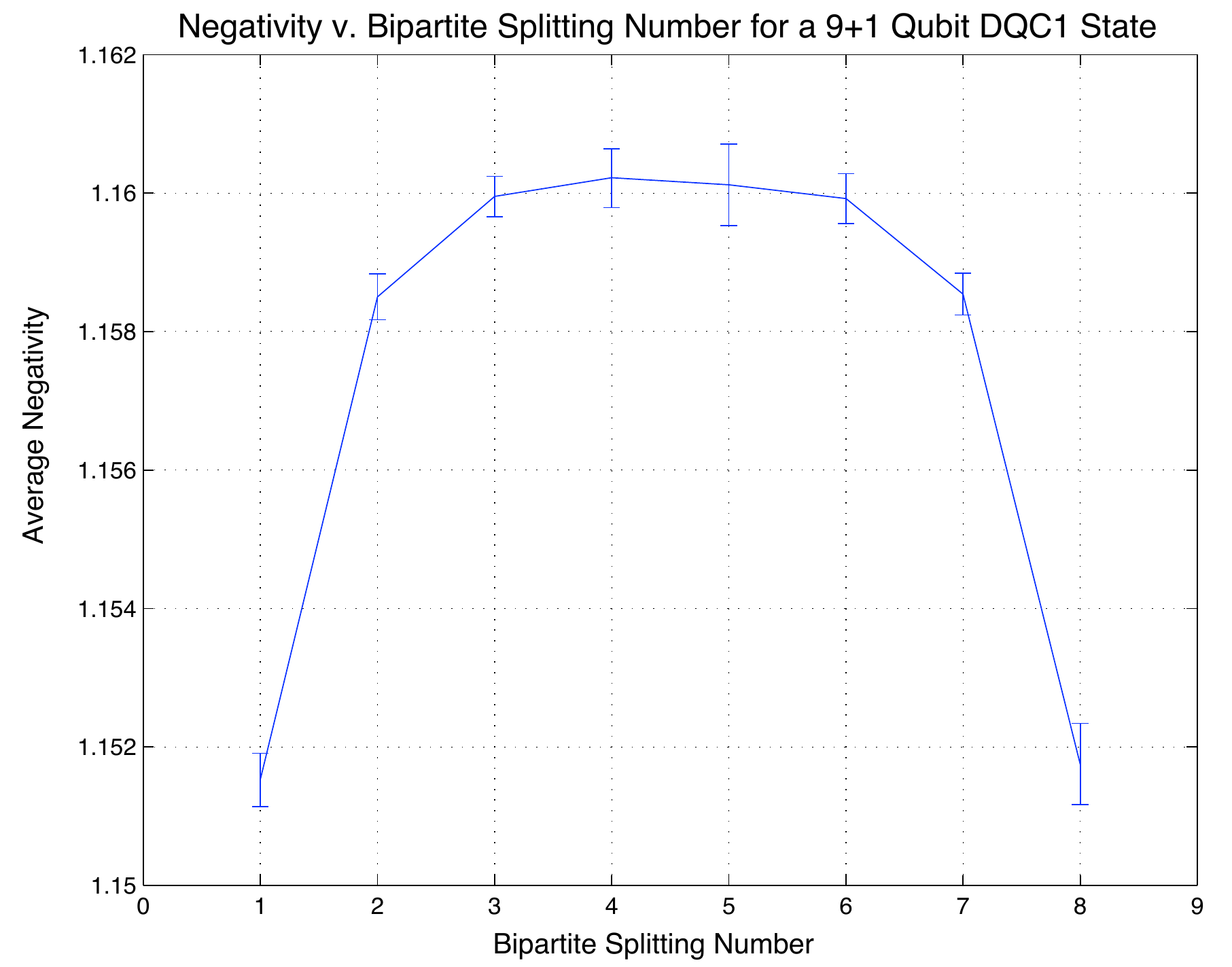}
\caption[Mean negativity across different splits in a 9 qubit DQC1
circuit]{Average negativity of the state $\rho_{10}$ of
Eq.~(\protect\ref{E:rhoout}) ($\alpha=1$) for a randomly chosen
unitary $U_9$ as a function of bipartite splitting number $k$ for
bipartite splittings $(10-k,k)$.  The error bars give the standard
deviations. The function attains a maximum when the bipartite
split is made between half of the qubits on which the unitary
acts.} \label{F:allsplits}
\end{figure}
For $n+1=5,\ldots,10$, the average negativity for the roughly
equal splitting lies between 1.135 and just above 1.16. The
standard deviation appears to converge exponentially to zero, as
in Ref.~\cite{scott03}, a behavior that is typical of
asymptotically equivalent matrices.  In addition, for $9+1$
qubits, we calculate the average negativity and standard deviation
for all nontrivial ($k\ne n$) bipartite splittings $(n+1-k,k)$,
and the results are shown in Fig.~\ref{F:allsplits}.

\section{Bounds on the negativity} \label{S:bounds}

In this section, we return to allowing the special qubit in the
circuit (\ref{E:circuitalpha}) to have initial
polarization~$\alpha$.  Since the value of $n$ is either clear
from context or fixed, we reduce the notational clutter by
denoting the state $\rho_{n+1}(\alpha)$ of
Eq.~(\ref{E:rhooutalpha}) as~$\rho_\alpha$.

Given a particular bipartite division, the partial transpose of
$\rho_\alpha$ with respect to the part that does not include the
special qubit is
\begin{equation}
    \breve \rho_\alpha = \frac{I_n+\alpha \breve C}{2N} \;\;\;\;
\mbox{where}\;\;\;\;
    \breve C \equiv
    \begin{pmatrix}
        0 & \breve U_n^\dag \\
        \breve U_n & 0
    \end{pmatrix} \;.
\end{equation}
Using the binomial theorem, we can expand
$\tr(\breve\rho^{\,s}_\alpha)$ in terms of $\tr(\breve C^k)$:
\begin{equation}
\label{E:binomialform}
    \tr(\breve\rho^{\,s}_\alpha) =
    \left({1\over2N}\right)^{\! \! s} \sum_{k=0}^s {s \choose k} \alpha^k \tr(\breve C^k) \;.
\end{equation}
When $k$ is odd, $\breve C^k$ is block off-diagonal, so its trace
vanishes.  When $k$ is even, we have
\begin{equation}
\tr(\breve C^k) = 2\,\tr\Bigl((\breve U_n \breve
U_n^\dag)^{k/2}\Bigr)\;.
\end{equation}
When $k=2$, this simplifies to $\tr(\breve C^2)=2\tr(\breve
U_n\breve U_n^\dag)=2\tr(U_n U_n^\dag)=2N$.  The crucial step here
follows immediately from the property $\tr(\breve A\breve
B)=\tr(AB)$, which we prove as a Lemma in
Appendix~\ref{A:appendix}.  Note that in general $\tr(\breve A_1
\breve A_2 \ldots \breve A_l) \not= \tr(A_1 A_2 \ldots A_l)$ if
$l>2$, so we cannot give a similar general calculation of
$\tr(\breve\rho^{\,s}_\alpha)$ for even $s \ge 4$, since it
involves terms of this form.

Using Eq.~(\ref{E:binomialform}), we can now obtain three
independent constraint equations on the eigenvalues
$\lambda_j=\lambda_j(\breve\rho_\alpha)$ of the partial transpose
$\breve\rho_\alpha$:
\begin{equation} \label{E:sumsofpowers}
    \sum_{j=1}^{2N}\lambda_j^s=
    \tr(\breve \rho^{\,s}_\alpha)
    = \frac{1}{2^s N^{s-1}}[(1+\alpha)^s + (1-\alpha)^s]\;,\qquad s=1,2,3.
\end{equation}
Since the negativity is given by
\begin{equation} \label{E:negdef}
    \sM(\rho_\alpha) = \sum_j \abs{\lambda_i} \;,
\end{equation}
we can find an upper bound on the negativity by maximizing
$\sum_j\abs{\lambda_j}$ subject to the
constraints~(\ref{E:sumsofpowers}).  If we consider only the
$s=1,2$ constraints, we obtain a nontrivial upper bound on the
negativity with little effort.  We find that adding the constraint
$s=3$ adds nothing asymptotically for large $N$, but for small $N$
yields a tighter bound than we get from the $s=1,2$ constraints,
although this comes at the cost of considerably more effort.  We
emphasize that these bounds apply to all bipartite divisions and
to all unitaries $U_n$.  Notice that we have no reason to expect
these bounds to be saturated, since the traces of higher powers of
$\breve\rho_\alpha$ impose additional constraints that we are
ignoring.  The one exception is the case of three qubits, where
the $s=1,2,3$ constraints are a complete set, and indeed, in this
case, the $s=1,2,3$ bound is $5/4$, which is saturated by the
unitary found in Sec.~\ref{S:examples}.

\begin{figure}
\includegraphics[scale=.75]{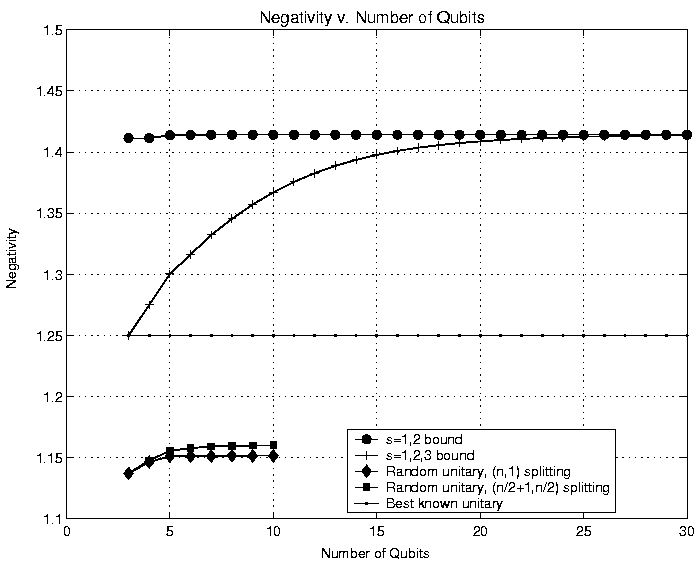}
\caption[Various bounds on the negativity of the DQC1
circuit]{Plot of the bounds on the negativity of states of the
form~(\ref{E:rhoout}), i.e., for a pure-state input in the zeroth
register ($\alpha=1$). The uppermost plot is the simple analytic
bound $\sM_{1,2}=\sqrt2$, obtained using the $s=1,2$~constraint
equations; the next largest plot is the numerically constructed
$s=1,2,3$ bound. One can see that the $s=1,2,3$ bound asymptotes
to the $s=1,2$ bound. As noted in the text, these bounds are
independent of the unitary $U_n$ and the bipartite division.  The
flat line shows the negativity $5/4$ for the state constructed in
Sec.~\ref{S:examples}, currently the state of the
form~(\ref{E:rhoout}) with the largest demonstrated negativity;
notice that for $n+1=3$, this state attains the $s=1,2,3$ bound.
The lowest two sets of data points display the expected
negativities for a randomly chosen unitary using the bipartite
splittings $(n,1)$ and $\left(\left\lfloor n/2 \right\rfloor +1,
\left\lceil n/2 \right\rceil\right)$, which were also plotted in
Figs.~\ref{F:randomm} and \ref{F:randomv}.} \label{F:totalfig}
\end{figure}

The remainder of this section is devoted to calculating the
$s=1,2$ and $s=1,2,3$ upper bounds.  A graphical summary of our
results for $\alpha=1$ is presented in Fig.~\ref{F:totalfig}.

\subsection{The $s=1,2$ Bound}

We can use Lagrange multipliers to reduce the problem to
maximizing a function of one variable, but first we must deal with
the absolute value in Eq.~(\ref{E:negdef}).  To do so, we assume
that $t$ of the eigenvalues are negative and the $2N-t$ others are
nonnegative, where $t$ becomes a parameter that must now be
included in the maximization. We want to maximize
\begin{equation}
\sM_{1,2} = -\sum_{i=1}^t \lambda_i + \sum_{j=t+1}^{2N}
\lambda_j\;,
\end{equation}
 subject to the constraints
\begin{equation} \label{E:cons}
    \sum_{k=1}^{2N} \lambda_k = 1
    \quad \mbox{and} \quad \sum_{k=1}^{2N}
    \lambda_k^2 = \frac{1+\alpha^2}{2N} \;.
\end{equation}
The notation we adopt here for the indices is that $i$ labels
negative eigenvalues and $j$ labels nonnegative eigenvalues, while
$k$ can label either.  This serves to remind us of the sign of an
eigenvalue just by looking at its index.

Introducing Lagrange multipliers $\mu$ and $\nu$, the function we
want to maximize is
\begin{equation}
    f(\lambda_k,t) =
    -\sum_{i=1}^t \lambda_i + \sum_{j=t+1}^{2N} \lambda_j +
    \mu\!\left(\sum_{k=1}^{2N} \lambda_k -1\right)
    + \nu\!\left(\sum_{k=1}^{2N} \lambda_k^2 -\frac{1+\alpha^2}{2N}\right)\;.
\end{equation}
Differentiating with respect to $\lambda_i$ and then $\lambda_j$,
we find
\begin{eqnarray}
    -1+\mu+2\nu \lambda_i = 0 \;, \\
    +1+\mu+2\nu \lambda_j = 0 \;.
\end{eqnarray}
We immediately see that in the maximal solution, all the negative
eigenvalues are equal, and all the nonnegative eigenvalues are
equal. We can now reformulate the problem in the following way.
If we call the two eigenvalues $\lambda_-$ and $\lambda_+$, our
new problem is to maximize
\begin{equation} \label{E:newneg}
    \sM_{1,2} = \sum_k \abs{\lambda_k}= -t\lambda_-+(2N-t)\lambda_+\;,
\end{equation}
subject to the constraints
\begin{eqnarray} \label{E:constraint1}
    t \lambda_- + (2N-t) \lambda_+ &=& 1\;,\\
    t \lambda_-^2 + (2N-t) \lambda_+^2 &=& \frac{1+\alpha^2}{2N}\;.
    \label{E:constraint2}
\end{eqnarray}
We can now do the problem by solving the constraints for
$\lambda_-$ and $\lambda_+$ in terms of $t$, plugging these
results into $\sM_{1,2}$, and then maximizing over $t$.

Before continuing, we note two things.  First, $t$ cannot be $2N$,
for if it were, then all the eigenvalues would be negative, making
it impossible to satisfy Eq.~(\ref{E:constraint1}).  Second,
unless $\alpha=0$, $t$ cannot be 0, for if it were, then all the
eigenvalues would be equal to $1/2N$ by Eq.~(\ref{E:constraint1}),
a situation Eq.~(\ref{E:constraint2}) says can occur only if
$\alpha=0$.  Since we are not really interested in the case
$\alpha=0$, for which $\rho_\alpha$ is always the maximally mixed
state, we assume $\alpha>0$ and $0<t<2N$ in what follows.

Solving Eqs.~(\ref{E:constraint1}) and (\ref{E:constraint2}) and
plugging the solutions into Eq.~(\ref{E:newneg}), we get the two
solutions
\begin{equation}
\label{E:ellipse}
    \sM_{1,2} = \frac{N-t \pm \alpha \sqrt{t(2N-t)}}{N} \;.
\end{equation}
We choose the positive branch, since it contains the maximum.
Maximizing with respect to $t$ treated as a continuous variable,
we obtain the upper bound,
\begin{equation} \label{E:maxneg}
    \sM_{1,2} = \sqrt{1+\alpha^2}\, \stackrel{\alpha \to 1}{=}
    \sqrt 2 \simeq 1.414\;,
\end{equation}
which occurs when the degeneracy parameter is given by
\begin{equation}
\label{E:degeneracy}
    t = N\!\left(1 -\frac{1}{\sqrt{1+\alpha^2}} \right)
    \stackrel{\alpha \to 1}{\simeq} 0.292\,N \;.
\end{equation}
The numbers on the right are for the case $\alpha = 1$,
corresponding to the special qubit starting in a pure state.
Notice that the upper bound~(\ref{E:maxneg}) allows a negativity
greater than 1 for all $\alpha$ except $\alpha=0$.

Since we did not yet enforce the condition that $t$ be a positive
integer, the bound~(\ref{E:maxneg}) can be made tighter for
specific values of $N$ and $\alpha$ by calculating $t$ and
checking which of the two nearest integers yields a larger
$\sM_{1,2}$.  Asymptotically, however, the ratio $t/N$ can
approach any real number, so this bound for continuous $t$ is the
same as the bound for integer $t$ in the limit $N\to\infty$.

\subsection{The $s=1,2,3$ Bound}

To deal with this case, we again make the assumption that $t$ of
the eigenvalues are negative and $2N-t$ are nonnegative and thus
write
\begin{equation}
\sM_{1,2,3} = -\sum_{i=1}^t \lambda_i + \sum_{j=t+1}^{2N}
\lambda_j\;,
\end{equation}
as before.  In addition to the constraints~(\ref{E:cons}), we now
have a third constraint
\begin{equation}
\label{E:constraint3}
    \sum_{k=1}^{2N} \lambda_k^3 = \frac{1+3 \alpha^2}{4N^2} \;.
\end{equation}
We specialize to the case $\alpha = 1$ for the remainder of this
subsection, because it is our main interest, and the algebra for
the general case becomes difficult.

Introducing three Lagrange multipliers, we can write the function
we want to maximize as
\begin{equation}
\label{E:lagrange123}
    f(\lambda_k,t) =
    -\sum_{i=1}^t \lambda_i + \sum_{j=t+1}^{2N} \lambda_j +
    \mu\!\left( \sum_{k=1}^{2N} \lambda_k - 1 \right)+
    \nu\!\left( \sum_{k=1}^{2N} \lambda_k^2 - \frac{1}{N} \right) +
    \xi\!\left( \sum_{k=1}^{2N} \lambda_k^3 - \frac{1}{N^2} \right)\;.
\end{equation}
Differentiating with respect to $\lambda_i$ and then $\lambda_j$
gives
\begin{eqnarray}
    -1+\mu+2\nu \lambda_i + 3 \xi \lambda_i^2 &=& 0 \;, \\
    +1+\mu+2\nu \lambda_j + 3 \xi \lambda_j^2 &=& 0 \;.
\end{eqnarray}
These equations being quadratic, we see that there are at most two
distinct negative eigenvalues and at most two distinct nonnegative
eigenvalues.  Since the sum of the two solutions of either of
these equations is $-2\nu/3\xi$, however, we can immediately
conclude either that one of the potentially nonnegative solutions
is negative or that one of the potentially negative solutions is
positive.  Hence, we find that at least one of the four putative
eigenvalues has the wrong sign, implying that there are at most
three distinct eigenvalues, though we don't know whether one or
two of them are negative.

Labelling the three eigenvalues by $A$, $B$, and $C$, we can
reduce the problem to solving the three constraint equations,
\begin{eqnarray} \label{E:123}
    u A + v B + w C & = & 1\;,\nonumber \\
    u A^2 + v B^2 + w C^2 & = & 1/N\;,\\
    u A^3 + v B^3 + w C^3 & = & 1/N^2\;,\nonumber
\end{eqnarray}
for $A$, $B$, and $C$ and then maximizing $\sM_{1,2,3}$ over the
degeneracy parameters $u$, $v$, and $w$, which are nonnegative
positive integers satisfying the further constraint
\begin{equation}
u+v+w=2N\;.
\end{equation}
We do not associate any particular sign with $A$, $B$, and $C$;
the signs are determined by the solution of the equations.

One might hope that the symmetry of Eqs.~(\ref{E:123}) would allow
for a simple analytic solution, but this appears not to be the
case. In solving the three equations, one is inevitably led to a
sixth-order polynomial in one of the variables, with the
coefficients given as functions of $u$, $v$, and $w$.  Rather than
try to solve this equation, which appears intractable, we elected
to do a brute force optimization for any given value of $2N$ by
solving Eqs.~(\ref{E:123}) for each possible value of $u$, $v$,
and $w$.  Picking the solution that has the largest negativity
then yields the global maximum.  We did this for each $N$ up to
$2N=78$.  The values of $u$, $v$, and $w$ that maximize the
negativity are always
\begin{equation} \label{E:uvw}
    u=\left[N\left(1-\frac{1}{\sqrt 2}\right)\right] \;,
    \ v=1 \;,
    \ w = 2N-1-u \;,
\end{equation}
where $[x]$ denotes the integer nearest to $x$.  The unique
eigenvalue corresponding to $v=1$ is the largest positive
eigenvalue, $w$ is the degeneracy of another positive eigenvalue,
and $u$ is the degeneracy of the negative eigenvalue.  Notice that
the degeneracy of the negative eigenvalue is exactly what was
found in the $s=1,2$ case.  Using the results~(\ref{E:uvw}) as a
guide, we did a further numerical calculation of the maximum for
larger values of $N$, by considering only the area around the
degeneracy values given by Eq.~(\protect\ref{E:uvw}).  While this
is not a certifiable global maximum, the perturbation expansion
described below matches so well that the two are indistinguishable
if they are plotted together for $n+1>7$.  This gives us
confidence that the numerically determined upper bound
$\sM_{1,2,3}$, which we plot in Fig.~\ref{F:totalfig}, is indeed a
global maximum for all $N$.

We have used the numerical work to help formulate a perturbation
expansion that gives the first correction to the $N\to\infty$
behavior of the $s=1,2,3$ bound.  Defining $x=1/N$, we rewrite the
constraint equations~(\ref{E:123}) as
\begin{eqnarray} \label{E:con}
    a A + b B + c\,C & = & x\;,\nonumber \\
    a A^2 + b B^2 + c\,C^2 & = & x^2\;,\\
    a A^3 + b B^3 + c\,C^3 & = & x^3\;,\nonumber
\end{eqnarray}
where $a=u/N$, $b=v/N$, and $c=w/N$.  We also have the constraint
\begin{equation} \label{E:con2}
a+b+c=2\;.
\end{equation}
As $x$ is the variable that is asymptotically small, we seek an
expansion in terms of it.

Our numerical work tells us that there are two positive
eigenvalues, one of which is larger and nondegenerate.  In
formulating our perturbation expansion, we let $B$ and $C$ be the
positive eigenvalues, with $B$ being the larger one, having
degeneracy $v=b_1\ge1$.  We do not assume that $b_1$ is 1, as the
numerics show, but rather let the equations force us to that
conclusion.  With this assumption, the form of the
constraints~(\ref{E:con}) shows that the variables have the
following expansions to first order beyond the $N\to\infty$ form:
\begin{eqnarray} \label{E:degs}
    a & = & a_0 + a_1 x^{1/3}\;,\nonumber \\
    b & = & b_1x\;,\\
    c & = & c_0 + c_1x^{1/3}\;,\nonumber
\end{eqnarray}
and
\begin{eqnarray} \label{E:evs}
    A & = & A_0 x + A_1 x^{4/3}\;,\nonumber \\
    B & = & B_1 x^{2/3}\;,\\
    C & = & C_0 x + C_1 x^{4/3}\;.\nonumber
\end{eqnarray}
We see that we are actually expanding in the quantity $y=x^{1/3}$.
In terms of these variables, the negativity is given by
\begin{eqnarray}
 \sM_{1,2,3}\!\!&=&\!\!\!\!
 \frac{-a A + b B + c C}{x}\nonumber\\
 &=&\!\!\!\!-a_0A_0+c_0C_0+(-a_0A_1-a_1A_0+c_0C_1+c_1C_0)x^{1/3}+O(x^{2/3}),
\end{eqnarray}
which we now endeavor to maximize.

Substituting Eqs.~(\ref{E:degs}) and (\ref{E:evs}) into the
constraints~(\ref{E:con}) and (\ref{E:con2}) and equating terms
with equal exponents of $x$, we obtain, to zero order,
\begin{eqnarray}
    a_0 + c_0 & = & 2\;,\nonumber \\
    a_0 A_0 + c_0 C_0 & = & 1\;,\\
    a_0 A_0^2+ c_0 C_0^2 & = & 1 \;.\nonumber
\end{eqnarray}
Solving for $a_0$, $c_0$, and $C_0$ in terms of $A_0$ and
substituting the results into the zero-order piece of
$\sM_{1,2,3}$ gives
\begin{equation}
 \label{E:neg1}
 \sM_{1,2,3}=\frac{1-4A_0+2A_0^2}{1-2A_0+2A_0^2}\;.
\end{equation}
Maximizing Eq.~(\ref{E:neg1}) gives $A_0^2=1/2$ and, hence,
$A_0=-1/\sqrt{2}$, since $A$ is the negative eigenvalue.  This
leads to $a_0=1-1/\sqrt{2}$, $c_0 = 1+1/\sqrt{2}$, and
$C_0=1/\sqrt{2}$, and the resulting $N\to\infty$ upper bound is
$\sM_{1,2,3}=\sqrt2$, as expected.

If we carry this process out to first order beyond the
$N\to\infty$ behavior, we obtain, after some algebraic
manipulation, $\sM_{1,2,3} =
\sqrt2-b_1^{1/3}x^{1/3}/2^{7/6}+O(x^{2/3})$.  Maximizing this
simply means making $b_1$ as small as possible, i.e., choosing
$b_1=1$, whence we obtain the following asymptotic expression for
the $s=1,2,3$ upper bound:
\begin{equation}
\sM_{1,2,3} = \sqrt2 - \frac{1}{2^{7/6}N^{1/3}} +
O\!\left(\frac{1}{N^{2/3}}\right)\;.
\end{equation}
This shows that the upper bound of $\sqrt 2$ is approached
monotonically from below in the asymptotic regime.  In addition,
the procedure verifies that in the maximum solution, the largest
positive eigenvalue is nondegenerate.  For the case of qubits we
have $N=2^n$, implying that the approach to the $N\to\infty$ bound
is exponentially fast.

\section{Conclusion} \label{S:conclusion}

The mixed-state quantum circuit in Fig.~(\ref{E:circuitalpha})
provides an efficient method for estimating the normalized trace
of a unitary operator, a task that is thought to be exponentially
hard on a classical computer.  If one believes that global
entanglement is the essential resource for the exponential speedup
achieved by quantum computation, then the question begging to be
answered is whether there is any entanglement in the circuit's
output state in Eq.~(\ref{E:rhooutalpha}). The purpose of this
chapter was to investigate this question.

A notable feature of the circuit in Fig.~(\ref{E:circuitalpha}) is
that it provides an efficient method for estimating the normalized
trace no matter how small the initial polarization $\alpha$ of the
special qubit in the zeroth register, as long as that polarization
is not zero. Since all the other qubits are initially completely
unpolarized, we are led to characterize the computational power of
this circuit as the ``power of even the tiniest fraction of a
qubit.''  We provide preliminary results regarding the
entanglement that can be achieved for $\alpha<1$. Our results are
consistent with, but certainly do not demonstrate the conclusion
that separable states cannot provide an exponential speedup and
that entanglement is possible no matter how small $\alpha$ is. The
question of entanglement for subunity polarization of the special
qubit deserves further investigation.

Our key conclusions concern the case where the special qubit is
initially pure ($\alpha=1$).  We find that the circuit in
Fig.~(\ref{E:circuit}) typically does produce global entanglement,
but the amount of this entanglement is quite small. Using
multiplicative negativity to measure the amount of entanglement,
we show that as the number of qubits becomes large, the
multiplicative negativity in the state in Eq.~(\ref{E:rhoout}) is
a vanishingly small fraction of the maximum possible
multiplicative negativity for roughly equal splittings of the
qubits.  This hints that the key to computational speedup might be
the global character of the entanglement, rather than the amount
of the entanglement.  In the spirit of the pioneering contribution
of Wyler~\cite{wyler74}, what happier motto can we find for this
state of affairs than {\it Multum ex Parvo}, or A Lot out of A
Little.

\chapter{Classical simulation of quantum computation \label{chap:mps}}

\hfill \textsl{I looked for an answer to my question. But reason
could not give me an answer- reason is incommensurable with the
question.}

\hfill  -Leo Tolstoy in \textsl{Anna Karenina}

\vskip1.0cm

    Progress in our understanding of what makes quantum evolutions
computationally more powerful than a classical computer has been
scarce. A step forward, however, was achieved by identifying
entanglement as a {\em necessary} resource for quantum
computational speed-ups. Indeed, a speed-up is only possible if in
a quantum computation, entanglement spreads over an adequately
large number of qubits \cite{jozsa03a}. In addition, the amount of
entanglement, as measured by the Schmidt rank of a certain set of
bipartitions of the system, needs to grow sufficiently with the
size of the computation \cite{vidal03a}. Whenever either of these
two conditions is not met, the quantum evolution can be
efficiently simulated on a classical computer. These conditions
(which are particular examples of subsequent, stronger classical
simulation results based on tree tensor networks (TTN)
\cite{sdv06,vandernest06}) are only necessary, and thus not
sufficient, so that the presence of large amounts of entanglement
spreading over many qubits does not guarantee a computational
speed-up, as exemplified by the Gottesman-Knill theorem
\cite{nielsen00a}.

    The above results refer exclusively to quantum computations with
pure states. As shown in Chapter \ref{chap:dqc1}, the scenario for
mixed-state quantum computation is rather different. The
intriguing {\em deterministic quantum computation with one quantum
bit} (DQC1 or `the power of one qubit') \cite{kl98} involves a
highly mixed state that does not contain much entanglement
\cite{datta05a} and yet it performs a task, the computation with
fixed accuracy of the normalized trace of a unitary matrix,
exponentially faster than any known classical algorithm. This also
provides an exponential speedup over the best known classical
algorithm for simulations of some quantum processes \cite{pklo04}.
Thus, in the case of a mixed-state quantum computation, a large
amount of entanglement does not seem to be necessary to obtain a
speed-up with respect to classical computers.

        A simple, unified explanation for the pure-state and
mixed-state scenarios is possible \cite{vidal03a} by noticing that
the decisive ingredient in both cases is the presence of {\em
correlations}. Indeed, let us consider the Schmidt decomposition
of a vector $\ket{\Psi}$, given by
\begin{equation}
    \ket{\Psi} = \sum_{i=1}^{\chi} \lambda_{i} \ket{i_A}\otimes \ket{i_B},
\end{equation}
where $\braket{i_A}{j_A} = \braket{i_B}{j_B} = \delta_{ij}$ and
$\chi$ is the rank of the reduced density matrices $\rho_A \equiv
\tr_B[\proj{\psi}]$ and $\rho_B \equiv \tr_A[\proj{\psi}$]; and
the (operator) Schmidt decomposition of a density matrix $\rho$
given by \cite{zv04}
\begin{equation}
    \rho = \sum_{i=1}^{\chi^{\sharp}} \lambda^{\sharp}_{i} ~O_{iA} \otimes O_{iB},
\label{eq:Schmidt_rho}
\end{equation}
where $\tr (O_{iA}^{\dagger}O_{jA}) = \tr (O_{iB}^{\dagger}O_{jB})
= \delta_{ij}$. The Schmidt ranks $\chi$ and $\chi^{\sharp}$ are a
measure of correlations between parts $A$ and $B$, with
$\chi^{\sharp} = \chi^2$ if $\rho=\proj{\Psi}$. Let the density
matrix $\rho_t$ denote the evolving state of the quantum computer
during a computation. Notice that $\rho_t$ can represent both pure
and mixed states. Then, as shown in Refs. \cite{vidal03a} and
\cite{sdv06,vandernest06}, the quantum computation can be
efficiently simulated on a classical computer using a TTN
decomposition if the Schmidt rank $\chi^{\sharp}$ of $\rho$
according to a certain set of bipartitions $A:B$ of the qubits
scales polynomially with the size of the computation. In other
words, a necessary condition for a computational speed-up is that
correlations, as measured by the Schmidt rank $\chi^{\sharp}$,
grow super-polynomially in the number of qubits. In the case of
pure states (where $\chi = \sqrt{\chi^{\sharp}}$) these
correlations are entirely due to entanglement, while for mixed
states they may be quantum or classical.

    Our endeavor in this chapter is to study the DQC1 model of quantum
computation following the above line of thought. In particular, we
elucidate whether DQC1 can be efficiently simulated with any
classical algorithm, such as those in
\cite{vidal03a,sdv06,vandernest06} (and, implicitly, in
\cite{jozsa03a}), which exploit limits on the amount of
correlations, in the sense of a small $\chi^{\sharp}$ according to
certain bipartitions of the qubits. We will argue here that the
state $\rho_t$ of a quantum computer implementing the DQC1 model
displays an exponentially large $\chi^{\sharp}$, in spite of its
containing only a small amount of entanglement \cite{datta05a}. We
will conclude, therefore, that none of the simulation techniques
mentioned above can be used to efficiently simulate `the power of
one qubit'.

    On the one hand, our result indicates that entanglement is not
behind the (suspected) computational speed-up of DQC1. On the
other hand, by showing the failure of a whole class of classical
algorithms to efficiently simulate this mixed-state quantum
computation, we reinforce the conjecture that DQC1 leads indeed to
an exponential speed-up. We note, however, that our result does
\emph{not} rule out the possibility that this circuit could be
simulated efficiently using some other classical algorithm.

    Before we move on to the presentation of our results and
conclusions, we will present a short review of tree tensor
networks(TTN) for the sake of completeness and continuity. This
will be the subject of the next section.

\section{Tree Tensor Networks (TTN)}
\label{ttn}

The discussion in this section is not original and a more detailed
compilation can be found in~\cite{ms05}. For simplicity, let us
deal with pure states of $n$ qudits, denoted that by
 \be
\ket{\Psi} = \sum_{i_1=1}^d\cdots\sum_{i_n=1}^d c_{i_1\cdots
i_n}\ket{i_1}\otimes\cdots\otimes\ket{i_n}.
 \ee
This state is characterized by $d^n$ complex amplitudes
$c_{i_1\cdots i_n}.$ We can think of all these coefficients as one
entity, a rank-$n$ tensor. Such a tensor can be represented
graphically, with vertices labelled by $c$, and connected by $n$
open wires, each of which is labelled by a distinct index. One may
represent a tensor \emph{network} by starting with such graphical
representation of its tensors, and then connecting wires
corresponding to the same index. As quantum gates are performed on
the state $\ket{\Psi}$, the graph (or the network) gets more
involved.

    In the quantum circuit model, result of the quantum computation is
obtained by taking trace of the final state against measurement
operators. In our present picture, this translates to contracting
vertices of the tensor networks, that is, removing the edges
between two vertices and replacing them by a single one. This
operation is known as tensor contraction. The complexity of
simulating a quantum computation thus reduces to that of
simulating tensor contractions. For graphs which have small
treewdiths, this operation can be approximated classically with a
high accuracy in polynomial time. Without going into its formal
definition, the treewidth of a graph tells us how different it is
from a tree. A tree has treewidth of 1. Single cycles of length at
least 3 have a treewidth of 2.

        For any graph $G=(V,E)$, the line graph $G^*$ is defined as
follows: $V(G^*)\equiv E(G)$ and
 $$
 E(G^*)\equiv\left\{\{e_1,e_2\} \subseteq E(G): e_1\neq e_2,\exists\; v \in V(G)\; \mbox{such that $e_1$ and $e_2$ are incident on
$v$}\right\}.
 $$
The contraction complexity of a graph $G$, $\mathrm{cc}(G)$ is
equal to the the treewidth of the line graph $G^*$,
$\mathrm{tw}(G^*)$~\cite{ms05}. Although determining the treewidth
of a general graph is $\mathrm{NP}$-hard~\cite{acp87}, for trees
it is trivial. For this simple case then, $\mathrm{cc}(G)$ is
trivial. Additionally, if the graph underlying the quantum circuit
is such that $\mathrm{tw}(G)=d$, then the circuit can be simulated
in time $e^{O(d)}.$

        This is the inspiration for defining quantum circuits as tree tensor
networks. The closer the underlying line graph for a quantum
circuit is to a tree, easier it is to simulate the corresponding
quantum circuit efficiently classically. This, not surprisingly
then, is the motivation behind all the present simulation
techniques of quantum computation like matrix product states
(MPS), projected entangled-pair states (PEPS),
Affleck-Kennedy-Leib-Tasaki (AKLT) states. If one is able to show
that a certain quantum computation \emph{cannot} be represented as
a TTN, then a large class of techniques are excluded in providing
an efficient classical simulation for it. This will be our line of
approach in the rest of this chapter, with reference to the DQC1
model.

\section{DQC1 and Tree Tensor Networks (TTN) }
\label{dqcttn}

The DQC1 model, represented in Eq. (\ref{E:circuit}), provides an
estimate of the normalized trace $\tr(U_n)/2^{n}$ of a $n$-qubit
unitary matrix $U_n\in\mathbb{U}(2^n)$ with fixed accuracy
efficiently \cite{kl98}. This quantum circuit transforms the
highly-mixed initial state $\rho_0 \equiv \proj{0}\otimes I_n/2^n$
at time $t=0$ into the final state $\rho_T$ at time $t=T$,
 \be
\rho_T = \frac{1}{2^{n+1}}\left(%
\begin{array}{cc}
  I_n & U^{\dg}_n \\
  U_n & I_n \\
\end{array}%
\right), \label{eq:rhoT}
 \ee
through a series of intermediate states $\rho_t$, $t\in [0,T]$.
The simulation algorithms relevant in the present discussion
\cite{jozsa03a,vidal03a,sdv06,vandernest06} require that $\rho_t$
be efficiently represented with a TTN \cite{sdv06,vandernest06}
(or a more restrictive structure, such as a product of $k$-qubit
states for fixed $k$ \cite{jozsa03a} or a matrix product state
\cite{vidal03a}) at all times $t\in [0,T]$. Here we will show that
the final state $\rho_T$, henceforth denoted simply by $\rho$,
cannot be efficiently represented with a TTN. This already implies
that none of the algorithms in
\cite{jozsa03a,vidal03a,sdv06,vandernest06} can be used to
efficiently simulate the DQC1 model.

Storing and manipulating a TTN requires computational space and
time that grows linearly in the number of qubits $n$ and as a
small power of its rank $q$. The rank $q$ of a TTN is the maximum
Schmidt rank $\chi^{\sharp}_i$ over all bipartitions $A_i:B_i$ of
the qubits according to a given tree graph whose leaves are the
qubits of our system (see \cite{sdv06,vandernest06} for details).
The key observation of this chapter is that for a {\em typical}
unitary matrix $U_n$, the density matrix $\rho$ in Eq.
(\ref{eq:rhoT}) is such that any TTN decomposition has
exponentially large rank $q$. By {\em typical}, here we mean a
unitary matrix $U_n$ efficiently generated through a (random)
quantum circuit. That is, $U_n$ is the product of poly($n$)
one-qubit and two-qubit gates. In the next section we present
numerical results that unambiguously suggest that, indeed, {\em
typical} $U_n$ necessarily lead to TTN with exponentially large
rank $q$.

We notice that the results of the next section do not exclude the
possibility that the quantum computation in the DQC1 model can be
efficiently simulated with a TTN for particular choices of $U_n$.
For instance, if $U_n$ factorizes into single-qubit gates, then
$\rho$ can be seen to be efficiently represented with a TTN of
rank 3, and we can not rule out an efficient simulation of the
power of one qubit for that case. Of course, this is to be
expected, given that the trace of such $U_n$ can be computed
efficiently in the first place.

\section{Exponential growth of Schmidt ranks}

        In this section we study the rank $q$ of any TTN for the final state $\rho$ of the DQC1
circuit, Eq. (\ref{eq:rhoT}). We numerically determine that a
lower bound to such a rank grows exponentially with the number of
qubits $n$.

\begin{figure}
\includegraphics[scale=.75]{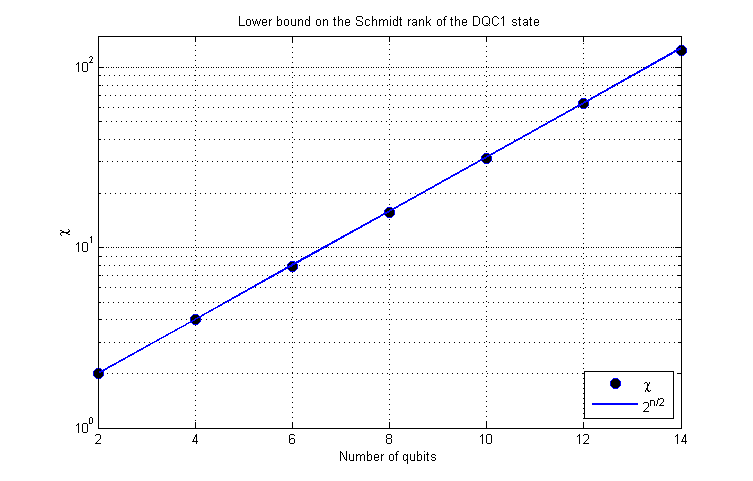}
\caption[Schmidt rank of the DQC1 state for any
equipartition]{Lower bound for the operator Schmidt rank
$\chi^{\sharp}$ of the DQC1 state for any equipartition $n/2:n/2$,
as given by the Schmidt rank $\chi$ of the pure state in Eq.
(\ref{purestate}). The dots are for even numbers of qubits, and
the fit is the line $2^{n/2}$. $\chi$ is calculated for a pure
state obtained by applying $2n$ random 2-qubit gates on the state
$\ket{0}^{\otimes n}$. This is evidence that for a {\em typical}
unitary $U_n$, the rank $q$ of any TTN for the DQC1 state $\rho$
in Eq. (\ref{eq:rhoT}) grows exponentially with $n$.}
 \label{14qubits}
\end{figure}

The Schmidt rank $\chi$ of a pure state
$\ket{\rho_{\phi_A\psi_B}}$ is
\begin{equation}
\label{purestate} \ket{\rho_{\phi_A\psi_B}} \equiv \rho
\ket{\phi_A}\ket{\psi_B} =  \sum_{i=1}^{\chi^{\sharp}}
\lambda^{\sharp}_{i} ~O_{iA}\ket{\phi_A} \otimes
O_{iB}\ket{\psi_B},
\end{equation}
obtained by applying the density matrix $\rho$ onto a product
state $\ket{\phi_A}\ket{\psi_B}$ is a lower bound on the operator
Schmidt rank $\chi^{\sharp}$ of $\rho$, i.e., $\chi^{\sharp} \geq
\chi$. For the purpose of our numerics, we consider the pure state
$U_n\ket{0}^{\otimes n}$. We build $U_n$ as a sequence of $2n$
random two-qubit gates, applied to pairs of qubits, also chosen at
random. The random two-qubit unitaries are generated using the
mixing algorithm presented in \cite{emerson03a}. Note that
applying $2n$ gates means that the resulting unitary is
efficiently implementable, a situation for which the DQC1 model is
valid. For an even number of qubits $n$, we calculate the smallest
Schmidt rank $\chi$ over all $n/2:n/2$ partitions of the qubits
(similar results can be obtained for odd $n$). The resulting
numbers are plotted in Fig (\ref{14qubits}).

        The above numerical results strongly suggest that
the final state $\rho$ in the DQC1 circuit has exponential Schmidt
rank for a {\em typical} unitary $U_n$. We are not able to provide
a formal proof of this fact. This is due to a general difficulty
in describing properties of the set $\mathbb{U}_{qc}(2^{n})$ of
unitary matrices that can be efficiently realized through a
quantum computation. Instead, the discussion is much simpler for
the set $\mathbb{U}(2^n)$ of generic $n$-qubit unitary matrices,
where it is possible to prove that $\rho$ cannot be efficiently
represented with a TTN for a Haar generated $U_n\in
\mathbb{U}(2^n)$, as discussed in the next section. Notice that
Ref. \cite{emerson05a} shows that random (but efficient) quantum
circuits generate random $n$-qubit gates $U_n\in
\mathbb{U}_{qc}(2^n)$ according to a measure that converges to the
Haar measure in $\mathbb{U}(2^n)$. Combined with the theorem in
the next section, this would constitute a formal proof of the
otherwise numerically evident exponential growth of the rank $q$
of any TTN for the DQC1 final state $\rho$.

\section{A formal proof for the Haar-distributed case}

    Our objective in this section is to analyze the Schmidt rank $\chi^{\sharp}$ of
the density matrix $\rho$ in Eq. (\ref{eq:rhoT}) for certain
bipartitions of the $n+1$ qubits, assuming that $U_n \in
\mathbb{U}(2^n)$ is Haar-distributed. As an aside, we calculate
the negativity of a random pure state in
Appendix~\ref{A:randomneg}.

It is not difficult to deduce that for any tree of the $n+1$
qubits, there exists at least one edge that splits the tree in two
parts $A$ and $B$, with $n_A$ and $n_B$ qubits, where
$n_0=\min(n_A,n_B)$ fulfills $n/5 \leq n_0 \leq 2n/5$. In other
words, if a rank-$q$ TTN exists for the $\rho$ in Eq.
(\ref{eq:rhoT}), then there is a bipartition of the $n+1$ qubits
with $n_0$ qubits on either $A$ or $B$ and such that the Schmidt
rank $\chi^{\sharp} \leq q$. Theorem \ref{t1}, our main technical
result, shows that if $U_n$ is chosen randomly according to the
Haar measure, then the Schmidt rank of any such bipartition
fulfills $\chi^{\sharp} \geq O(2^{n_0})$. Therefore for a randomly
generated $U_n \in \mathbb{U}(2^n)$, a TTN for $\rho$ has rank $q$
(and computational cost) exponential in $n$, and none of the
techniques of \cite{jozsa03a,vidal03a,sdv06,vandernest06} can
simulate the outcome of the DQC1 model efficiently.

Consider now any bipartition $A:B$ of the $n+1$ qubits, where $A$
and $B$ contain $n_A$ and $n_B$ qubits, with the minimum $n_0$ of
those restricted by $n/5\leq n_0 \leq 2n/5$. Without loss of
generality we can assume that the top qubit lies in $A$. Actually,
we can also assume that $A$ contains the top $n_A$ qubits. Indeed,
suppose $A$ does not have the $n_A$ top qubits. Then we can use a
permutation $P_n$ on all the $n$ qubits to bring the $n_A$ qubits
of $A$ to the top $n_A$ positions. This will certainly modify
$\rho$, but since
 \be
\left(%
\begin{array}{cc}
  P_n & 0 \\
  0 & P_n \\
\end{array}%
\right)\left(%
\begin{array}{cc}
  I_n & U^{\dg}_n \\
  U_n & I_n \\
\end{array}%
\right)\left(%
\begin{array}{cc}
  P^{T}_n & 0 \\
  0 & P^{T}_n \\
\end{array}%
\right)=\left(%
\begin{array}{cc}
  I_n & V^{\dg}_n \\
  V_n & I_n \\
\end{array}%
\right),
 \ee
where $V_n = P_n U_n P^T_n$ is another Haar-distributed unitary,
we obtain that the new density matrix is of the same form as
$\rho$. Finally, in order to ease the notation, we will assume
that $n_A=n_0$ (identical results can be derived for $n_B=n_0$).
Thus $n/5 \leq n_A \leq 2n/5$.

 We note that
    \be
    \left(%
\begin{array}{cc}
  I_n & U^{\dg}_n \\
  U_n & I_n \\
\end{array}%
\right) = \mathbb{I}_2 \otimes \mathbb{I}_n +  \left(%
\begin{array}{cc}
  0 & 1 \\
  0 & 0 \\
\end{array}%
\right) \otimes U_n^{\dg} + \left(%
\begin{array}{cc}
  0 & 0 \\
  1 & 0 \\
\end{array}%
\right) \otimes U_n,
    \ee
so that if we multiply $\rho$ by the product state
\begin{equation} \label{eq:prod}
 \ket{\phi_{\vec{\alpha}}} \equiv \ket{t,i,j} \equiv \ket{t,i_A}\ket{j_B},
\end{equation}
where $\vec{\alpha} \equiv (t,i,j)$, $t=0,1$; $i=1,\dots d_A$;
$j=1,\dots d_B$, we obtain $\ket{\psi_{\vec{\alpha}}} \equiv \rho
\ket{\phi_{\vec{\alpha}}}$ where
\begin{eqnarray}
\label{pstate}
 \ket{\psi_{\vec{\alpha}}} = \left\{
\begin{array}{c}
\frac{1}{2^{n+1}} (\ket{0,i,j} + \ket{1}\otimes U_n\ket{i,j}) ~~~ \mbox{if} ~t=0\\
\frac{1}{2^{n+1}} (\ket{1,i,j} + \ket{0}\otimes
U_n^{\dagger}\ket{i,j}) ~~~ \mbox{if} ~t=1.
\end{array} \right.
\end{eqnarray}
This also justifies our choice of the pure state used in the
numerical calculations in the previous section.

Let us consider now the reduced density matrix
    \ben
    \sigma^B_{\vec{\alpha}} &\equiv&
    \tr_A[\proj{\psi_{\vec{\alpha}}}] \nonumber\\
    &=&\frac{1}{2^{n+1}}\left(\ket{j}\bra{j} + \tr_A[U_n\ket{i,j}\bra{i,j} U_n^{\dg}]\right)
    \een
for $t=0$ (for $t=1$, $U_n$ and $U_n^\dagger$ need to be
exchanged). For a unitary matrix $U_n$ randomly chosen according
to the Haar measure on $\mathrm{U}(n)$, $U_n\ket{i,j}$ is a random
pure state on $A\otimes B$. Here, and henceforth $A$ is the space
of the first $n_A$ qubits without the top qubit. It follows from
\cite{hlw06} that the operator
\begin{equation} \label{eq:Q}
    Q = \tr_A [U_n\ket{i,j}\bra{i,j}
U_n^{\dg}]
\end{equation}
has rank $d_A$. Therefore the rank of $\sigma^{B}_{\vec{\alpha}}$
(equivalently, the Schmidt rank $\chi$ of
$\ket{\psi_{\vec{\alpha}}}$) is at least $2^{n_0}$. From Eq.
(\ref{purestate}) we conclude that the Schmidt rank of $\rho$
fulfills $\chi^{\sharp} \geq 2^{n_0} \geq 2^{n/5}$. We can now
collate these results into

\begin{theorem}
 \label{t1}
Let $U_n$ be an $n$-qubit unitary transformation chosen randomly
according to the Haar measure on $\mathrm{U}(2^n)$, and let $A:B$
denote a bipartition of $n+1$ qubits into $n_A$ and $n_B$ qubits,
where $n_0\equiv\min(n_A,n_B)$. Then $n/5\leq n_0\leq 2n/5$ and
the Schmidt decomposition of $\rho$ in Eq. (\ref{eq:rhoT})
according to bipartition $A:B$ fulfills $\chi^{\sharp} \geq
2^{n/5}$.
\end{theorem}

We have seen that we cannot efficiently simulate DQC1 with an
algorithm that relies on having a TTN for $\rho$ with low rank
$q$. However, in order to make this result robust, we need to also
show that $\rho$ cannot be well approximated by another
$\tilde{\rho}$ accepting an efficient TTN. This is what we do
next.

        We now explore the robustness of the statement of Theorem \ref{t1}. To
this end, we consider the Schmidt rank $\tilde{\chi}^{\sharp}$ for
a density matrix $\tilde{\rho}$ that approximates $\rho$ according
to a fidelity $F(O_1,O_2)$ defined in terms of the natural inner
product on the space of linear operators,
    $$
F(O_1,O_2)\equiv\tr({O_1^{\dg}O_2})\Bigg/{\sqrt{\tr(O_1^{\dg}O_1)}\sqrt{\tr(O_2^{\dg}O_2)}}\;,
    $$
where $F = 1$ if and only if $O_1=O_2$ and $F =
|\braket{\psi_1}{\psi_2}|^2$ for projectors $O_i = P_{\psi_i}$ on
pure states $\ket{\psi_i}$. We will show that if $\tilde{\rho}$ is
close to $\rho$, then $\tilde{\chi}^{\sharp}$ for a bipartition as
in Theorem 1 is also exponential. To prove this, we will require a
few lemmas which we now present.

\begin{lemma}
\label{maxsch}
 Let $\ket{\Psi}$ be a bipartite vector with $\chi$
terms in its Schmidt decomposition,
$$
\ket{\Psi}  = N_{\Psi}\! \sum_{i=1}^{\chi} \lambda_{i}
\ket{i_{A}}\ket{i_{B}},~~ \lambda_{i}\geq \lambda_{i+\!1} \geq 0,
~\sum_{i=1}^{\chi}\lambda_{i}^2 = 1,
$$
where $N_{\Psi} \equiv \sqrt{\braket{\Psi}{\Psi}}$, and let
$\ket{\Phi}$ be a bipartite vector with norm $N_{\Phi}$ and
Schmidt rank $\chi'$, where $\chi' \leq \chi$. Then,
\begin{equation}
\max_{\ket{\Phi}} |\braket{\Psi}{\Phi}| =
N_{\Psi}N_{\Phi}\sqrt{\sum_{i=1}^{\chi'} \lambda_i^{2}}\;.
\end{equation}
\end{lemma}

{\it Proof: } Let $\mu_i$ denote the Schmidt coefficients of
$\ket{\Phi}$. It follows from Lemma 1 in \cite{vjn00} that $
 \max_{\ket{\Phi}}|\braket{\Psi}{\Phi}| =
 N_{\Psi}N_{\Phi}\sum_{i=1}^{\chi'}\lambda_i\mu_i,
 $
and the maximization over $\mu_i$ is done next. A straightforward
application of the method of Lagrange multipliers provides us with
$\mu_i = c\lambda_i,~i=1,2,\dots,\chi'$ for some constant $c$.
Since $\sum_{i=1}^{\chi'}\mu_i^2=1=c^2\sum_{i=1}^{\chi'}
\lambda_i^2$, $c=1/\sqrt{\sum_{i=1}^{\chi'} \lambda_i^2}.$ Thus,
$$
 \max_{\ket{\Phi}}|\braket{\Psi}{\Phi}| =
c N_{\Psi}N_{\Phi}\sum_{i=1}^{\chi'}\lambda_i^2,
 $$
 and the result follows.
\qed

     We will also use two basic results related to majorization theory.  Recall that, by
definition, a decreasingly ordered probability distribution
$\vec{p} = (p_1,p_2,\dots,p_d)$, where $p_{\alpha} \geq
p_{\alpha+1}\geq 0$, $\sum_{\alpha} p_{\alpha}=1$, is {\em
majorized} by another such probability distribution $\vec{q}$,
denoted $\vec{p} \prec \vec{q}$, if $\vec{q}$ is more ordered or
concentrated than $\vec{p}$ (equivalently, $\vec{p}$ is flatter or
more mixed than $\vec{q}$) in the sense that the following
inequalities are fulfilled:
\begin{equation}
    \sum_{\alpha=1}^k p_{\alpha} \leq \sum_{\alpha=1}^k q_{\alpha}~~~\forall~k=1,\dots,d
\end{equation}
with equality for $k=d$.
The following result can be found in Exercise II.1.15 of
\cite{bhatia}:

\begin{lemma}
    Let $\rho_{\vec{x}}$ and $\rho_{\vec{y}}$ be density matrices with
eigenvalues given by probability distributions $\vec{x}$ and
$\vec{y}$. Let $\sigma(M)$ denote the decreasingly ordered
eigenvalues of hermitian operator $M$. Then
$$
    \sigma(\rho_{\vec{x}} + \rho_{\vec{y}}) \prec \vec{x} + \vec{y}.
$$
\end{lemma}
The next result follows by direct inspection.
\begin{lemma}
Let coefficients $\delta_i$, $1\leq i \leq d$, be such that
$-\delta \leq \delta_i \leq \delta$ for some positive $\delta \leq
1$ and $\sum_i \delta_i = 1$, and consider the probability
distribution $\vec{p}(\{\delta_i\})$,
 $$
    \vec{p}(\{\delta_i\}) \equiv \left(\frac{1}{2} + \frac{1+\delta_1}{2d}, \frac{1+\delta_2}{2d},\cdots, \frac{1+\delta_d}{2d}\right).
 $$
Then
$$
    \vec{p}(\{\delta_i\}) \prec \vec{p}(\{\delta_i^*\}),
$$ where
$$
    \delta_i^* \equiv \left\{
    \begin{array}{c}
    ~~\delta~~~ i \leq d/2 \\
    -\delta ~~~ i > d/2,
    \end{array} \right.
$$ and we assume $d$ to be even.
\end{lemma}

Finally, we need a result from \cite{hlw06}:
\begin{lemma}
With probability very close to 1,
\begin{eqnarray}
    \label{hayden}
&&\mathrm{Pr}\Big[(1-\delta)\frac{\Upsilon}{d_A} \leq
Q \leq (1+\delta)\frac{\Upsilon}{d_A} \Big] \nonumber\\
&& \geq1-\left(\frac{10~
d_A}{\delta}\right)^{2d_A}2^{(-d_B\;\delta^2/14 \ln2)} \nonumber\\
&&\geq 1- O\left(\frac{1}{\exp(\delta^2\exp(n))}\right),
\end{eqnarray}
where $d_A = 2^{n_A} = 2^{n_0}$ and $d_B = 2^{n_B} = 2^{n-n_0
+1}$, and the operator $Q$ defined in Eq. (\ref{eq:Q}) is within a
ball of radius $\delta$ of a (unnormalized) projector
$\Upsilon/d_A$ of rank $d_A$ [provided $d_B$ is a large multiple
of $d_A\log d_A/\delta^2$ \cite{hlw06}, which is satisfied for
large $n$, given that $n/5 \leq n_0 \leq 2n/5$].
\end{lemma}

Our second theorem uses the fact that the Schmidt decomposition of
$\rho$ does not only have exponentially many coefficients, but
that these are roughly of the same size.


\begin{theorem}
Let $\rho$, $U_n$, and $A\!:\!B$ be defined as in Theorem
\ref{t1}. If $F(\rho,\tilde{\rho}) \geq 1-\epsilon$, then with
probability $p(\delta,n) = 1 - O(\exp(-\delta^2\exp(n)))$, the
Schmidt rank for $\tilde{\rho}$ according to bipartition $A\!:\!B$
satisfies $\tilde{\chi}^{\sharp} \geq (1-4\epsilon-\delta)
2^{n/5}$.
\end{theorem}

\textit{Proof:}  For any product vector of Eq. (\ref{eq:prod}) we
have
    \ben
    \label{eq:hello}
    |\bra{tij}\rho\tilde{\rho}\ket{tij}|
    &\leq& ~N_{\vec{\alpha}}~\tilde{N}_{\vec{\alpha}}~\sqrt{\sum_{k=1}^{\tilde{\chi}^{\sharp}}(\lambda_k^{ij})^2}
\\\nonumber
    &\leq& N_{\vec{\alpha}}~\tilde{N}_{\vec{\alpha}}\;g(\tilde{\chi}^{\sharp}/d_A),
    \een
where
\begin{equation}
g(x)\equiv \sqrt{\frac{1+(1+\delta)x}{2}},
\end{equation}
and $N_{\vec{\alpha}} \equiv \sqrt{\bra{tij}\rho^2\ket{tij}}$,
$\tilde{N}_{\vec{\alpha}} \equiv
\sqrt{\bra{tij}\tilde{\rho}^2\ket{tij}}$. The first inequality in
(\ref{eq:hello}) follows from Lemma 1, whereas the second one
follows from the fact that the spectrum $\vec{p}$ of
 $$
\rho_B \equiv (N_{\vec{\alpha}})^{-2}\tr_A[\rho\proj{tij}\rho] =
\frac{1}{2}(\proj{j} + Q ),
 $$
where $Q$ has all its $d_A$ non-zero eigenvalues $q_{i}$ in the
interval $2^{-n_0}(1-\delta) \leq q_i \leq 2^{-n_0}(1+\delta)$, is
majorized by $\vec{p}(\{\delta_i^*\})$, as follows from Lemmas 2
and 3. Then,
\begin{eqnarray}
&&1-\epsilon \leq \frac{\tr\rho\tilde{\rho}}{\sqrt{\tr\rho^2}\sqrt{\tr \tilde{\rho}^2}} \label{eq:th1} \nonumber\\
&=& \frac{\sum_{\vec{\alpha}} \bra{\vec{\alpha}}\rho\tilde{\rho}
\ket{\vec{\alpha}}}{ \sqrt{\sum_{\vec{\alpha}'}
\bra{\vec{\alpha}'}\rho^2 \ket{\vec{\alpha}'}
\sum_{\vec{\alpha}''} \bra{\vec{\alpha}''}
\tilde{\rho}^2\ket{\vec{\alpha}''}}
}\label{eq:th2}\nonumber\\
&\leq& g(\tilde{\chi}^{\sharp}/d_A) \frac{\sum_{\vec{\alpha}}
N_{\vec{\alpha}}\tilde{N}_{\vec{\alpha}}}{\sqrt{\sum_{\vec{\alpha}'}
(N_{\vec{\alpha}'})^{2}
 \sum_{\vec{\alpha}''} (\tilde{N}_{\vec{\alpha}''})^{2}} }\nonumber\\
&\leq& g(\tilde{\chi}^{\sharp}/d_A)\nonumber,
\end{eqnarray}
where in the last step we have used the Cauchy-Schwarz inequality,
$|\braket{x}{y}| \leq \sqrt{\braket{x}{x}}\sqrt{\braket{y}{y}}$.
The result of the theorem follows from
$g(\tilde{\chi}^{\sharp}/2^{n_0}) \geq 1-\epsilon $. \qed

\section{Conclusions}
\label{disc}

The results in this chapter show that the algorithms of
\cite{jozsa03a,vidal03a,sdv06,vandernest06} are unable to
efficiently simulate a DQC1 circuit. The efficiency of a quantum
simulation using these algorithms relies on the possibility of
efficiently decomposing the state $\rho$ of the quantum computer
using a TTN. We have seen that for the final state of the DQC1
circuit no efficient TTN exists.

        It is also interesting to note that the numerics and Theorems 1
and 2 in this chapter can be generalized for any fixed
polarization $\tau$ ($0<\tau\leq 1$) of the initial state
$\tau\proj{0} + (1-\tau)\mathbb{I}/2$ of the top qubit of the
circuit in Eq (\ref{E:circuit}), implying that the algorithms of
\cite{jozsa03a,vidal03a,sdv06,vandernest06}
 are also unable to efficiently simulate the
power of even the \emph{tiniest} fraction of a qubit.

\chapter{Quantum discord in the DQC1 model}\label{chap:discord}

\hfill \textsl{We are coming now rather into the region of
guesswork! \\
Say, rather, into the region where we balance probabilities and
choose the most likely. It is the scientific use of imagination,
but we always have some material basis on which to start our
speculation.}

\hfill  -Arthur Conan Doyle in \textsl{The Hound of the
Baskervilles}

\vskip1.0cm

        We have seen in the last two chapters (Chapters
\ref{chap:dqc1}, \ref{chap:mps}) that the DQC1 model presents to
us a very intriguing challenge. It is a mixed-state quantum
computation scheme \cite{kl98} that has very little entanglement
\cite{datta05a}, yet it cannot be simulated by matrix product
state techniques \cite{datta07a}. This suggests that entanglement
cannot be the sole resource that drives mixed-state quantum
computation. Although in pure state quantum computation
entanglement can be shown to be an essential resource
\cite{jozsa03a,vidal03a}, mixed-state quantum computation is a
different story. It is evident that there is a gap in the resource
based accounting for quantum computational speedups. It is this
gap that we hope to fill with the quantum discord
\cite{ollivier01a}.

    Quantum discord, introduced by Ollivier and
Zurek~\cite{ollivier01a}, captures the nonclassical correlations,
including but not limited to entanglement, that can exist between
parts of a quantum system. Some details about quantum discord are
presented in Sec~\ref{introdiscord}. One way of studying the
quantum nature of a computational process is to investigate the
nonclassical correlations in the quantum state at various stages
during the computation. We investigate the effectiveness of
discord in characterizing the performance of the model of quantum
information processing introduced by Knill and Laflamme
in~\cite{kl98}, which is often referred to as the {\em power of
one qubit}, or DQC1.  In this model, information processing is
performed with a collection of qubits in the completely mixed
state coupled to a single control qubit that has some nonzero
purity. Such a device can perform efficiently certain
computational tasks for which there is no known efficient method
using classical information processors.

In this thesis, thus far we have seen how discord can be used to
characterize the nonclassical nature of the correlations in
quantum states and studied its role and place in quantum
information theory. We now apply these ideas to the DQC1 or {\em
power-of-one-qubit\/} model~\cite{kl98} of mixed-state quantum
computation, which accomplishes the task of evaluating the
normalized trace of a unitary matrix efficiently. The quantum
circuit corresponding to this model has a collection of $n$ qubits
in the completely mixed state, $I_n/2^n$, coupled to a single pure
control qubit.  A generalized version of this quantum circuit,
with the control qubit having sub-unity polarization is shown
below in
Fig~\ref{E:circuit}. 
This circuit evaluates the normalized trace of $U_n$,
$\tau=\tr(U_n)/2^n$, with a polynomial overhead going as
$1/\alpha^2$.

The problem of evaluating $\tau$ is believed to be hard
classically. Quantum mechanically, the circuit provides an
estimate of $\tau$ up to a constant accuracy in a number of trials
that does not scale exponentially with $n$.  It does so by making
$X$ and $Y$ measurements on the top qubit.  The averages of the
obtained binary values provide estimates for $\tau_R \equiv
\mathrm{Re}(\tau)$ and $\tau_I \equiv \mathrm{Im}(\tau)$.  The top
qubit is completely separable from the bottom mixed qubits at all
times.  The final state has vanishingly small entanglement, as
measured by the negativity~\cite{datta05a} across any split that
groups the top qubit with some of the mixed qubits.  Nonetheless,
there is evidence that the quantum computation performed by this
model cannot be simulated efficiently using classical
computation~\cite{datta07a}.

The DQC1 circuit transforms the highly-mixed initial state $\rho_0
\equiv \proj{0}\otimes I_n/2^n$ into the final state $\rho_{n+1}$
given by Eq~\ref{E:rhoout}.
Within this model the only place to look for nonclassical
correlations is in this state.

Everything about the DQC1 setup, including the measurements on the
control qubit, suggests a bipartite split between the control
qubit $M$ and the mixed qubits $S$.  Relative to this split, we
turn to computing the quantum discord for the state
$\rho_{SM}=\rho_{n+1}$. The joint state $\rho_{n+1}$ has
eigenvalue spectrum
\begin{equation*}
\label{fullspectrum} {\bm
\lambda}(\rho_{n+1})=\frac{1}{2^{n+1}}(\underbrace{1-\alpha,\cdots,1-\alpha
} _ { 2^n
\mathrm{times}},\underbrace{1+\alpha,\cdots,1+\alpha}_{2^n
\mathrm{times}}),
\end{equation*}
which gives a joint entropy $H(S,M)=n+H_2[(1-\alpha)/2]$, where
$H_2[\cdot]$ is the binary Shannon entropy. The marginal density
matrix for the control qubit at the end of the computation is
 \begin{equation}
 \rho_M = \frac{1}{2}\left(%
\begin{array}{cc}
  1 & \alpha\;\tau^* \\
  \alpha \;\tau & 1 \\
\end{array}%
\right), \label{eq:rhoM}
 \end{equation}
which has eigenvalues $(1\pm \alpha|\tau|)/2$ and entropy $H(M) =
H_2[(1-\alpha|\tau|)/2 ]$.

The evaluation of the quantum conditional entropy involves a
minimization over all possible one-qubit projective measurements.
The projectors are given by $ \Pi_\pm =
\frac{1}{2}(I_1\pm\bm{a\cdot\sigma})$, with $\bm{a\cdot
a}=a_1^2+a_2^2 + a_3^2 =1.$  The post-measurement states are
 \begin{equation}
 \rho_{S|\pm}=\frac{1}{p_\pm 2^{n+1}}\bigg(
I_n \pm \alpha\frac{a_1-i a_2}{2} U_n \pm \alpha\frac{a_1 + i
a_2}{2} U_n^\dg \bigg) ,
 \end{equation}
occurring with outcome probabilities $p_\pm = [1\pm\alpha(a_1
\tau_R + a_2 \tau_I)]/2$. The post-measurement states are
independent of $a_3$, so without loss of generality, let $a_3=0$,
$a_1= \cos\phi$, and $a_2=\sin\phi$.  The corresponding
post-measurement states are
\begin{equation}
\rho_{S|\pm} = \frac{1}{p_\pm 2^{n+1}} \bigg( I_n \pm
\alpha\frac{e^{-i\phi} U_n + e^{i \phi} U_n^{\dagger}}{2} \bigg).
\end{equation}

To find the discord of the state at the end of the computation, we
need the spectrum of $\rho_{S|\pm}$ so that we can compute
$H(\rho_{S|\pm})$.  The eigenvalues of any unitary operator $U_n$
are phases of the form $e^{i\theta_k}$, so we have
\begin{equation}
\lambda_k \bigg( \frac{e^{-i\phi} U_n + e^{i\phi} U_n^{\dg}}{2}
\bigg) =\cos(\theta_k-\phi), \quad k=1,\cdots, 2^n,
\end{equation}
and
\begin{equation}
\label{postspectrum} \lambda_k(\rho_{S|\pm})= \frac{1}{2^n}
\frac{1\pm\alpha\cos(\theta_k-\phi)}{1\pm\alpha(\tau_R\cos\phi+\tau_I\sin\phi)}
\equiv q_{k\pm}.
\end{equation}
We also have $\tau_R = 2^{-n} \sum_k \cos \theta_k$ and $\tau_I =
2^{-n} \sum_k \sin\theta_k$. All this gives
$H(\rho_{S|\pm})=H({\bm q}_\pm)$ and thus
\begin{eqnarray}
\label{condent}
\tilde H_{\Pi_\pm}& =& p_+ H(\rho_{S|+})+p_- H(\rho_{S|-})\nonumber \\
&=&\frac{1}{2}[H({\bm q}_+)+H({\bm q}_-)] + \frac{\alpha}{2}
(\tau_R\cos\phi + \tau_I\sin\phi)[H({\bm q}_+)-H({\bm q}_-)].
\end{eqnarray}

We now use the fact that we are interested in the behavior of the
quantum discord of the DQC1 state for a typical unitary. By
typical, we mean a unitary chosen randomly according to the (left
and right invariant) Haar measure on $\mathbb{U}(2^n)$.  For such
a unitary, it is known that the phases $\theta_k$ are almost
uniformly distributed on the unit circle with large
probability~\cite{diaconis03a}.  Thus for typical unitaries
$\sum_k e^{i\theta_k}$ is close to zero.  Hence both $\tau_R$ and
$\tau_I$ are small, and we can ignore the second term on the
right-hand side in Eq.~(\ref{condent}).  In addition, the phases
$\theta_k$ can be taken to be placed at (with large probability)
the $2^n$th roots of unity, i.e., $\theta_k = 2\pi k/2^n$.  It
follows that the spectra $\lambda_k(\rho_{S|\pm})$ are independent
of $\phi$.  Hence the entropies we are interested in computing are
also independent of $\phi$, and we can set $\phi$ to zero without
loss of generality.  This choice for $\phi$ corresponds to
measuring the pure qubit $M$ along $X$. The $X$ measurement gives
the real part of the normalized trace of $U_n$, and it is one of
the two measurements discussed in the original proposal by Knill
and Laflamme. Setting $\phi=\pi/2$ yields the other measurement,
along $Y$, which gives the imaginary part of the normalized trace
of $U_n$.

In the limit of large $n$, we can simplify Eq.~(\ref{condent}) as
follows:
\begin{eqnarray}
\label{eq:sum}
\tilde H&=&  \frac{1}{2}[H({\bm q}_+)+H({\bm q}_-)] \nonumber \\
&=& -\frac{1}{2^{n+1}}\sum_{k=1}^{2^n}\bigg[(1+\alpha
 \cos\theta_k) \log\bigg(\frac{1+\alpha \cos\theta_k}{2^n}\bigg)\nonumber \\
&& + \;\;\;(1-\alpha \cos\theta_k)\log \bigg( \frac{1-\alpha
\cos \theta_k}{2^n} \bigg) \bigg]\nonumber \\
&=& n - \frac{1}{2^{n+1}} \sum_{k=1}^{2^n} \bigg[ \log
\big(1-\alpha^2 \cos^2\theta_k \big) + \alpha \cos\theta_k\log
\bigg( \frac{1+\alpha \cos\theta_k}{1-\alpha \cos\theta_k}\bigg)
\bigg].
\end{eqnarray}
Furthermore, when $n$ is large, we can replace the sum in the
above equation with an integral to obtain
\begin{eqnarray}
\tilde H&=&n-\frac{1}{4\pi}\bigg[\int_0^{2\pi}\log(1-\alpha^2
\cos^2x) \mathrm{d}x +  \alpha \int_0^{2\pi} \cos x\log \bigg(
\frac{1+\alpha \cos
x}{1-\alpha \cos x} \bigg) \mathrm{d}x \bigg] \nonumber \\
&=& n +1 -
\log\Big(1+\sqrt{1-\alpha^2}\Big)-\Big(1-\sqrt{1-\alpha^2}\Big)\log
e.
\end{eqnarray}
Note that when the sums are replaced by integrals, $H({\bm
q}_+)-H({\bm q}_-)=0$, providing further justification for
ignoring the second term in Eq.~(\ref{condent}).

\begin{figure}[!ht]
\begin{center}
\includegraphics[scale=0.75]{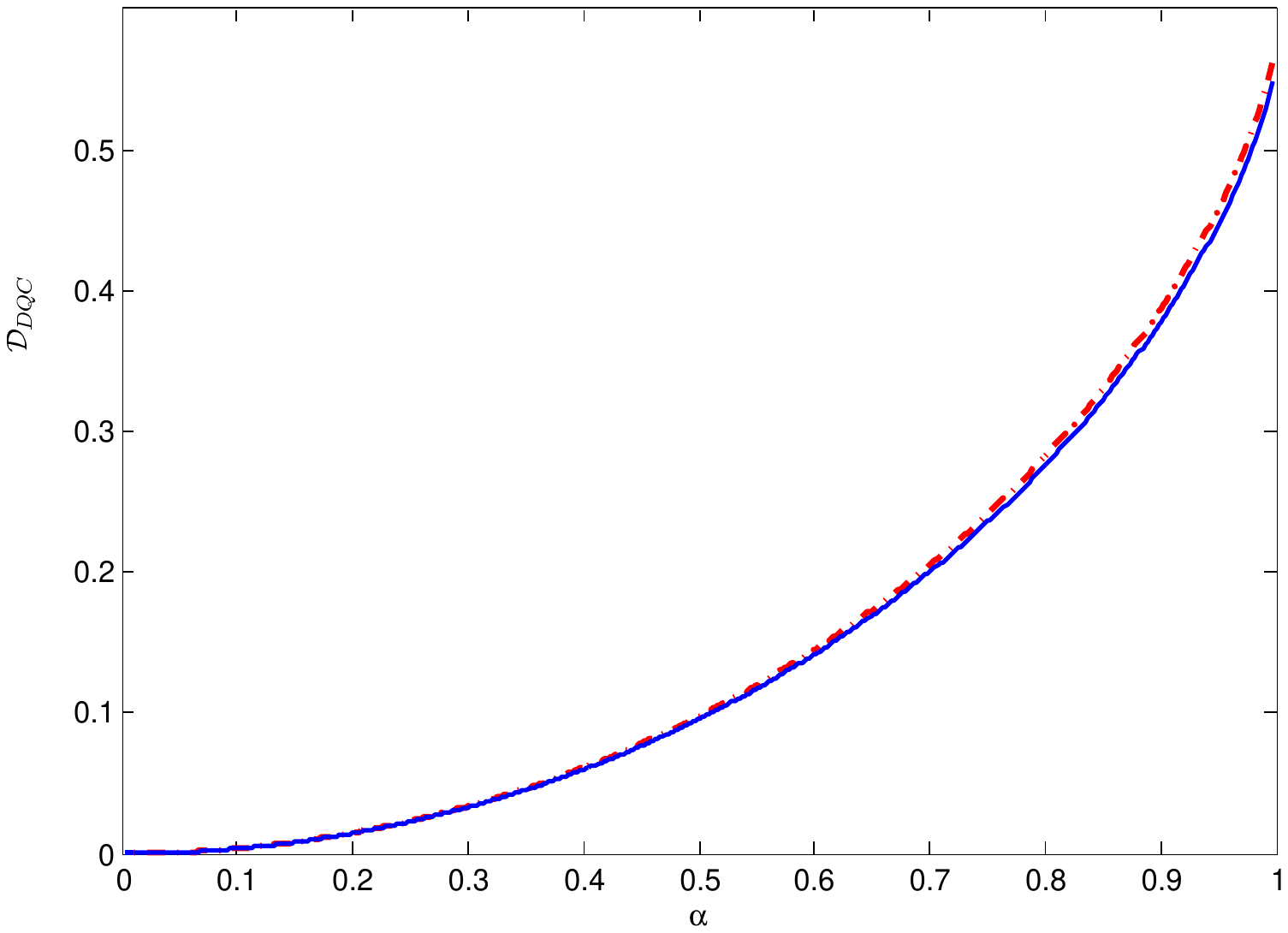}
\caption[Discord in the DQC1 circuit]{The dashed (red) line shows
the average discord in a DQC1 circuit with five qubits in the
mixed state ($n=5$) coupled to a qubit with purity $\alpha$.  The
average is taken over 500 instances of pseudo-random unitary
matrices.  The discord is shown as a function of the purity of the
control qubit. The solid (green) line shows the analytical
expression in Eq.~(\ref{randomdiscord}), which grows monotonically
from 0 at $\alpha=0$ (completely mixed control qubit) to $2-\log
e=0.5573$ at $\alpha=1$ (pure control qubit). These values of
discord should be compared with a maximum possible discord of 1
when $M$ is a single qubit.} \label{discorddqc}
\end{center}
\end{figure}

When $|\tau|$ is small, $H(M)\simeq 1$, and the quantum discord
for the DQC1 state is then given by the simple expression
\be
\label{randomdiscord}
 \mathcal{D}_{\rm DQC1} = 2 - H_2
\Big(\frac{1-\alpha}{2} \Big) - \log\Big(1+\sqrt{
1-\alpha^2}\Big)-\Big(1-\sqrt{1-\alpha^2}\Big)\log e.
 \ee
Note that the above expression for the discord, valid for large
$n$. is independent of $n$. Figure~\ref{discorddqc} compares the
discord from Eq.~(\ref{randomdiscord}) with the average discord in
a DQC1 circuit having five qubits in the mixed state ($n=5)$
coupled to a control qubit with purity $\alpha$.  The average is
taken over 500 instances of pseudo-random unitary matrices,
generated using the efficient algorithm presented
in~\cite{emerson03a}. The convergence of this ensemble to the Haar
measure on the unitary group is shown in~\cite{emerson05a}. We see
that in spite of the approximations made in obtaining
Eq.~(\ref{randomdiscord}), the analytic expression provides a very
good estimate of the discord even when $n$ is as low as five.

There is no entanglement between the control qubit and the mixed
qubits in the DQC1 circuit at any point in the computation, yet
there are nonclassical correlations, as measured by the discord,
between the two parts at the end of the computation for any
$\alpha > 0$. Other bipartite splittings of $\rho_{n+1}$ can
exhibit entanglement, but it was shown in~\cite{datta05a} that the
partial transpose criterion failed to detect entanglement in
$\rho_{n+1}$ for $\alpha \leq 1/2$. In this domain, several other
tests for entanglement, including the first level of the scheme of
Doherty~\textit{et~al.}~\cite{doherty04a}, which is based on
semi-definite programming, also failed to detect entanglement. The
above expression is thus the first signature of nonclassical
correlations in the DQC1 circuit for $\alpha \leq 1/2$.

In conclusion, we calculated the discord in the DQC1 circuit and
showed that nonclassical correlations are present in the state at
the end of the computation even if there is no detectable
entanglement. Thus for some purposes, quantum discord might be a
better figure of merit for characterizing the quantum resources
available to a quantum information processor.  We present evidence
of the presence of nonclassical correlations in the DQC1 circuit
when $\alpha \leq 1/2$. For qubits quantum discord is known to be
a true measure of nonclassical correlations~\cite{hamieh04a}.
This suggests that nonclassical correlations other than
entanglement, as quantified by the discord, might explain the
(sometimes exponential) speed-up in the DQC1 circuit and perhaps
the speedup in other quantum computational circuits. For pure
states, discord becomes a measure of entanglement.  Therefore,
using discord to connect quantum resources to the advantages
offered by quantum information processors has the additional
advantage that it works well for both pure- and mixed-state
quantum computation.

\chapter{Conclusion\label{sec:conclusion}}

\hfill \textsl{If you want a happy ending, that depends, of
course, on where you stop your story.}

\hfill  -Orson Welles

\vskip1.0cm

        The intent of this chapter is eponymous. The attempt will
be to present the results of this thesis in a well-rounded manner,
put them all in the perspective of a `big picture', and end with a
few suggestions for open problems.

        A one line conclusion of this thesis would be the following:

\emph{There is more to quantum information science than just
entanglement}.

\noindent However, this statement needs qualifications. I
presented evidence that entanglement fails to explain exponential
speedups in mixed-state quantum computation, at least in the DQC1
model of Knill and Laflamme, the reason being that the amount of
entanglement in the system is minimal, not scaling with the system
size. This was the content of Chapter~\ref{chap:dqc1}. Then I
showed that this lack of entanglement does not mean that the
system is classically simulatable. This is important as the
simulation techniques that have their roots in renormalization
group theory are very powerful and among the most versatile in
many-body physics. Thus Chapter~\ref{chap:mps} leaves us with a
deep foreboding about the role of entanglement in mixed-state
quantum computation and the criterion for classical simulatability
of quantum systems that involve entanglement. In
Chapter~\ref{chap:discord}, I was able to show that there is a
non-trivial amount of quantum discord in the DQC1 circuit. It is
thus possible that quantum discord is the resource that drives
mixed-state quantum computation.

        Quantum discord can be thought of as a generalization of
entanglement. For pure states, they are identical. For mixed
quantum states, discord captures nonclassical correlations beyond
entanglement. As I have presented in Section~\ref{S:cirac}, the
scope of quantum discord is wider than just mixed-state quantum
computation. I discussed the role of discord in distribution of
entanglement without investing any entanglement.

        Every PhD thesis is a compilation of results and solutions
to problems that often seem disparate. Yet, it is rarely the case
that these problems are not steps towards comprehending a
fundamental problem in the field of study. In my case, the results
dealt with the DQC1 model, its entanglement content, its classical
simulatablity or resource behind its exponentially enhanced
operation. I also proposed quantum discord as a substitute for
entanglement for the role of the resource in mixed-state quantum
computation. I also studied how discord seems to be relevant in
other information processing tasks and the major part that the
DQC1 model can play in bringing together different areas of
physics, mathematics and computer science. The fundamental problem
I have attempted to attack is the following:

\emph{Is entanglement a key resource for computational power? }

\noindent This is what I would call the `big picture.' Jozsa and
Linden have spoken most lucidly on this. The answer to this
proceeds as follows: The significance of entanglement for
pure-state computations is derived from the fact that unentangled
pure states of $n$ qubits have a description involving only
$poly(n)$ parameters (in contrast to $O(2^n)$ parameters for a
general pure state). But this special property of unentangled
states (of having a `small' descriptions) is contingent on a
particular mathematical description, as amplitudes in the
computational basis. If we were to adopt some other choice of
mathematical description for quantum states (and their evolution),
then, although it will be mathematically equivalent to the
amplitude description, there will be a different class of states
which will now have a polynomially sized description; i.e. two
formulations of a theory which are mathematically equivalent (and
hence equally logically valid) need not have their corresponding
mathematical descriptions of elements of the theory being
interconvertible by a polynomially bounded computation. With this
in mind we see that the significance of entanglement as a resource
for quantum computation is not an intrinsic property of quantum
physics itself, but is tied to a particular additional (arbitrary)
choice of mathematical formalism for the theory.

        Thus, suppose that instead of the amplitude description we choose some other
mathematical description $\mathcal{D}$ of quantum states (and
gates). Indeed, there is a rich variety of possible alternative
descriptions. Then there will be an associated property of states,
$prop(\mathcal{D})$, which guarantees that the
$\mathcal{D}$-description of the quantum computational process
grows only polynomially with the number of qubits (e.g. if
$\mathcal{D}$ is the amplitude description, then
$prop(\mathcal{D})$ is just the notion of entanglement). Thus,
just as for entanglement, we can equally well claim that
$prop(\mathcal{D})$ for any $\mathcal{D}$ is an essential resource
for quantum-computational speed-up! Entanglement itself appears to
have no special status here.

        Explicit examples of $prop(\mathcal{D})$ would include the
dimension of the Hilbert space in which the state resides, the
purity of a quantum state or the stabilizer formalism and even
quantum discord. Thus, in a fundamental sense, the power of
quantum computation over classical computation ought to be derived
simultaneously from all possible classical mathematical formalisms
for representing quantum theory, not any single such formalism and
associated quality (such as entanglement), i.e. we have arrived at
the enigmatic prospect of needing a representation of quantum
physics that does not single out any particular choice of
mathematical formalism.

            We now seem to have realized the fundamental problem
we ought to be tackling is perhaps too hard to solve. Yet, this
should not be a detriment, but rather a challenge to be at least
working on the right problem, hard or not . In this thesis, my
attempt has been to start chipping at a small corner of this
massive monolith. The sculpture is far from being done, but the
work has begun.

        As is with any work of scientific research, there always
remain fairly immediate questions whose answers hold the
possibility of further enlightening the topic at hand. This thesis
is no exception. There remain open questions that could have been
addressed in the course of my research. It is my belief that given
enough time and effort, these could be resolved. Some of these
concern the DQC1 model, others the nature and properties of
quantum discord. As the last segment of this thesis, I sketch some
of these prospective problems:

    \begin{enumerate}

 \item \textbf{DQC1 and knot theory}: Evaluation of the Jones
polynomial of the trace closure of a braid is DQC1 complete. Knot
invariants are related to partition functions of the Ising model,
and this can be used to find a physical interpretation of the
complexity class DQC1. As a continuation of the discussion in
Section~\ref{introdiscord}, the actual task would be to cast the
complexity of approximating partition functions with different
boundary conditions as graph theoretic problems whose complexity
is well studied.

    \item \textbf{Classical simulation of quantum systems}:
In Chapter~\ref{chap:mps}, I found that the DQC1 system is not
efficiently simulatable classically using matrix product states
(MPS) which had successfully simulated other quantum systems with
little entanglement~\cite{vidal03b,zv04}. It is known that pair
entangled projected states (PEPS) can simulate any quantum
computation efficiently \cite{swvc07}. My next aim in this regard
is to study the simulatability of the DQC1 model using some
intermediate simulation scheme, lying between MPS and PEPS, and if
possible, find a measurement-based quantum computation system that
emulates DQC1. This will tell us more about the limits of quantum
computation and the boundaries of classical computation.

     \item \textbf{DQC1 and discrete Wigner functions}: Quantum
computational speedup has been related to the positivity of
discrete Wigner functions \cite{galvao05}. Such Wigner functions
$W$ can be defined so that the only pure states having
non-negative $W$ for all such functions are stabilizer states.
Also, the unitaries preserving non-negativity of $W$ for all
definitions of $W$ in the class form a subgroup of the Clifford
group. This means pure states with non-negative $W$ and their
associated unitary dynamics are classical in the sense of
admitting an efficient classical simulation scheme using the
stabilizer formalism \cite{cggpp06}. It is not hard to calculate
discrete Wigner functions for mixed states. The hindrance in
drawing conclusions similar to the pure state case arises from the
non-uniqueness of discrete Wigner functions as they now are
defined~\cite{gfw04}. One must minimize over all such definitions
to draw conclusions about the non-negativity of Wigner functions
as is done in \cite{cggpp06}. Further thought in this direction
might be worthwhile and provide new insights into the power of the
DQC1 model.

    \item \textbf{Computation of quantum discord}: The minimization involved
in the computation of quantum discord is non-trivial. It is a
minimization of an entropic quantity over the set of POVMS. This
quantity can be recast as the Holevo quantity in quantum
information theory. We know that the computation of Holevo
capacity in general is NP-Hard \cite{bs07a}, but the quantum
conditional entropy is a very special case. An interesting
question would therefore be the complexity of calculating the
quantum conditional entropy. More practically, attempts to reduce
the minimization to a convex optimization technique, like
semi-definite or semi-infinite programming, would be helpful and
worthwhile.

 \item \textbf{Quantum discord in quantum communication}: Beyond
the results of Section~\ref{S:cirac}, the most important
contribution would be to prove the role of discord in the general
protocol of entanglement distribution. The protocol is given in
\cite{cvdc03}, but the challenge is that the general protocol
involves a mediating particle which is higher than two
dimensional, making the explicit evaluation of discord
challenging. In this context, the previous task is related to this
one.

    \end{enumerate}

            The above list is by no means exhaustive. The endeavor
has been to present a few problems that I myself have been
interested in. These might lead us to understanding certain
aspects of the DQC1 model and quantum discord. It is likely that
their study would lead to more open problems. All these and many
more challenging and profound questions can arise in the study of
mixed-state quantum computation and quantum discord.

\chapter*{Appendices}

\appendix
\chapter[DQC1 and Knot Theory]{DQC1 and Knot Theory}{\label{A:knot}}

\section{Quadratically Signed Weight Enumerators}
\label{S:qswe}

A general quadratically signed weight enumerator is of the form
 $$
S(A,B,x,y)= \sum_{b: Ab=0}(-1)^{b^T Bb}x^{|b|}y^{n-|b|},
  $$
where $A$ and $B$ are be matrices over $\mathbb{Z}_2$ with $B$ of
dimension $n$ by $n$ and $A$ of dimension $m$ by $n$. The variable
$b$ in the summand ranges over column vectors in $\mathbb{Z}_2$ of
dimension $n$, $b^T$ denotes the transpose of $b$, $|b|$ is the
weight of $b$ (the number of ones in the vector $b$), and all
calculations involving $A$, $B$ and $b$ are modulo 2. The absolute
value of $S(A,B, x, y)$ is bounded by $(|x| + |y|)^n$. In general,
one can consider the computational problem of evaluating these
sums. In particular, it is known that for integers $k,l$,
evaluating $S(A,B,k,l)$ exactly is $\# \mathbf{P}$ complete. Now,
let $A$ be square, of size $n$ by $n$, and let $\overleftarrow{A}$
denote the lower triangular part of $A$, which is the matrix
obtained from $A$ by setting to zero all the entries on or above
the diagonal. Let $diag(A)$ denote the diagonal matrix whose
diagonal is the same as that of $A$ and $\mathbb{I}$ denote the
identity matrix. For matrices $C$ and $D$ with the same number of
columns, let $[C;D]$ denote the matrix obtained by placing $C$
above $D$. Then, the following two theorems hold:
 \begin{theorem}
 Given $diag(A)=\mathbb{I}$, $k,l$ positive integers, and the
promise that \\$|S(A,\overleftarrow{A},k,l)|\geq
(k^2+l^2)^{n/2}/2$, determining the sign of
$S(A,\overleftarrow{A},k,l)$ is BQP-complete.
 \end{theorem}

 \begin{theorem}
Given $diag(A)=\mathbb{I}$, $k,l$ positive integers, and the
promise that \\$|S([A;A^T],\overleftarrow{A},k,l)|\geq
(k^2+l^2)^{n/2}/2$, the sign of $S([A;A^T],\overleftarrow{A},k,l)$
can be efficiently determined by the DQC1 model.
 \end{theorem}
In these two problems, the integers $k$ and $l$ can be restricted
to 4 and 3, respectively, without affecting their hardness with
respect to polynomial reductions (using classical deterministic
algorithms). Note that in the second one, the question of DQC1
completeness is still open.

        Lidar has made the first connection between QSWEs and partition
functions of spin models on certain classes of graphs. In that
case, $A$ denotes the adjacency matrix of the relevant graph.
Though these classes are highly restricted at the present,
attempts at broadening their scope are ongoing. Through the
connection to QSWEs, DQC1 can be applied to analyze problems in
coding theory, like weight generating functions of binary codes.
Studying the complexity of such problems will certainly give us a
better understanding of the complexity classes P, BQP, and DQC1
and their relative inclusions. QWSEs relate the DQC1 model to
Ising model partition functions in a way entirely independent from
that to be outlined in Section~\ref{S:jones}. On the other hand,
as shown in Fig~\ref{relations}, the DQC1 model can be employed to
tackle problems in graph and knot theory, disciplines that have
provided scores of problems for theoretical computer science in
general and quantum computation in particular. Identifying
problems that are DQC1 complete and studying then in relation to
BQP, P and NP complete graph and knot theoretic problems might
unravel the underlying mathematical structures that separate P,
BQP, DQC1. Such an understanding might illuminate greatly the
mathematical framework behind the power of quantum computation,
just as studying the complexity of partition functions might
unravel the physical foundations for that power. The diagram in
Fig~\ref{relations} outlines prospective avenues for exploring
such problems in the context of mixed-state quantum computation,
and has the potential of opening up whole new paradigms in
mixed-state quantum computation research.

\section{Jones Polynomials}
\label{S:jones}

It has been known for quite some time that evaluating the
partition functions of certain spin models, like the Ising model,
is related to knot invariants, in particular, the Jones polynomial
\cite{welsh93}. This is independent of the connections arrived at
via  Kauffman brackets and QSWEs \cite{lidar04,kl01} as we
discussed in Appendix~\ref{A:knot}. The difference between trace
and plat closure is in the way the ends of the braids are joined
to make a knot. This has a counterpart in the physical world. Plat
closures correspond to open boundary conditions, while trace
closures correspond to periodic boundary conditions~\cite{j89}.
The partition function of a system depends on its boundary
conditions. Therefore, one would expect that its evaluation would
behave analogously. Of course, when we talk of evaluation, we
actually mean approximation to its true value with a reasonable
accuracy. The notion of `reasonable accuracy' is a fairly
technical one. Put simply, we aim for an approximation whose error
scales only polynomially with the problem size. The exact
evaluation of partition function of the Ising model on a general
graph is $\# \mathbf{P}$-hard even when all the interactions are
unity and there are no external magnetic fields \cite{welsh93}. We
believe that the exact evaluation of the normalized trace of a
unitary is $ \# \mathbf{P}$-complete (See
Section~\ref{S:classical}). As the DQC1 model provides a
polynomial algorithm for an approximate solution to this hard
problem, it might be possible to approximate partition functions
through the DQC1 algorithm and uncover physical reasons for the
success of the power of one qubit model.

\chapter[Proof of the Lemma]{Proof of the Lemma}{\label{A:appendix}}

\textbf{Lemma:} $\tr(\breve A \breve B) = \tr(A B) $.\\

\textbf{Proof:} Define two operators $A$ and $B$ by
\begin{equation}
A=\sum_{i,j,k,l} a_{ij,kl} \ket{ij}\!\bra{kl}\;, \quad
B=\sum_{m,n,p,q} b_{mn,pq} \ket{mn}\!\bra{pq} \;.
\end{equation}
Taking the partial transpose with respect to the second subsystem,
we find
\begin{equation}
\breve A=\sum_{i,j,k,l} a_{ij,kl} \ket{il}\!\bra{kj}\;, \quad
\breve B=\sum_{m,n,p,q} b_{mn,pq} \ket{mq}\!\bra{pn} \;.
\end{equation}
Calculating the quantities of interest, we find that they are
indeed equal.
\begin{eqnarray}
    \tr(\breve A \breve B) &=&
        \sum_{\scriptsize\begin{array}{l}i,j,k,l, \\ m,n,p,q \end{array}}
        a_{ij,kl} b_{mn,pq} \braket{pn}{il} \braket{kj}{mq}
    = \sum_{i,j,k,l} a_{ij,kl} b_{kl,ij} \;, \\
    \tr\left(A  B\right) &=&
        \sum_{\scriptsize \begin{array}{l}i,j,k,l, \\ m,n,p,q \end{array}}
        a_{ij,kl} b_{mn,pq} \braket{pq}{ij} \braket{kl}{mn}
    = \sum_{i,j,k,l} a_{ij,kl} b_{kl,ij}\;.
\end{eqnarray}

\chapter[Negativity of a random pure state]{Negativity of a random pure state}{\label{A:randomneg}}

        An eigenvector of a random $n$-qubit unitary can be considered to
be a random pure state in the Hilbert space of $n$-qubits,
$\mathcal{H}$. Any such state can be Schmidt decomposed for a
given bipartition $\mathcal{H}_A\otimes \mathcal{H}_B$ with
dimensions $\mu$ and $\nu$ ($\mu \leq \nu$) and the distribution
of the Schmidt coefficients is given by (for $\mu = \nu$)
    \be
P(\mathbf{p}) \mathrm{d}\mathbf{p} = N \delta(1-\sum_{i=1}^\mu
p_i) \prod_{1\leq i < j \leq \mu} (p_i-p_j)^2 \prod_{k=1}^\mu
\mathrm{d}\mathbf{p}_k.
    \ee
Since the negativity for pure states is
    \be
\mathcal{N} = \left(\sum_{i=1}^\mu\sqrt{p_i}\right)^2 = 1 +
\mathop{\sum_{i,j=1}}_{i\neq j}^{\mu}\sqrt{p_i p_j}
    \ee
    \be
\avg{\mathcal{N}}=1 + \int \mathop{\sum_{i,j=1}}_{i\neq
j}^{\mu}\sqrt{p_i p_j}P(\mathbf{p}) \mathrm{d}\mathbf{p}.
    \ee
At this point, it helps to change variables such that $p_i=r q_i$
which removes the hurdle of integrating over the probability
simplex, whereby
    \be
Q(\mathbf{q})\mathrm{d}\mathbf{q}\equiv \prod_{1\leq i<j\leq
\mu}\left(q_i-q_j\right)^2
\prod_{k=1}^\mu e^{-q_k}\,\mathrm{d}q_k \label{Q}\\
=N\,e^{-r}r^{\mu\nu-1}P(\mathbf{p})\,\mathrm{d}\mathbf{p}\,\mathrm{d}r\;.
    \ee
The new variables $q_i$ take on values independently in the range
$[0,\infty)$.  Integrating over all the values of the new
variables, we find that the normalization constant is given by
$N=\overline Q/\Gamma(\mu\nu)$, where $\overline{Q}\equiv\int
Q(\mathbf{q})d\mathbf{q}$.  Similarly, we find that
\begin{equation}
\int \sqrt{q_i q_j}Q(\mathbf{q})\mathrm{d}\mathbf{q} = \overline
Q\, \frac{\Gamma(\mu^2+1)}{\Gamma(\mu^2)} \int \sqrt{p_i
p_j}P(\mathbf{p})\,\mathrm{d}\mathbf{p}\;. \label{QtoP}
\end{equation}

    Notice that the first product in Eq.~(\ref{Q}) is the square of
the Van der Monde determinant \cite{scott03,s96}
\begin{equation}
\Delta(\mathbf{q}) \,\equiv\, \prod_{1\leq i<j\leq
\mu}\left(q_i-q_j\right) = \left| \begin{array}{ccc}
 1 & \ldots & 1 \\
 q_1 & \ldots & q_\mu \\
 \vdots & \ddots & \vdots \\
 q_1^{\mu-1} & \ldots & q_\mu^{\mu-1}
 \end{array} \right|
 = \left| \begin{array}{ccc}
 r_0^{\alpha}(q_1) & \ldots & r_0^{\alpha}(q_\mu) \\
 r_1^{\alpha}(q_1) & \ldots & r_1^{\alpha}(q_\mu) \\
 \vdots & \ddots & \vdots \\
 r_{\mu-1}^{\alpha}(q_1)  & \ldots & r_{\mu-1}^{\alpha}(q_\mu)
 \end{array} \right|\;.
 \label{Van2}
\end{equation}
The second determinant in Eq.~(\ref{Van2}) follows from the basic
property of invariance after adding a multiple of one row to
another, with $\alpha\equiv\nu-\mu$ and the polynomials
$r^\alpha_k(q)\equiv k!L^\alpha_k(q)$ judiciously chosen to be
rescaled Laguerre polynomials \cite{gradshteyn}, satisfying the
recursion relation
\begin{equation}
r^\alpha_k(q) = r^{\alpha+1}_k(q)-kr^{\alpha+1}_{k-1}(q) =
\sum_{i=0}^j(-1)^i{\,j\,\choose i} k(k-1)\dots(k-i+1)
r^{\alpha+j}_{k-i}(q), \label{L3}
\end{equation}
and having the orthogonality property
\begin{equation}
\int_0^\infty dq\,e^{-q}q^\alpha r^\alpha_k(q)r^\alpha_l(q) =
\Gamma(k+1)\Gamma(\alpha+k+1)\delta_{kl}\;. \label{L1}
\end{equation}

These facts in hand, we can evaluate
\begin{eqnarray}
\overline{Q} &=&
\int\Delta(\mathbf{q})^2\prod_{k=1}^\mu e^{-q_k}q_k^\alpha\,dq_k \nonumber\\
&=&\mathop{\sum_{i_1,\dots,i_\mu}}_{j_1,\dots,j_\mu}
\epsilon_{i_1\dots i_\mu}\epsilon_{j_1\dots j_\mu}
\prod_{k=1}^\mu\int dq_k\,e^{-q_k}q_k^\alpha
r^\alpha_{i_k-1}(q_{k})r^\alpha_{j_k-1}(q_k) \nonumber \\
&=&\sum_{i_1,\dots,i_\mu}\epsilon_{i_1\dots i_\mu}^2
\prod_{k=1}^\mu\Gamma(i_k)\Gamma(\alpha+i_k) \nonumber \\
&=&\mu!\prod_{k=1}^\mu\Gamma(k)\Gamma(\alpha+k)\;.
\end{eqnarray}

        For $\mu=\nu,\; \alpha = 0$ and we can simplify the
algebra considerably. Under these conditions,
    \be
\mathop{\sum_{i,j=1}}_{i\neq j}^\mu \int \sqrt{q_i q_j}
Q(\mathbf{q})\,\mathrm{d}\mathbf{q} = \overline{Q}
\sum_{k,l=0}^{\mu-1}\left[
I_{kk}^{(1/2)}I_{ll}^{(1/2)}-\left(I_{kl}^{(1/2)}\right)^2\right]\;,
\label{S2}
    \ee
where
    \be
I_{kl}^{(1/2)} \equiv
\int_0^{\infty}e^{-q}\sqrt{q}\,L_k(q)L_l(q)\;\mathrm{d}q.
    \ee
We thus have
    \be
 \label{E:avgneg}
\avg{\mathcal{N}}=1 + \frac{1}{\mu^2}\sum_{k,l=0}^{\mu-1}\left[
I_{kk}^{(1/2)}I_{ll}^{(1/2)}-\left(I_{kl}^{(1/2)}\right)^2\right]\;,
    \ee
except that the integral needs to be evaluated. For that, we use
the generating function for Laguerre polynomials\footnote{I thank
C. Chandler for reminding me this trick.} \cite{gradshteyn}
    \be
 (1-z)^{-1} e^{xz/z-1} = \sum_{l=0}^{\infty}
 L_l(x)z^l\;\;\;\;\;\;\;\;  |z|\leq1,
    \ee
and
    \be
\int_0^{\infty}e^{-st}t^{\beta}\,L_n^{\alpha}(t)\;\mathrm{d}t=
\frac{\Gamma(\beta+1)\,\Gamma(\alpha+n+1)}{n!\,\Gamma(\alpha+1)}s^{-\beta-1}F\left(-n,\beta+1;\alpha+1,\frac{1}{s}\right),
    \ee
$F$ being the hypergeometric function such that
    \be
F(a,b;c;z)= \sum_{n=0}^{\infty}
\frac{(a)_n\,(b)_n}{(c)_n}\frac{z^n}{n!},
    \ee
and $(a)_n = a(a+1)(a+2)...(a+n-1)$ is the Pochhammer symbol. Note
that if $a$ is a negative integer, $(a)_n = 0$ for $n > |a|$ and
the hypergeometric series terminates. Then,
    \ben
\sum_{l=0}^{\infty} I_{kl}^{(1/2)} z^l
&=&\int_0^{\infty}e^{-x}\sqrt{x}\,L_k(x)(1-z)^{-1}
e^{xz/z-1}\;\mathrm{d}x  \nonumber\\
&=&
s\int_0^{\infty}e^{-sx}\sqrt{x}\,L_k(x)\;\mathrm{d}x\;\;\;\;\;\;\;\;\;\;\;\;\;\;\;\;\;\;s=1/(1-z) \nonumber\\
&=&s
\left[\frac{\Gamma(3/2)\,\Gamma(k+1)}{k!\,\Gamma(1)}\right]s^{-3/2}F\left(-k,\frac{3}{2};1;\frac{1}{s}\right)\nonumber\\
&=&\frac{\sqrt{\pi}}{2}\sum_{t=0}^{k}\frac{(-k)_t\,(3/2)_t}{(1)_t}\frac{1}{t!}(1-z)^{t+1/2}\nonumber\\
&=&\frac{\sqrt{\pi}}{2}\sum_{l=0}^{\infty}\sum_{t=0}^{k}\frac{(-1)^l}{l!}\frac{(-k)_t\,(3/2)_t}{(t!)^2}(t+\frac{1}{2})_{\underline{l}}\,z^l,
    \een
whereby
    \be
I_{kl}^{(1/2)}=\frac{\sqrt{\pi}}{2}\frac{(-1)^l}{l!}\sum_{t=0}^{k}\frac{(-k)_t\,(3/2)_t}{(t!)^2}(t+\frac{1}{2})_{\underline{l}}\;,
    \ee
and $(a)_{\underline{n}}=a(a-1)(a-2)...(a-n+1)$ is the `falling
factorial'. Using the following identities for the Pochhammer
symbols
    \bes
    \ben
(x)_{\underline{n}}&=&(-1)^n (-x)_n,\\
(-x)_{n}&=& (-1)^n(x-n+1)_n, \\
(x)_n  &=& \Gamma(x+n)/\Gamma(x),
    \een
    \ees
we have
   \ben
 \label{E:Ikl}
 I_{kl}^{(1/2)}&=&\frac{(-1)^l}{l!}\sum_{t=0}^{k}
\left(\begin{array}{c}
  k \\
  t \\
\end{array}
   \right)\frac{[\Gamma(t+3/2)]^2}{t!\,\Gamma(t-l+3/2)}\nonumber\\
&=&\frac{\pi}{4}\frac{(-1)^l}{\Gamma[\frac{3}{2}-l] l!} \;
_3F_2\left(\frac{3}{2},\frac{3}{2},-k;1,\frac{3}{2}-l;1\right).
 \een

\begin{figure}[!h]
\begin{center}
\includegraphics{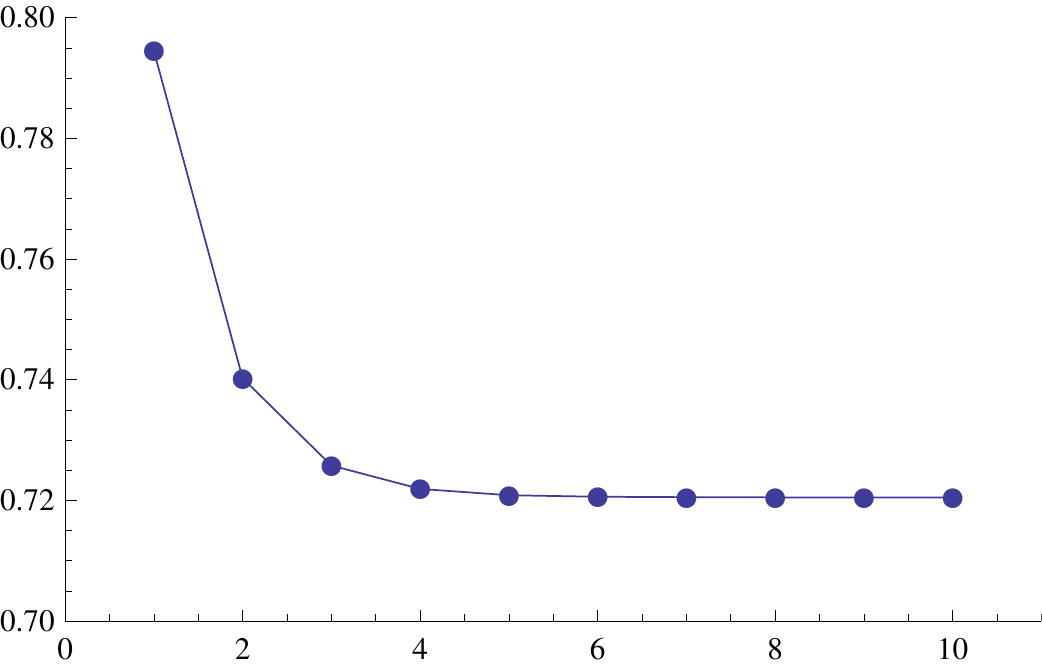}
\caption[Negativity of random pure states]{The value of the
normalized negativity $\avg{\mathcal{N}}/2^{n/2}$ of a random
(Haar-distributed) pure state}
 \label{negativitylimit}
\end{center}
\end{figure}

        To get the final expression for the negativity in
Eq~\ref{E:avgneg}, we substitute the expression for the integrals
from Eq~\ref{E:Ikl}. The expressions are not very illuminating,
and for the lack of an asymptotic expression, we plot the
numerical value in Fig~\ref{negativitylimit}. As we see, the
negativity also scales exponentially with the system size in the
asymptotic limit. In fact, it goes as a constant multiple
($0.720507$) of the maximum possible negativity, just as the
Schmidt rank of the DQC1 circuit that has a random unitary in it.


\end{document}